\documentclass[fleqn,usenatbib]{mnras}

\usepackage{newtxtext,newtxmath}

\usepackage[T1]{fontenc}
\usepackage{ae,aecompl}

\usepackage{graphicx}	
\usepackage{amsmath}	
\usepackage{amssymb}	
\usepackage{verbatim}

\title[Galactic potential constraints from streams]{Galactic potential constraints from clustering in action space of combined stellar stream data}

\author[Reino et al.]{
Stella Reino,$^{1}$
Elena M. Rossi,$^{1}$
Robyn E. Sanderson,$^{2,3}$
Elena Sellentin,$^{1}$
\newauthor \ Amina Helmi,$^{4}$
Helmer H. Koppelman,$^{4}$
Sanjib Sharma$^{5}$
\\
$^{1}$Leiden Observatory, Leiden University, Niels Bohrweg 2, 2333 CA Leiden, The Netherlands\\
$^{2}$Department of Physics and Astronomy, University of Pennsylvania, 209 S 33rd St., Philadelphia, PA 19104, USA\\
$^{3}$Center for Computational Astrophysics, Flatiron Institute, 162 5th Ave., New York, NY 10010, USA\\
$^{4}$Kapteyn Astronomical Institute, University of Groningen, P.O. Box 800, 9700 AV Groningen, The
Netherlands\\
$^{5}$Sydney Institute for Astronomy, School of Physics, The University of Sydney, NSW 2006, Australia
}

\date{Accepted 2021 January 29. Received 2021 January 29; in original form 2020 June 29}

\pubyear{2019}

\begin{document}
\label{firstpage}
\pagerange{\pageref{firstpage}--\pageref{lastpage}}
\maketitle

\begin{abstract}
Stream stars removed by tides from their progenitor satellite galaxy or globular cluster act as a group of test particles on neighboring orbits, probing the gravitational field of the Milky Way. While constraints from individual streams have been shown to be susceptible to biases, combining several streams from orbits with various distances reduces these biases. We fit a common gravitational potential to multiple stellar streams simultaneously by maximizing the clustering of the stream stars in action space. We apply this technique to members of the GD-1, Pal 5, Orphan and Helmi streams, exploiting both the individual and combined data sets. We describe the Galactic potential with a St\"ackel model, and vary up to five parameters simultaneously. We find that we can only constrain the enclosed mass, and that the strongest constraints come from the GD-1, Pal 5 and Orphan streams whose combined data set yields $M(< 20\ \mathrm{kpc}) = 2.96^{+0.25}_{-0.26} \times 10^{11} \ M_{\odot}$. When including the Helmi stream in the data set, the mass uncertainty increases to $M(< 20\ \mathrm{kpc}) = 3.12^{+3.21}_{-0.46} \times 10^{11} \ M_{\odot}$.

\end{abstract}

\begin{keywords}
dark matter, Galaxy: kinematics and dynamics, Galaxy: structure, Galaxy: fundamental parameters, methods: numerical
\end{keywords}



\section{Introduction}

The outer reaches of the Milky Way, known as the ``halo'', are dominated by dark matter. Knowledge of the mass and shape of the halo is required for placing strong constraints on the formation history of the Milky Way, testing the nature of dark matter and modified gravity models \citep[e.g.][]{Mao2015, Thomas2018}.
Some of the most promising dynamical tracers of the Galactic potential in the halo region are stellar steams. Stellar streams form when stars are torn from globular clusters or dwarf galaxies due to Galactic tidal forces. The stars in the ensuing debris gradually stretch out in a series of neighboring orbits. This property makes stellar streams superb probes of the underlying gravitational potential, allowing us to constrain the mass distribution within the extent of their orbits \citep{Johnston1999}. In addition, density variations and gaps within a stream can potentially provide information about past encounters with small-scale substructure and therefore an opportunity to detect the presence of dark matter subhaloes \citep{Carlberg2012, Sanders2016b, Erkal2017, Bonaca2019, Banik2019, Bonaca2020}. 

The first detections of streams included the discovery of the tidally distorted Sagittarius Dwarf Galaxy by \citet{Ibata1994}, the tidal tails around multiple globular clusters by \citet{Grillmair1995} and the Helmi streams by \citet{Helmi1999}. Since then, the number of known streams has grown rapidly owing to the high-quality data from wide-field surveys. The first surge in discoveries came with the arrival of the Sloan Digital Sky Survey, where among others, the GD-1 \citep{GD2006}, Orphan \citep{Grillmair2006, Belokurov2006}, Palomar 5 \citep{Odenkirchen2001} and NGC 5466 streams \citep{GJ2006} were found. More discoveries from other surveys, such as PAndAS, Pan-STARRS1 and the Dark Energy Survey, followed \citep{Martin2014, Bernard2014, Koposov2014, Bernard2016, Shipp2018}. 

Despite the abundance of known streams \citep[see e.g.][]{NewbergCarlin2016, Mateu2018}, full six-dimensional phase space maps of stream members, crucial for obtaining accurate constraints on the Galactic potential, have only been made for a few cases. Recently, the second data release of Gaia \citep[Gaia DR2,][]{Gaia2018} expanded our ability to make such maps by several orders of magnitude, by measuring proper motions for more than a billion Milky Way stars. This phenomenal wealth of data has already facilitated the discovery of many new streams \citep{Malhan2018, Ibata2019, Meingast2019} and prompted further investigations of the previously known ones \citep{PW2018, PW2019, Koposov2019, Koppelman2019}. To gain a full 6D view of more distant streams, Gaia data must be combined with radial-velocity measurements for faint stars from current or future wide-field spectroscopic surveys such as RAVE \citep{RAVE2017}, $S^5$ \citep{S52019}, WEAVE \citep{weave2012}, 4MOST \citep{deJong2019}, SDSS-V \citep{Kollmeier2017} etc. 

Perhaps the most intuitive approach for constraining the Galactic potential with stellar streams is the orbit-fitting technique, where orbits integrated in different potentials are compared with the tracks of observed streams \citep[e.g.][]{Koposov2010, Newberg2010}. However, the oversimplification that streams perfectly follow the original progenitor's orbit has been shown to lead to systematic biases when used to constrain the Galactic potential \citep{SandersBinney2013a}. More realistic stream modelling involves creating either full N-body simulations of disruptions of stellar clusters (the most accurate but also most computationally expensive option) or particle-spray models, where the stream is created by ejecting stars from the Lagrange points of an analytical model of the progenitor at specific times \citep{Bonaca2014, Kupper2015, Erkal2019}.

All these methods compare models to observed streams in 6-dimensional phase space, or some subset of measured positions and velocities. It is, however, possible to simplify the behaviour of streams considerably by switching to action-angle coordinates \citep{McMillanBinney2008,SandersBinney2013b, Bovy2016}. In this work we follow the action-space clustering method of \cite{Sanderson2015}. Actions are integrals of motion that, save for orbital phase, completely define the orbit of a star bound in a static or adiabatically time-evolving potential. Converting the 6-dimensional phase-space position of a star to action space essentially compresses the entire orbit of a star to just three numbers. Stream stars move along similar orbits and thus should cluster tightly in action space. However, since action calculation requires knowledge of the Galactic potential, clustering occurs only if the actions are calculated  with something close to the true potential \citep{Penarrubia2012, Magorrian2014, Yang2020}. Therefore, our strategy to find the true Galactic potential is to identify the potential that produces the most clumpy distribution of stars in action space.

The method outlined in \cite{Sanderson2015} quantifies the degree of action space clustering with the Kullback-Leibler divergence, which is used to determine both the best-fit potential and its associated uncertainties. In that work we demonstrated the effectiveness of this procedure by successfully recovering the input parameters of a potential using mock streams evolved in that potential. Here, we apply the same technique to real stellar streams.

Stellar streams can possess a variety of morphologies: some appear as long narrow arcs, others shells, while still others are fully phase-mixed and no longer easily distinguishable as single structures \citep{Hendel2015, Amorisco2015}. However, even if phase-mixing has induced a lack of apparent spatial features, the fact that these stars still follow similar orbits causes them to condense into a single cluster in action-space. Our method is thus applicable to streams in any evolutionary stage. 

Another important advantage of this technique is that it can be applied to multiple streams simultaneously. Combining multiple streams is crucial since it helps counteract the biases to which single-stream fits have been shown to be susceptible \citep{Bonaca2014}. Single-stream fits that account for only statistical uncertainties are severely limited by systematics, both in the insufficiency of the potential model (compare for example the \citet{LawMajewski2010} and \citet{VeraCiroHelmi2013} fits to the Sagittarius stream in the era before Gaia) and in the limited range of orbits explored. Only \emph{simultaneous} fitting of multiple streams can begin to probe the extent and nature of these systematic uncertainties by consolidating several independent measurements of the mass profile over a range of Galactic distances.

This paper is organised as follows. In Section \ref{sec:method}, we explain the theoretical background and the details of our procedure. In particular, the St\"ackel potential used to model the Milky Way is introduced in Section \ref{sec:St\"ackel}, the calculation of actions from the observed phase space is described in Section \ref{sec:calc_actions}, and Sections \ref{sec:kld1} and \ref{sec:kld2} discuss how we determine the best-fit potential (see also Appendix \ref{app:B}) and confidence intervals, respectively. In Section \ref{sec:catalogue}, we introduce the four streams in our sample, giving a brief overview of their properties and an outline of our data sets (for more detail see Appendix \ref{app:A}). Our results, for both individual and combined streams, are presented in Section \ref{sec:results1} for the one-component model and in Section \ref{sec:5par} for the two-component model. Section \ref{sec:validation} is dedicated to validating our results and presenting the predicted orbits. In Section \ref{sec:discussion}, we compare our results with other potential models of the Milky Way and discuss the implications, and in Section \ref{sec:conclusions} we summarize our main conclusions.

\section{Method}
\label{sec:method}

To constrain the Milky Way's gravitational potential, we exploit the idea that stream stars' action-space distributions bear the memory of their progenitor's orbit. We describe the Galactic potential with a one- or two-component St\"ackel model (\S\ref{sec:St\"ackel}), converting the observed phase space coordinates of each star in our sample into action space coordinates (\S\ref{sec:calc_actions}) for a wide, \emph{astrophysically motivated} range of St\"ackel potential parameters. The model for the potential that produces the most clumped configuration of actions is selected as the best-fit potential (\S\ref{sec:kld1}). 
 Once the best-fit potential is identified, its confidence intervals are determined by quantifying the relative difference between the action distribution in the best-fit potential and those of all other considered potentials (\S\ref{sec:kld2}). 

\subsection{St\"ackel potential}
\label{sec:St\"ackel}

While there exist algorithms to estimate approximate actions for any gravitational potential \cite[see review by][]{Sanders2016}, analytical transformation from phase space coordinates to action-angle coordinates is possible only for a small set of potentials. The best suited of these to describe a real galaxy is the axisymmetric St\"ackel model \citep{Batsleer1994,deZeeuw1985}. This work exploits potentials of the St\"ackel form, enabling us to explore the relevant parameter space efficiently. In addition, using a potential with analytic actions avoids introducing additional numerical errors from action estimation, which are a function of the actions themselves, and are several orders of magnitude higher for radial than circular orbits \citep{Vasiliev2019}.

The Hamilton-Jacobi equation is a formalization of classical mechanics used for solving the equations of motion of mechanical systems \citep[see e.g.][]{Goldstein1950}. The St\"ackel potential, when expressed in ellipsoidal coordinates, allows the Hamilton-Jacobi equation to be solved by the separation of variables and therefore the actions to be calculated analytically. Here, we describe the St\"ackel potential using spheroidal coordinates: the limiting case of ellipsoidal coordinates that is used to describe an axisymmetric density distribution.

The transformation from cylindrical coordinates $R,\ z,\ \phi$ to spheroidal coordinates $\lambda,\ \nu, \ \phi$ is achieved using the equation
\begin{equation}
\label{eq:transformation}
\frac{R^2}{\tau -a^2} + \frac{z^2}{\tau -c^2} = 1 \ ,
\end{equation}
where $\tau = \lambda, \nu$. Hence, this is a quadratic equation for $\tau$ with roots $\lambda$ and $\nu$.
Parameters $a$ and $c$, which can be interpreted as the scale lengths on the equatorial and meridional planes, respectively, define the location of the foci $\Delta = \sqrt{a^2 - c^2}$ and therefore the shape of the coordinate system. We also define the axis ratio of the coordinate surfaces, $e \equiv \frac{a}{c}$.

An oblate density distribution has $a > c$, while a prolate density distribution has $a < c$. Further details about this coordinate system can be found in \cite{deZeeuw1985} and \cite{DZ1988}.
The St\"ackel potential, $\Phi$, in spheroidal coordinates has the form
\begin{equation}
\begin{aligned}
\label{eq:St\"ackel}
&\Phi (\lambda, \nu) = - \frac{f(\lambda) - f(\nu)}{\lambda -\nu} \ ,\\
&f(\tau) = (\tau - c^2) \mathcal{G} (\tau) \ ,
\end{aligned}
\end{equation}
where $\mathcal{G}(\tau)$ is the potential in the $z = 0$ plane, defined as
\begin{equation}
\mathcal{G} (\tau) = \frac{GM_{\mathrm{tot}}}{\sqrt{\tau} + c} \ ,
\end{equation}
with $M_{\mathrm{tot}}$ the total mass and G the gravitational constant.
Putting these elements together, we get 
\begin{equation}
\Phi (\lambda, \nu) = - \frac{GM_{\mathrm{tot}}}{\sqrt{\lambda} + \sqrt{\nu}} \ .
\end{equation}

It is possible to combine two St\"ackel potentials for a more realistic model of the Galaxy \citep{Batsleer1994}. In this case, we have two components in the full potential, $\Phi_{\mathrm{outer}}$ and $\Phi_{\mathrm{inner}}$, each following Equation~\ref{eq:St\"ackel}. In \citet{Batsleer1994} the inner component is intended to represent the disc, while the outer component is associated with the halo. However, for our work the individual components are not intended to represent specific structures of the Milky Way; their purpose is simply to add more flexibility to our model. 

The two components have different axis ratios and scale radii, defined by parameters $a_{\mathrm{outer}}$, $c_{\mathrm{outer}}$ and $a_{\mathrm{inner}}$, $c_{\mathrm{inner}}$. For the overall potential to retain the St\"ackel form (as defined by Equation~\ref{eq:St\"ackel}), and hence the separability of the Hamilton-Jacobi equation, the two components must share the same foci and therefore the coordinates must be related by
\begin{equation}
\lambda_{\mathrm{outer}} - \lambda_{\mathrm{inner}} = \nu_{\mathrm{outer}} - \nu_{\mathrm{inner}} = q \ , 
\end{equation}
and the parameters of the the two components' coordinate systems have to be linked by
\begin{equation}
\label{eq:coordlinks}
a_{\mathrm{outer}}^2 - a_{\mathrm{inner}}^2 = c_{\mathrm{outer}}^2 - c_{\mathrm{inner}}^2 = q \ ,
\end{equation}
where $q$ is a constant.
The total potential is then
\begin{equation}
\begin{aligned}
&\Phi (\lambda_{\mathrm{outer}}, \nu_{\mathrm{outer}}, q) =  \\
&-GM_{\mathrm{tot}}\Bigg[ \frac{1-k}{\sqrt{\lambda_{\mathrm{outer}}} + \sqrt{\nu_{\mathrm{outer}}}} + \frac{k}{\sqrt{\lambda_{\mathrm{outer}}-q} + \sqrt{\nu_{\mathrm{outer}} -q}}\Bigg]
\end{aligned}
\end{equation}
where $k$ is the ratio of the inner component mass to the outer component mass, and $M_{\rm tot}$ is the sum of the two component masses.

We set $q>0$, so that the scale of the outer component is larger than the scale of the inner component. We restrict our model to potentials where the inner component has an oblate shape, i.e. $e_{\mathrm{inner}} > 1$, suitable for the inner regions of the Milky Way. In addition, we require the inner component to be flatter than the outer component by restricting $e_{\mathrm{inner}}>e_{\mathrm{outer}}$. 
These choices force the outer component also to have $e_{\mathrm{outer}} > 1$, meaning the overall model is limited to quasi-spherical and oblate shapes. The cause for this final restriction becomes evident when we write down $q$ in the following from:
\begin{equation}
q = \frac{c_{\mathrm{inner}}^2(e_{\mathrm{inner}}^2 - e_{\mathrm{outer}}^2)}{e_{\mathrm{outer}}^2 - 1}
\end{equation}

Ultimately, the cause for this final restriction comes from the requirement that both the disc potential and the halo potential share the same foci, which means they have to be oriented in the same way.
We first consider a model that consists of a single component represented by a St\"ackel potential. The set of three parameters that defines a particular potential is $\boldsymbol{\zeta} = (M_{\mathrm{tot}}, a, e)$. We select trial potentials by drawing $40$ points from a uniform distribution in log space for each parameter over its prior range: [0.7, 1.8] in $ \log_{10}(a/ \rm kpc)$, [11.5, 12.5] in $\log_{10}(M/M_{\odot})$, and [$\log_{10}(0.5), \log_{10}(2.0)$] in $\log_{10}(e)$. Consequently, there are $40^3$ trial potentials for this model.

Next, we consider two-component St\"ackel potential, defined by a set of five parameters $\boldsymbol{\zeta} = (M_{\mathrm{tot}}, a_{\mathrm{outer}}, e_{\mathrm{outer}}, a_{\mathrm{inner}}, k)$. In this case the parameters are not all independent, but are constrained by Equations (\ref{eq:coordlinks}). Thus, we select the trial potentials by drawing $50$ points for the shape parameters, again from uniform distributions in log space, over the following ranges: [0.7, 1.8] in $ \log_{10}(a_{\mathrm{outer}}/ \rm kpc)$, [$\log_{10}(1.0), \log_{10}(2.0)$] in $\log_{10}(e_{\mathrm{outer}})$ and [$\log_{10}(0.), \log_{10}(0.7)$] in $\log_{10}(a_\mathrm{inner})$. Of these, we only use the ($\sim 8000$) parameter combinations that allow us to construct a mathematically valid potential; i.e., one where both $e_{\mathrm{inner}}$ and $e_{\mathrm{outer}}$ are larger than 1.
In addition, we draw $20$ points for the mass parameters in these parameter ranges: [11.5, 12.5] in $\log_{10}(M/M_{\odot})$ and [$\log_{10}(0.01), \log_{10}(0.3)$] in $\log_{10}(k)$.
For each of these potentials we find the mass enclosed within $r = R^2 + z^2$ by calculating 
\begin{equation}
\begin{aligned}
\label{eq:encmass}
M(< r) = 2 \pi \int^r_0 \int^{\sqrt{r^2 - R^2}}_{-\sqrt{r^2 - R^2}} \rho(R,z)\,R\,dz\,dR \ ,
\end{aligned}
\end{equation}
where the density $\rho (R,z)$ is found through the Poisson equation. Therefore, rather than determining the mass enclosed within an isodensity contour, we integrate the density profile out to a spherical $r$ in order to compare with previous work.

\subsection{Actions}
\label{sec:calc_actions}

In this work, we analyse stellar data in action-angle coordinates.
The actions are integrals of motion that uniquely define a bound stellar orbit and the angles are periodic coordinates that define the phase of the orbit. For a bound, regular orbit \footnote{An orbit for which the angle-action variables exist \citep{BinneyTremaine2008}.}, the actions $J_i$ are related to the coordinates $q_i$ and their conjugate momenta $p_i$ by
\begin{equation}
J_i = \frac{1}{2 \pi} \oint p_i dq_i \ ,
\end{equation}
where the integration is over one full oscillation in $q_i$. 
In the spheroidal coordinate system defined in \S\ref{sec:St\"ackel}, $q_1 = \lambda$, $q_2 = \nu$ and $q_3 = \phi$. The expressions for the actions in the St\"ackel potential are found by solving the Hamilton-Jacobi equation via separation of variables \citep[see e.g.][]{BinneyTremaine2008}. This leads to the definition of three integrals of motion: the total energy $E$, and the actions $I_2$ and $I_3$; and to the equations for the momenta.
The integral of motion $I_2$ is related to the angular momentum in the z-direction,
\begin{equation}
I_2 = \frac{L_z^2}{2} ,
\end{equation}
while $I_3$ can be seen as a generalization of $L - L_z$ \citep{DZ1988},
\begin{equation}
I_3 = \frac{1}{2} (L_x^2 + L_y^2) + (a^2 - c^2) \Big[ \frac{1}{2} v_z^2 - z^2 \frac{\mathcal{G} (\lambda) - \mathcal{G} (\nu)}{\lambda - \nu} \Big] \ .
\end{equation}
The momenta, $p_{\tau}$, are then expressed as a function of the $\tau$ coordinate and the three integrals of motion:
\begin{equation}
p_{\tau}^2 = \frac{1}{2 (\tau - a^2)} \Big[ \mathcal{G} (\tau)  - \frac{I_2}{\tau - a^2} - \frac{I_3}{\tau - c^2} + E \Big] \ ,
\end{equation}
from which the first two actions can be calculated as
\begin{equation}
\label{eq:st_actions}
J_{\tau} = \frac{1}{2 \pi} \oint p_{\tau} d\tau \ ,
\end{equation}
where the integral is over the full oscillation of the orbit in $\tau$, i.e. the limits are the roots of $p_{\tau}^2$.
As $p_{\tau}$ is only a function of $\tau$ and the three integrals of motion, it follows that $J_{\tau}$ is also an integral of motion. The third action, $J_{\phi}$, is equal to $L_z$ and therefore independent of the particular axisymmetric potential.

We calculate the actions for all stars in our sample for each trial potential. 
From the observed sky positions and proper motions in Gaia DR2, cross-matched with distance and radial velocity estimators from the various sources discussed in \S\ref{sec:catalogue}, we derive the Galactocentric phase-space coordinates $\boldsymbol{\omega} = (\boldsymbol{x}, \boldsymbol{v})$, where $\boldsymbol{x}$ is the three-dimensional position vector and $\boldsymbol{v}$ is the three-dimensional velocity vector. Details of this transformation are given in Appendix~\ref{app:A}. The phase-space coordinates are then used to calculate $(\tau, p_{\tau})$ and $E$, $I_2$ and $I_3$ for each star in each trial potential. As mentioned before, $J_{\phi} = L_z$ and does not vary from potential to potential. The other two actions, $J_{\lambda}$ and $J_{\nu}$ are found from Equation~\ref{eq:st_actions} by numerical integration. We discuss the influence of measurement errors in \S\ref{sec:validation}.

Some combinations of observed phase-space coordinates and trial potential can result in the star being unbound from the Galaxy, in which case its actions are undefined. In our analysis, we throw out any potential that produces unbound stars, a reasonable assumption given that the stars in our data set are all well within the Galaxy's expected virial radius and have velocities much less than estimates of the escape velocity. We comment on the impact of this choice on our results in \S\ref{sec:validation}.

\subsection{Determination of the best-fit potential}
\label{sec:kld1}

In the previous section, we transformed the phase space coordinates of stream stars to action space coordinates for particular trial potentials. Now, we analyse the resulting action distributions to measure their degree of clustering. We quantify the degree of clustering with the use of the Kullback-Leibler divergence following \cite{Sanderson2015}. The Kullback-Leibler divergence (KLD) measures the difference between two probability distributions $p(\boldsymbol{x})$ and $q(\boldsymbol{x})$ and is defined by
\begin{equation}
\mathrm{KLD} (p \ ||\ q) = \int p(\boldsymbol{x}) \log \frac{p(\boldsymbol{x})}{q(\boldsymbol{x})}\, d^n x \ .
\end{equation}
The larger the difference between the two probability distributions, the larger the value of the associated KLD. If the two distributions are identical the KLD value is 0.

For a discrete sample $[\boldsymbol{x}_i]$ with $\ i = 1,...\, ,N$ drawn from a distribution $p(\boldsymbol{x})$, the KLD can be calculated via Monte Carlo integration as
\begin{equation}
\mathrm{KLD} (p \ ||\ q) \approx \frac{1}{N} \sum_i^N \log \frac{p(\boldsymbol{x}_i)}{q(\boldsymbol{x}_i)}, \quad \mathrm{if} \ q(\boldsymbol{x}_i) \neq 0 \ \forall i.
\end{equation}
We now specify the distribution $q(\boldsymbol{x})$ to be a uniform distribution $u(\boldsymbol{J})$, in the actions. This uniform distribution corresponds to a fully unclustered action space. To test whether a trial potential parameterized by $\boldsymbol{\zeta}$ maps the observed phase space data $\boldsymbol{\omega}$ to a more clustered distribution than $u(\boldsymbol{J})$, we set $p(\boldsymbol{x})$ to $p(\boldsymbol{J} \mid \boldsymbol{\zeta}, \boldsymbol{\omega})$. $p(\boldsymbol{J} \mid \boldsymbol{\zeta}, \boldsymbol{\omega})$ is the probability distribution $p$ of actions $\boldsymbol{J}$, given parameter values $\boldsymbol{\zeta}$ and the phase space coordinates $\boldsymbol{\omega}$.

The Kullback-Leibler divergence is then
\begin{equation}
\label{eq:kld1}
\mathrm{KLD1}(\boldsymbol{\zeta}) = \frac{1}{N} \sum_i^N \log \frac{p(\boldsymbol{J} \mid \boldsymbol{\zeta}, \boldsymbol{\omega})}{u(\boldsymbol{J})}  \biggr\rvert_{\boldsymbol{J} = \boldsymbol{J}_{\zeta}^i} \ ,
\end{equation}
where $N$ is the total number of stars in our sample. $p$ is evaluated at $\boldsymbol{J}_{\zeta}^i = \boldsymbol{J} (\boldsymbol{\zeta}, \boldsymbol{\omega}_i)$, where $\boldsymbol{\omega}_i$ are the phase space coordinates of star $i$.
The potential closest to the true potential gives rise to the most clumped probability distribution; i.e., the distribution that is the most peaked and therefore most {\em dissimilar} to a uniform distribution. We therefore select as our best-fit potential parameters, $\boldsymbol{\zeta}_0$, the parameters that maximise the Kullback-Leibler divergence across all our trial potentials. This is equivalent to selecting the model that produces the most similar orbits for all stars in a given stream, by exploring all possible star orbits over a range of models given their current phase space coordinates. We label Equation~\ref{eq:kld1} as KLD1 because the identification of the best-fit model is the first step in our procedure.
Practically, we calculate KLD1 using Equation~\ref{eq:kld1} for all potentials that are not discarded for producing unbound stars. We obtain the numerator in Equation~\ref{eq:kld1} by constructing a three-dimensional probability density function $p(\boldsymbol{J} \mid \boldsymbol{\zeta} \,, \boldsymbol{\omega})$ using the Enlink algorithm developed by \cite{Sharma2009}. This algorithm computes a locally adaptive metric by making use of a binary space-partitioning tree scheme, where the partitioning criterion is determined by comparing the Shannon entropy or information along different dimensions. The density is then computed using the Epanechnikov kernel with the smoothing length determined by the given number of nearest neighbors identified by the tree. We use the code's default 10 nearest neighbours as recommended in \cite{Sharma2009}.

The denominator in Equation~\ref{eq:kld1}, $u(\boldsymbol{J})$, is a uniform distribution normalised over the maximum possible range of $\boldsymbol{J}$:
\begin{equation}
u = \Big[ \Big(J_{\lambda}^{\mathrm{max}} -J_{\lambda}^{\mathrm{min}}\Big) \Big(J_{\nu}^{\mathrm{max}} - J_{\nu}^{\mathrm{min}}\Big)\Big(J_{\phi}^{\mathrm{max}} - J_{\phi}^{\mathrm{min}}\Big) \Big]^{-1} \ ,
\end{equation}
where $\boldsymbol{J}^{\mathrm{max}}$ and $\boldsymbol{J}^{\mathrm{min}}$ are the extrema amongst all $\boldsymbol{J}$ calculated for our five-parameter search. This means $u(\boldsymbol{J})$ is constant for all $\boldsymbol{J}$ and all $\boldsymbol{\zeta}$ and as such does not have an impact on maximizing $\mathrm{KLD1}(\boldsymbol{\zeta})$.

The {\em standard} $\mathrm{KLD1}(\boldsymbol{\zeta})$, calculated using Equation~\ref{eq:kld1}, gives equal weight to each of the stars in the sample, and it is suited to cases where our data sample includes either a single stream or multiple streams with unknown stellar membership. However when combined stellar stream data are analysed and star membership {\em is} known, as it is in our case, we can exploit this extra information and modify Equation~\ref{eq:kld1} accordingly. Equation~\ref{eq:kld1} has the implicit property that streams with more stars exert a larger influence on the results compared to streams with fewer stars, since each star contributes equally to the KLD. While this is reasonable when membership is not known a priori, it is not the ideal use of the data since then the largest and hottest streams, which give the least sensitive constraints, dominate over thinner and colder streams with far fewer members. When membership information is available, we can instead introduce a scheme that gives equal weight to all \emph{streams}, by weighting the contribution of each star with:
\begin{equation}
\label{eq:w}
w_{j} = \frac{1}{N_{\rm s}} \times \frac{1}{N_{j}} \ ,
\end{equation}
where $N_{\rm s}$ is the number of streams in our sample and $N_{j}$ is the number of stars in stream ``$j$'' (and therefore $N = \sum_j^{N_{\rm s}} N_j$).
This {\em weighted} $\mathrm{KLD1}(\boldsymbol{\zeta})$ is thus calculated as follows:
\begin{equation}
\label{eq:kld1_w}
\mathrm{wKLD1}(\boldsymbol{\zeta}) = \sum_j^{N_{\rm s}} \sum_i^{N_{j}} \ w_j \log \frac{p(\boldsymbol{J} \mid \boldsymbol{\zeta}, \boldsymbol{\omega})}{u(\boldsymbol{J})} \biggr\rvert_{\boldsymbol{J} = \boldsymbol{J}_{\zeta}^{ij}}\,
\end{equation}
where $\boldsymbol{J}_{\zeta}^{ij} = \boldsymbol{J} (\boldsymbol{\zeta}, \boldsymbol{\omega}_{ij})$, where $\boldsymbol{\omega}_{ij}$ are the phase space coordinates for star i in stream j. 

The KLD works best if there is little overlap between the different streams in action-space. We showed in previous work \citep{Sanderson2015} that the clustering-maximization algorithm will still find a good model potential if the streams overlap, and does not crucially depend on knowing stream membership. However, in our case, stream membership is known, and the performance of the algorithm can be further improved by incorporating this information in the {\em weighted} KLD1 approach. The most straightforward way of doing this is simply to shift each stream by a constant in action space so that they are well-separated, since it is the \emph{clustering}, not the \emph{location} in action space, that drives the fit.\footnote{In fact, in previous papers we used the product of the marginal distributions of $p$ instead of the uniform distribution as the comparison distribution $q$; this form of the KLD is known as the mutual information (MI), and for a multivariate Gaussian it can be shown that the MI depends only on the off-diagonal elements of the covariance matrix -- or in other words, on the correlation between actions.} This tactic also helps avoid the spurious solution achieved by increasing the mass and decreasing the scale radius until all stars are clustered near the origin in action space, which sometimes can dominate over the true consensus fit for severely overlapping streams or in cases with a high fraction of interlopers from the thick disc. For this work we displace the streams from one another in $L_{\rm z}$, since this action is independent of the potential for our axisymmetric model. On the other hand, when performing our analysis with the standard KLD1 (Equation~\ref{eq:kld1}) this shift in action space will {\em not} be applied.

\subsection{Determination of confidence intervals}
\label{sec:kld2}

One interpretation of the KLD is that of an average log-likelihood ratio of a data set. The likelihood ratio,
\begin{equation}
\Lambda = \prod_i^N \frac{p(\boldsymbol{x}_i \mid \boldsymbol{\zeta})}{q(\boldsymbol{x}_i \mid \boldsymbol{\zeta})}, \,
\end{equation}
indicates how much more likely $\boldsymbol{x}$ is to occur under $p(\boldsymbol{x} \mid \boldsymbol{\zeta})$ than under $q(\boldsymbol{x} \mid \boldsymbol{\zeta})$, where we recall that $p(\boldsymbol{x} \mid \boldsymbol{\zeta})$ and $q(\boldsymbol{x} \mid \boldsymbol{\zeta})$ are probability density functions. Therefore, the average log-likelihood ratio for $x_{i}$ is 
\begin{equation}
\langle \log \Lambda \rangle= \frac{1}{N} \sum_i^N \log \frac{p(\boldsymbol{x}_i \mid \boldsymbol{\zeta})}{q(\boldsymbol{x}_i \mid \boldsymbol{\zeta})} \ ,
\end{equation}
which is equivalent to calculating the Kullback-Leibler divergence.
The interpretation of KLD as an average log-likelihood ratio allows us to draw confidence intervals on the best-fit parameters through Bayes' theorem, which states that the posterior probability of a model defined by its parameters $\boldsymbol{\zeta}$, given data $\boldsymbol{x}$, is equal to the likelihood of the data given the model times the prior probability of the model:
\begin{equation}
p(\boldsymbol{\zeta} \mid \boldsymbol{x}) \propto p(\boldsymbol{x} \mid \boldsymbol{\zeta}) p(\boldsymbol{\zeta}) \ .
\end{equation}
This indicates that the ratio of likelihoods is directly linked to the ratio of posterior probabilities:
\begin{equation}
\begin{aligned}
\mathrm{KLD} (p \ ||\ q) &= \frac{1}{N} \sum_i^N \log \frac{p(\boldsymbol{x}_i \mid \boldsymbol{\zeta})}{q(\boldsymbol{x}_i \mid \boldsymbol{\zeta})} \\
&= \frac{1}{N} \sum_i^N \log \frac{p(\boldsymbol{\zeta} \mid \boldsymbol{x}_i)}{q(\boldsymbol{\zeta} \mid \boldsymbol{x}_i)} - \log \frac{p(\boldsymbol{\zeta})}{q(\boldsymbol{\zeta})}
\end{aligned}
\end{equation}
Assuming that the prior distributions are flat or equal, the KLD is thus equal to the expectation value of the log of the ratio of posterior probabilities \citep[for more information, see][]{Kullback1959}.

This leads us to the second step in our procedure, where we compare the action distribution of the best-fit potential, $p(\boldsymbol{J} \mid \boldsymbol{\zeta}_0 \,, \boldsymbol{\omega})$, to the action distributions of the other trial potentials, $p(\boldsymbol{J} \mid \boldsymbol{\zeta}_{\mathrm{trial}} \,, \boldsymbol{\omega})$, by computing
\begin{equation}
\label{eq:kld2}
\mathrm{KLD2}(\boldsymbol{\zeta}) = \frac{1}{N} \sum_i^N \log \frac{p(\boldsymbol{J} \mid \boldsymbol{\zeta}_0 \,, \boldsymbol{\omega})}{p(\boldsymbol{J} \mid \boldsymbol{\zeta}_{\mathrm{trial}}  \,, \boldsymbol{\omega})} \biggr\rvert_{\boldsymbol{J} = \boldsymbol{J}_0^i}\ ,
\end{equation}
where both functions are evaluated at $\boldsymbol{J}_0 = \boldsymbol{J}(\boldsymbol{\zeta}_0 \,, \boldsymbol{\omega})$, i.e. at the actions computed with the best-fit potential parameters $\boldsymbol{\zeta_0}$ and phase space $\boldsymbol{\omega}$. 
In our procedure, we use equation \ref{eq:kld2} to calculate the $\mathrm{KLD2}(\boldsymbol{\zeta})$ for each trial potential. In contrast to the calculation of the $\mathrm{KLD1}(\boldsymbol{\zeta})$, two different sets of actions are used to obtain the probability density functions in the numerator and the denominator: both sets of actions are calculated using the same observed phase space coordinates $\boldsymbol{\omega}$ but two different potentials (the best-fit potential with parameters $\boldsymbol{\zeta}_0$ and another trial potential with parameters $\boldsymbol{\zeta}_{\mathrm{trial}}$). We use Enlink to estimate the probability densities for the two sets of actions $\boldsymbol{J}(\boldsymbol{\zeta}_0 \,, \boldsymbol{\omega})$ and $\boldsymbol{J}(\boldsymbol{\zeta}_{\mathrm{trial}} \,, \boldsymbol{\omega})$.

Analogous to $\mathrm{KLD1}(\boldsymbol{\zeta})$, we introduce an alternative version of $\mathrm{KLD2}(\boldsymbol{\zeta})$ that incorporates weights. The weighted $\mathrm{KLD2}(\boldsymbol{\zeta})$ is defined as follows:
\begin{equation}
\label{eq:kld2_w}
\mathrm{wKLD2}(\boldsymbol{\zeta}) = \sum_i^N \ w_i \log \frac{p(\boldsymbol{J} \mid \boldsymbol{\zeta}_0, \boldsymbol{\omega})}{p(\boldsymbol{J} \mid \boldsymbol{\zeta}_{\mathrm{trial}}, \boldsymbol{\omega})} \biggr\rvert_{\boldsymbol{J} = \boldsymbol{J}_0^i}\ ,
\end{equation}
where the weights are calculated using Equation~\ref{eq:w}.

As discussed above, the $\mathrm{KLD2}(\boldsymbol{\zeta})$ values can be interpreted as the relative probability of parameters $\boldsymbol{\zeta}_0$ and $\boldsymbol{\zeta}_{\mathrm{trial}}$, given the data. The confidence intervals on the best-fit potential are then derived by estimating the value of $\mathrm{KLD2}(\boldsymbol{\zeta})$ at which the posterior distributions become significantly different, i.e. $\mathrm{KLD2}(\boldsymbol{\zeta})$ becomes significantly different from $\mathrm{KLD2}(\boldsymbol{\zeta}_0)=0$.

We begin by assuming that the posterior probability distributions are approximately D-dimensional Gaussians with a covariance matrix $\Sigma$ that is equal to a $D \times D$ identity matrix, where D is the number of free parameters in our model. We calculate $\mathrm{KLD2}(\boldsymbol{\zeta})$ between two of these identical Gaussian distributions placed at different positions: one centred at $\boldsymbol{\zeta}_0$ representing the probability distribution of our best-fit model and the other centred at $\boldsymbol{\zeta}_{\mathrm{trial}}$ representing the probability distribution of a trial model. Since we are interested in determining which of our trial models are within $1\sigma$ of our best-fit model, we use here the limiting case for $\boldsymbol{\zeta}_{\mathrm{trial}}$ where the Gaussian function for the trial model is centred exactly at $1\sigma$ away from $\boldsymbol{\zeta}_0$. 
The $1\sigma$ contour in this situation is a D-dimensional sphere, centered at $\boldsymbol{\zeta}_0$ with radius $1\sigma$. The second Gaussian is then centered on a point on this sphere, i.e. centered anywhere on $\|\boldsymbol{\zeta}_0\| + 1$.

Calculation of the $\mathrm{KLD2}(\boldsymbol{\zeta})$ between these two distributions results in
\begin{equation}
\label{eq:confidence}
\mathrm{KLD2}(\boldsymbol{\zeta}) = \frac{1}{N} \sum_i^N \log \frac{e^{-(r_i - \|\boldsymbol{\zeta}_0\|)^2/2}} {e^{-(r_i - (\|\boldsymbol{\zeta}_0\| + 1))^2/2}} = 0.5 \ , 
\end{equation}
where $r$ are points drawn from the Gaussian centered on $\boldsymbol{\zeta}_0$.
This corresponds to a $1\sigma$ confidence interval. Similarly, the $\mathrm{KLD2}(\boldsymbol{\zeta})$ value that signifies the limiting edge of $2\sigma$ confidence can be calculated using a trial model that is centred at exactly $2\sigma$ away from $\boldsymbol{\zeta}_0$, etc. 

The single parameter $1\sigma$ confidence intervals are drawn as the full range of parameter values in the subset of potentials that are within $1\sigma$ from the best-fit potential, i.e. from the subset of potentials that have  $\mathrm{KLD2}(\boldsymbol{\zeta}) \leq 0.5$ (or $\mathrm{wKLD2}(\boldsymbol{\zeta}) \leq 0.5$ when the weighted case is used).

\begin{figure*}
\centering
\includegraphics[width=0.32\linewidth]{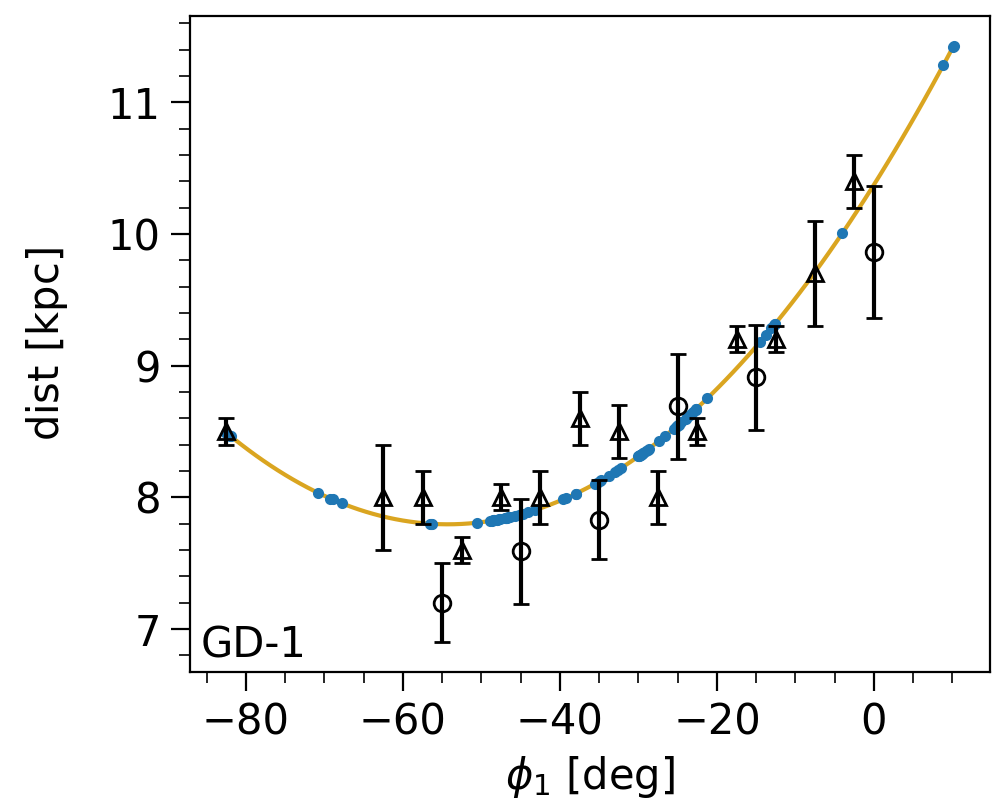}
\includegraphics[width=0.32\linewidth]{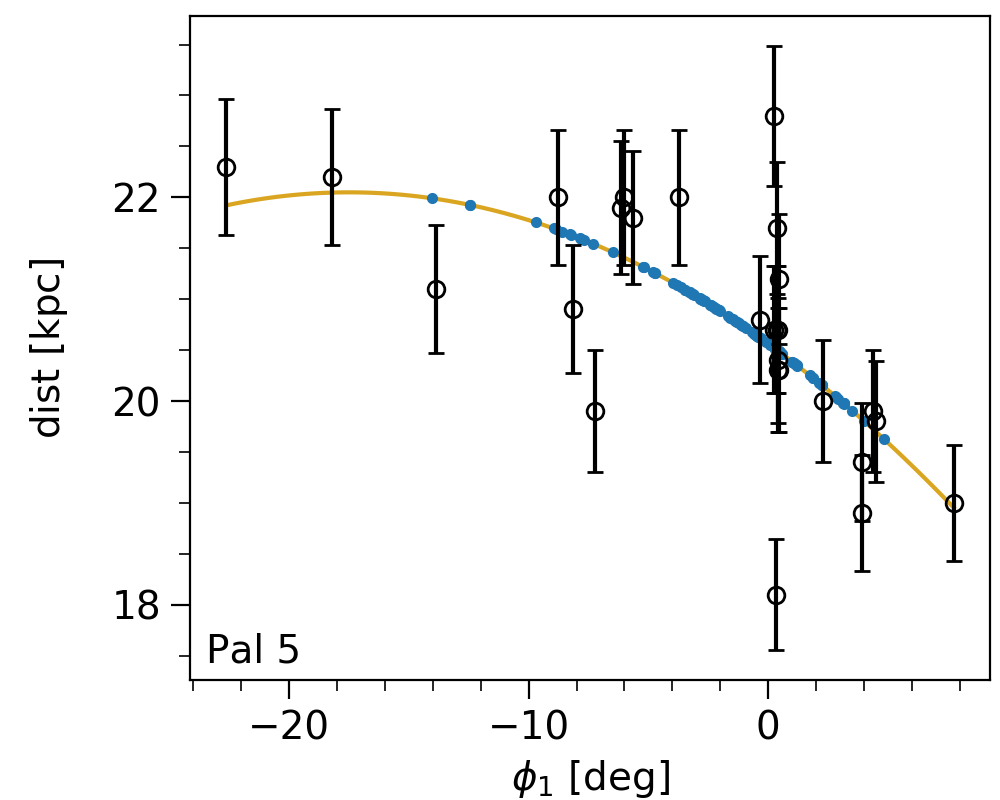}
\includegraphics[width=0.32\linewidth]{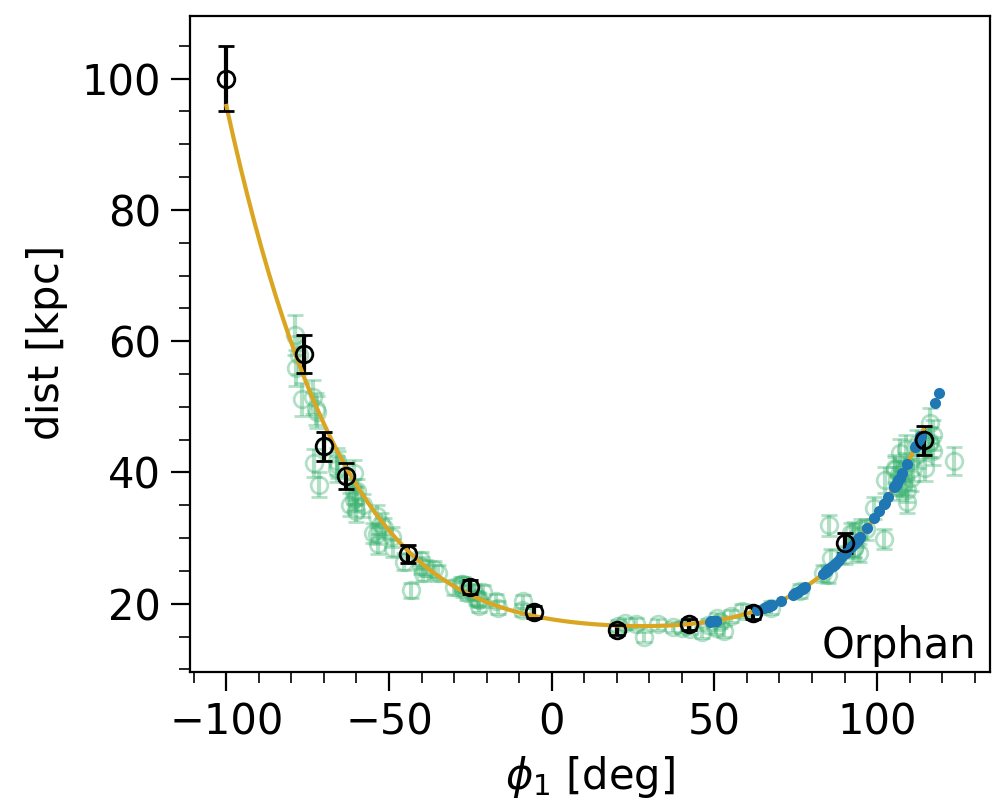}
\, \includegraphics[width=0.32\linewidth]{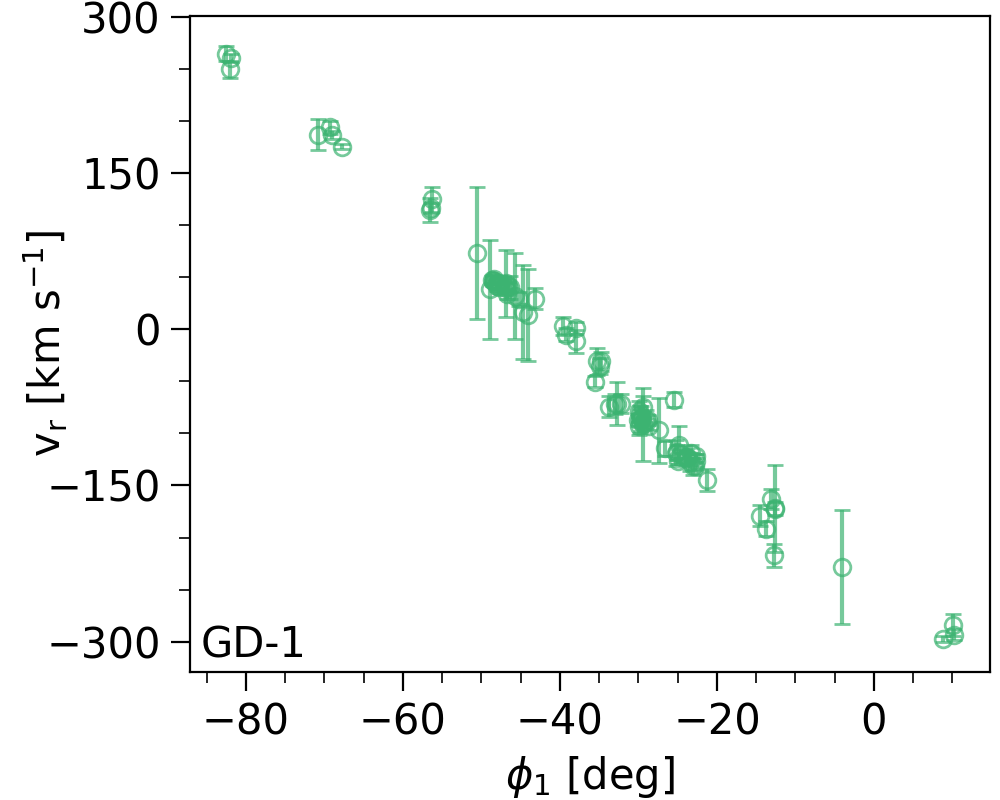}
\includegraphics[width=0.32\linewidth]{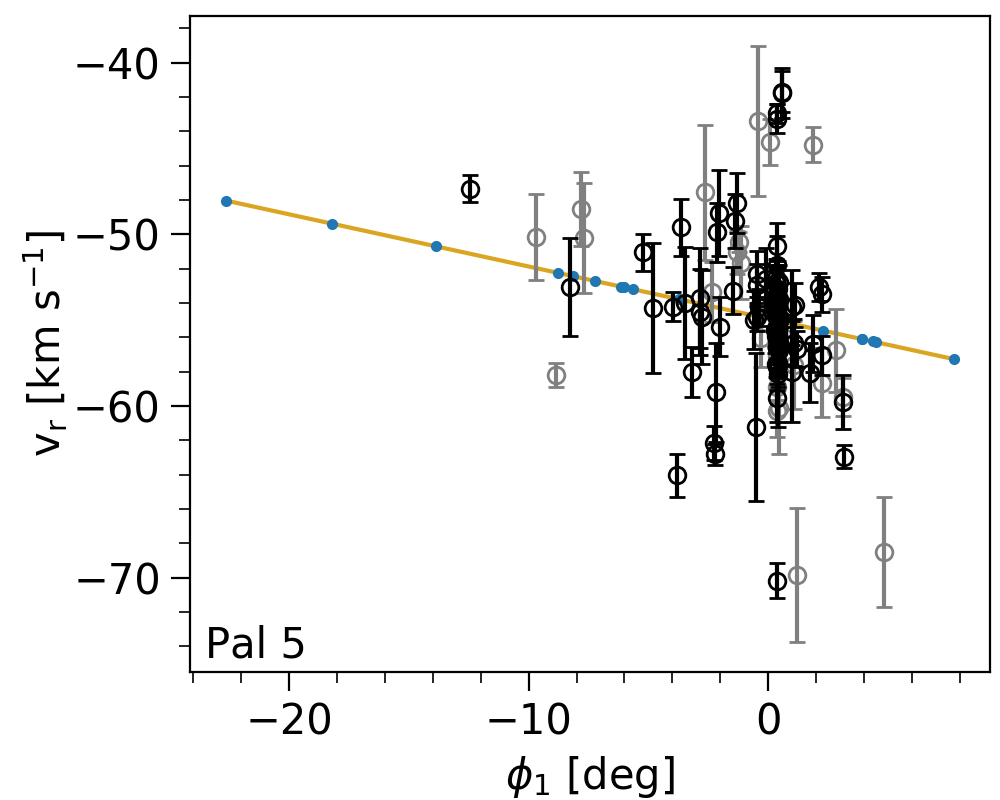}
\includegraphics[width=0.32\linewidth]{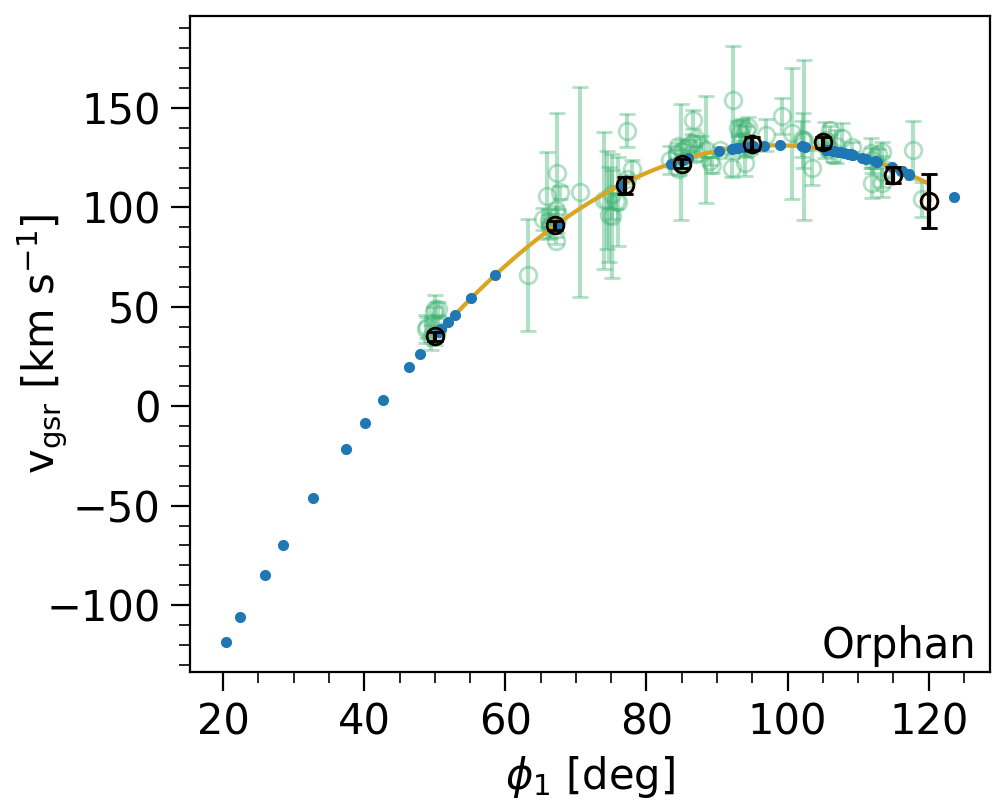}
\, \includegraphics[width=0.32\linewidth]{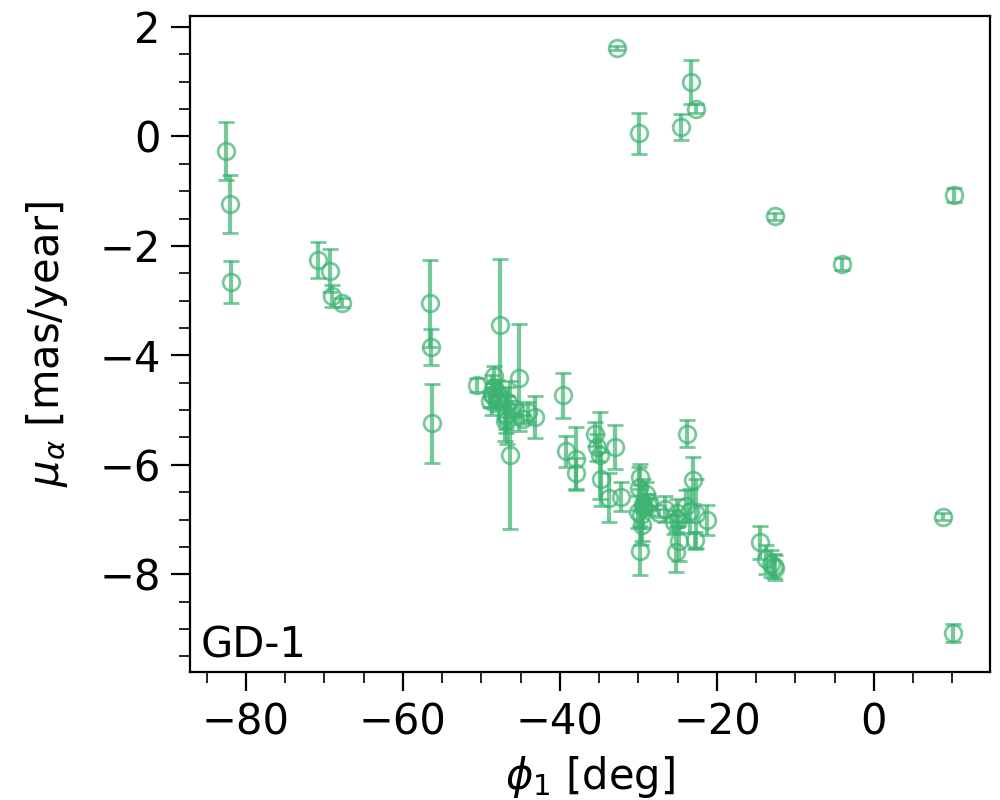}
\includegraphics[width=0.32\linewidth]{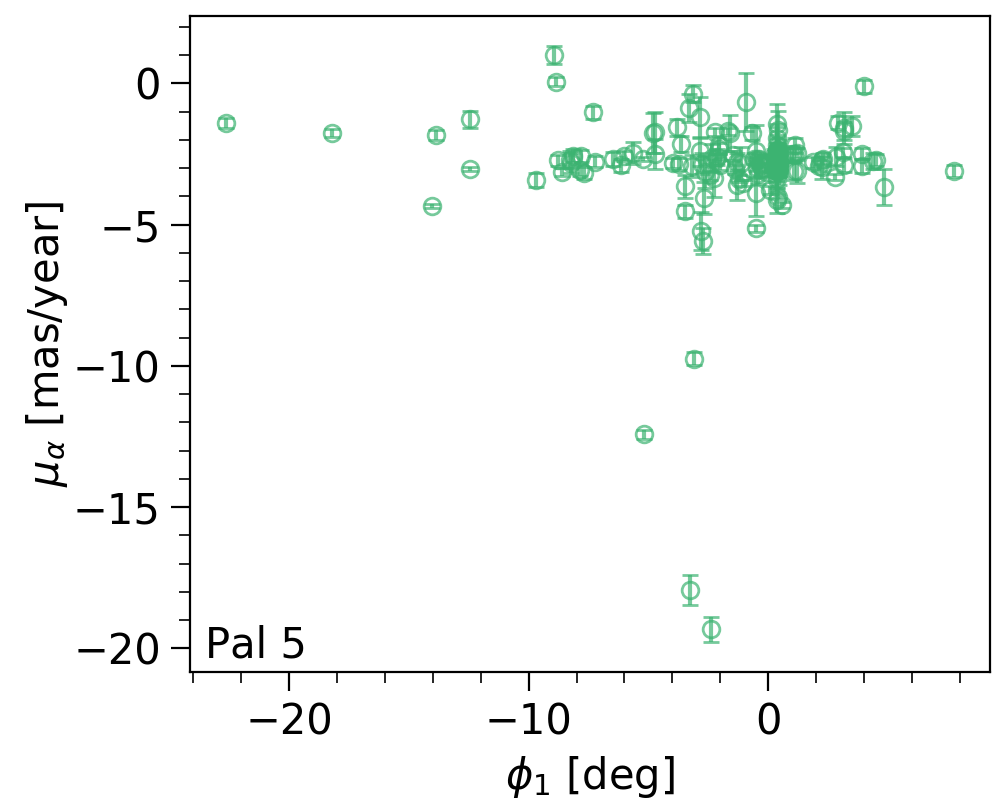}
\includegraphics[width=0.32\linewidth]{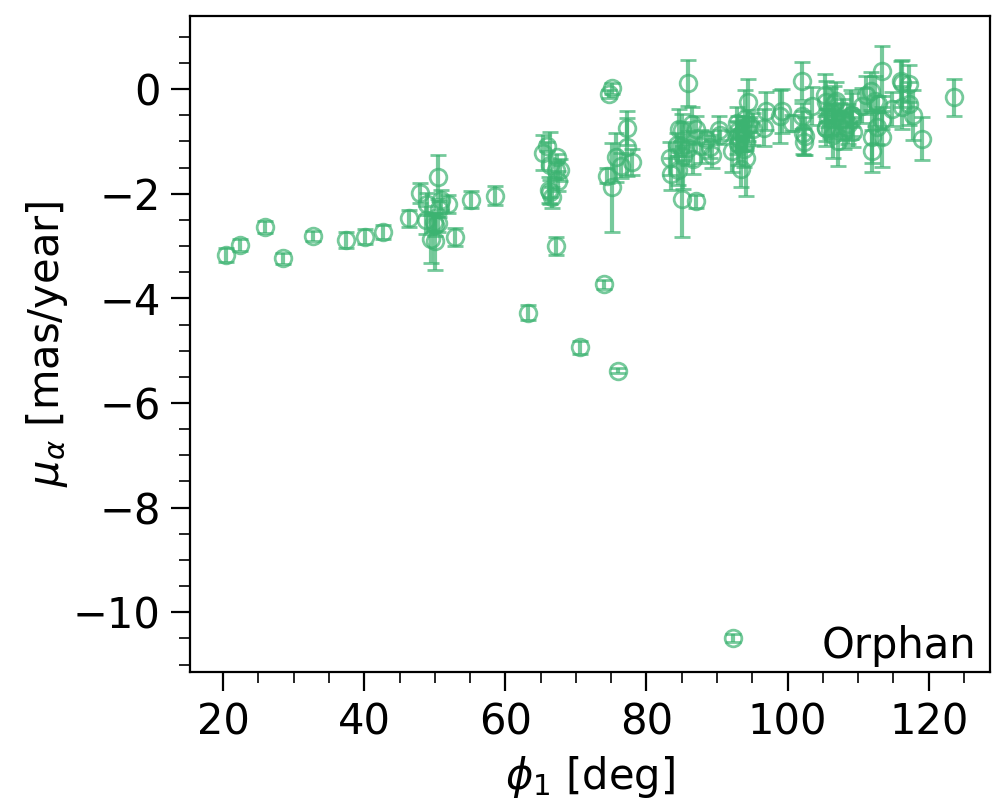}
\, \includegraphics[width=0.32\linewidth]{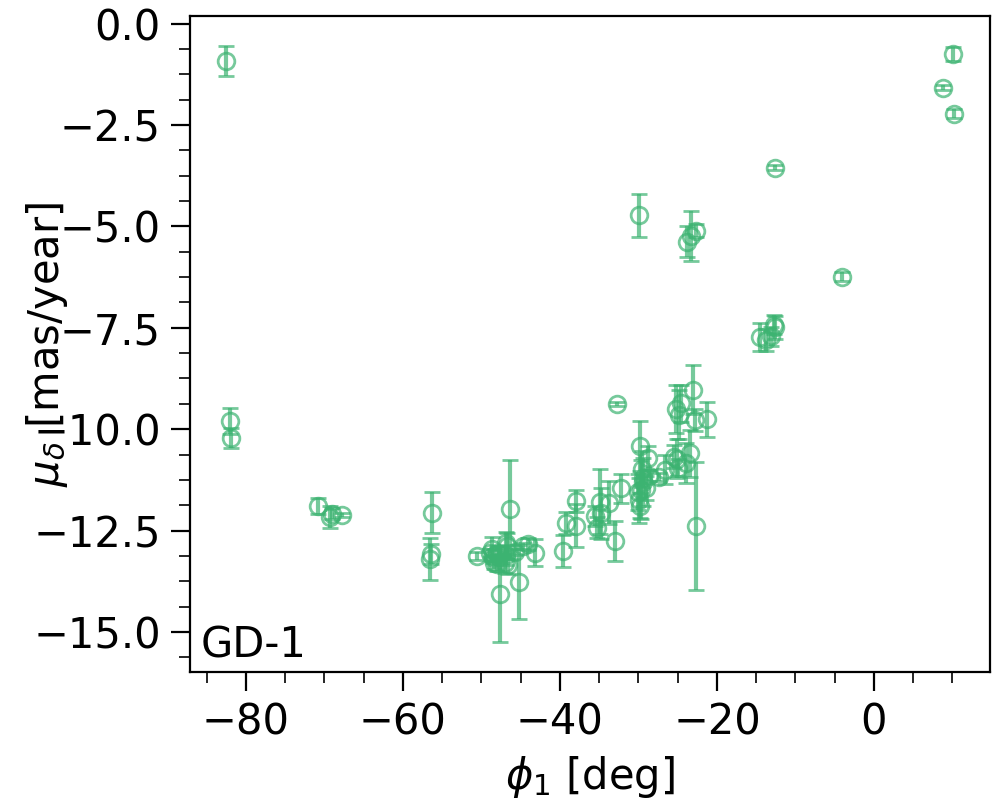}
\includegraphics[width=0.32\linewidth]{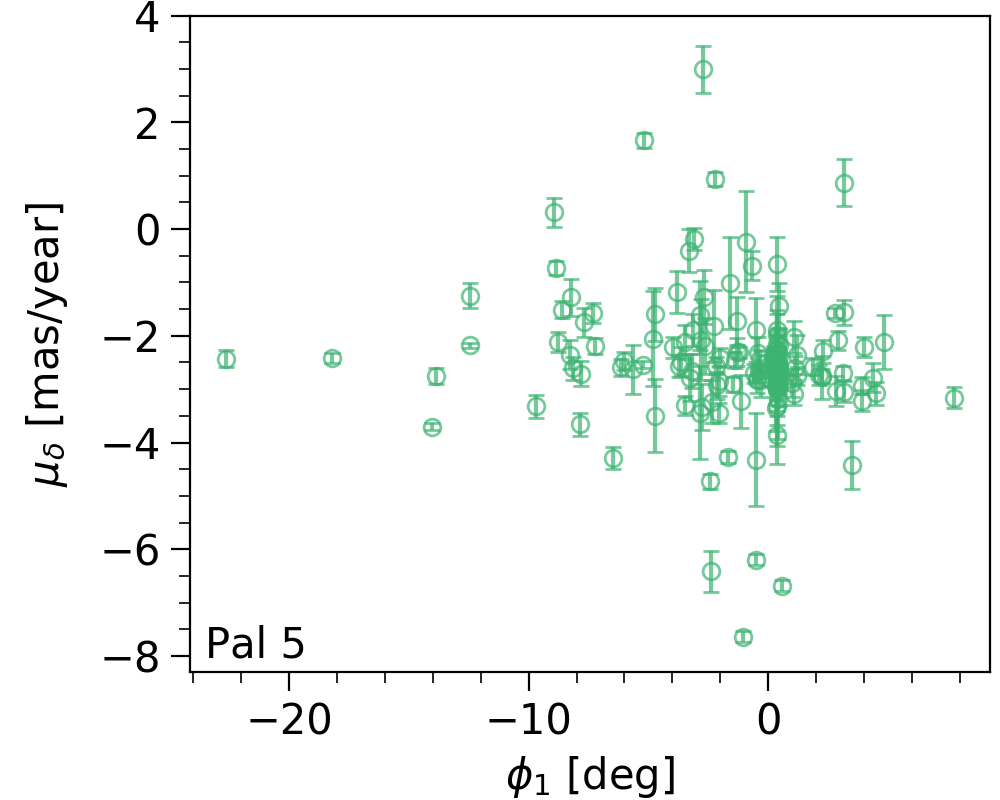}
\includegraphics[width=0.32\linewidth]{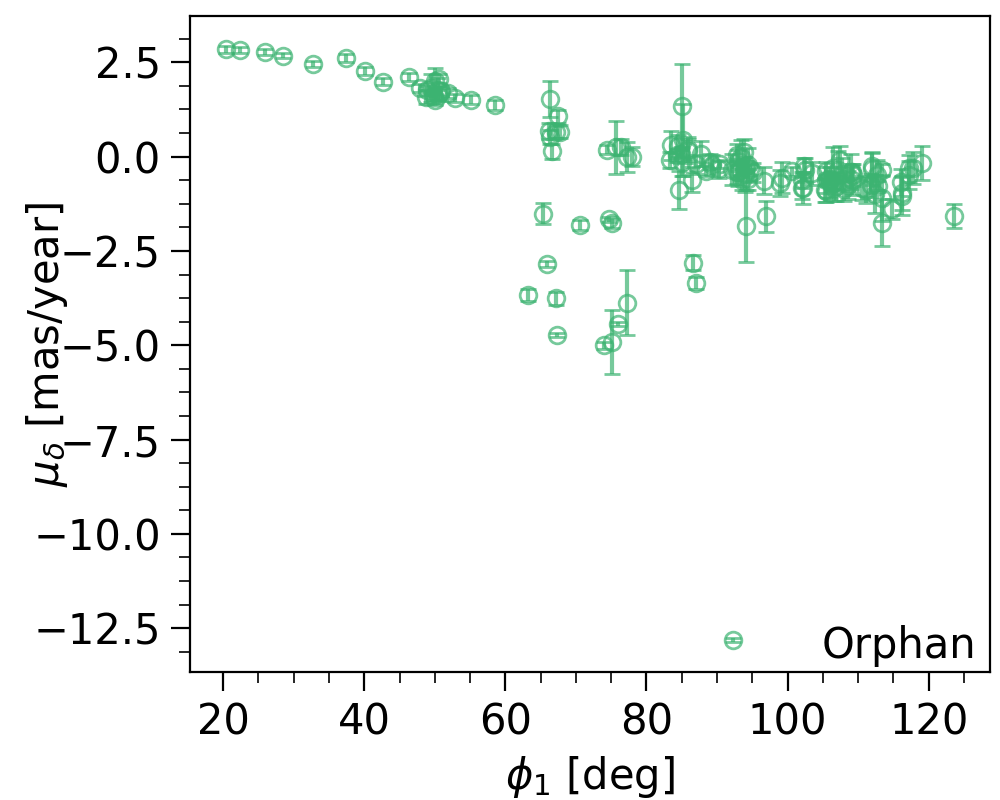}
\caption{Phase-space projections of our uncleaned stream data and fits used to interpolate missing distance and radial velocity (RV) data for stream stars. The coordinate $\phi_1$ on the x-axis is the stream-aligned coordinate for the stream portrayed on each particular panel. Measurements used for the polynomial fits are shown in black ($>0.5$ membership probability) or gray points ($<0.5$ membership probability); measurements for individual stars shown as green points are for comparison only. Polynomial fits are shown in yellow; estimated values we adopt for individual stars are shown as small blue points. Data sources are discussed in detail in Appendix \ref{app:A}. {\bf Top left panel:} fit to distance estimates along the track of GD-1 from \protect\cite{Koposov2010} (circles) and \protect\cite{Li2018} (triangles). {\bf Second row left panel}: RV measurements from \protect\cite{Koposov2010}, \protect\cite{Li2017} and \protect\cite{Willett2009}. {\bf Top centre panel}: fit to distances of individual Pal 5 members from \protect\cite{PW2019}. {\bf Second row centre panel}: fit to RVs of individual Pal 5 members from \protect\cite{Ibata2017}. {\bf Top right panel}: fit to distance estimates along the track of the Orphan Stream from \protect\cite{Koposov2019} Individual measurements from \protect\cite{Koposov2019} (green points) are shown for comparison. {\bf Second row right panel}: fit to RV estimates along the track of the Orphan stream from \protect\cite{Koposov2019}. Note that the RV estimates in this panel are in the Galactic standard of rest frame. Individual measurements from \protect\cite{Li2017} (green points) are shown for comparison.
{\bf Bottom two rows}: Proper motion measurements from Gaia DR2.}
\label{fig:tracks}
\end{figure*}

\section{Stream catalogue}
\label{sec:catalogue}
The goal of this work is to use the action space clustering method on real data of known stream stars. For this purpose we compiled data from 7 different literature sources \citep{Koposov2010, Willett2009, Li2017, Koposov2019, PW2019, Ibata2017, Koppelman2019} to obtain a data set containing full 6D phase-space information for stars in the GD-1, Helmi, Orphan and Palomar 5 streams. When complete 6D information for individual stars was not available, we made use of the stream's track: the measurements of the stream's mean phase-space position as a function of a coordinate aligned with the stream. We fit the tracks with a simple polynomial function in order to find the stars' missing 6D phase space components, based on their location along the stream. We do not assign membership probabilities to the stars in each stream, but rather treat all of them as certain members. A perfectly clean selection of stream stars is not crucial for our method, which can operate without any membership information at all since it relies on finding the most clustered \emph{total} action distribution. The addition of stars incorrectly classified as stream members---as long as such stars are in the minority---will simply result in slightly less clustered action-space for each of the trial potentials. We do not expect interlopers to bias the result: since stream membership is usually determined by making selections in some combination of positions, velocities, colors and magnitudes rather than in actions, it is unlikely that interlopers will cluster with the rest of the stream in action space near the best-fit potential. 

Nevertheless, to focus on the most informative stars, we perform cuts on some of the streams after visually inspecting them in $\mu_{\alpha}$ - $\mu_{\delta}$ or $L_z$ - $L_{\bot}$ space (we discuss the possible impact of this cut in section \ref{sec:validation}). Our full data set is available in electronic format online. In the following, we briefly review each stream's properties and compiled data set. For more details on our data assembly process we refer to Appendices~\ref{ap:gd-1}-\ref{ap:pal5}.

\subsection{The GD-1 stream}
GD-1 is a long and remarkably narrow stream first discovered by \cite{GD2006} in the SDSS data.
It lies at a distance of $\sim 15$ kpc from the Galactic center and $\sim 8$ kpc above the plane of the disc. Due to its thinness and location high above the Galactic disc it is thought to have formed from a tidally disrupted globular cluster, but no progenitor has yet been found. Orbits fitted to the available data have shown that GD-1 is moving retrograde with respect to the rotation of the Galactic disc, and is currently near pericentre (around 14 kpc), with apocentre 26--28 kpc from the Galactic centre \citep{Willett2009, Koposov2010}.

The GD-1 stream has seen considerable use in studies aiming to constrain the inner Galactic potential.
For example, using a single component potential \cite{Koposov2010} find that the orbit that best fits the GD-1 data corresponds to a potential with the circular velocity at the solar radius $V_{\rm c}(R_{\odot}) = 221^{+16}_{-20} \rm \ km \ s^{-1}$ and the flattening $q = 0.87^{+0.12}_{-0.03}$.
A more recent work by \cite{Malhan2019}, which uses a combination of Gaia DR2, SEGUE and LAMOST data, finds a circular velocity at the solar radius of $V_{\rm c}(R_{\odot}) = 244 \pm 4 \ \rm km \ s^{-1}$, the flattening of the halo $q = 0.82^{+0.25}_{-0.13}$ and the mass enclosed within 20 kpc $M(< 20 \ \mathrm{kpc}) = 2.5 \pm 0.2 \times 10^{11} \ M_{\odot}$.

We compiled a list of GD-1 members with measured radial velocities from \cite{Koposov2010}, \cite{Li2017} and \cite{Willett2009}. These stars' measurements are then supplemented with positions and proper motions from Gaia DR2. Finally, we fit a polynomial to the stream track distance information from \cite{Koposov2010} and \cite{Li2018} and use the resulting function to predict distances to each of our stream members based on their location along the stream. In total our GD-1 data set consists of 82 member stars with full 6D phase space information. We further clean this sample by discarding 13 stars that are not part of the central clump in either $\mu_{\alpha}$ - $\mu_{\delta}$ or $L_z$ - $L_{\bot}$ space, leaving 69 stars (see Appendix~\ref{ap:gd-1}).

\subsection{Orphan Stream}
\label{sec:orphan_data}
The Orphan stream was discovered by \cite{Grillmair2006} and \cite{Belokurov2006}  as a broad stream of stars extending $\sim 60$ degrees in the Northern Galactic hemisphere. Although thought to be the remnant of a small dwarf galaxy \citep{Grillmair2006}, no suitable progenitor for the stream has so far been found. Using SDSS DR7 data, \cite{Newberg2010} obtained a well-defined orbit to the stream and showed that the stars are on a prograde orbit with respect to disc rotation with a pericentre of 16 kpc and an apocentre of 90 kpc. At the time, the detected portion of the stream ranged from $\sim 20$ to $\sim 50$ kpc in the Galactic frame.

Recently, several discoveries regarding the Orphan stream were made by \cite{Koposov2019} who traced the track of the Orphan stream using RR Lyrae in the Gaia DR2 catalogue. They found that the stream is much longer than previously thought and showed that it also extends to the Southern Galactic hemisphere: the stream was found to extend from $\sim 50$ kpc in the North to $\sim 50$ kpc in the South, going through its closest approach at $\sim 15$ kpc from the Galactic Centre. They noticed, however, that the stream track behaviour changes between the two hemispheres. First, a twist in the stream track emerges soon after the stream crosses the Galactic plane from south to north. Second, the motion of the stars in the southern hemisphere is not aligned with the stream track. \cite{Erkal2019} show that these effects can be reproduced by adding the contribution of the Large Magellanic Cloud (LMC) into the Milky Way potential. These results demonstrate that the assumption that the Orphan stream stars orbit in a static Milky Way potential would lead to a bias.

Using the Orphan stream RR Lyrae from \cite{Koposov2019} and including the perturbation from the LMC \cite{Erkal2019} find that the best fit Milky Way potential has a mass enclosed within 50 kpc of $3.80 ^ {+0.14}_{-0.11} \times 10^{11} \ M_{\odot} $ and scale radius of the NFW halo of $17.5 ^ {+2.2}_{-1.8}\ \rm kpc$. As a comparison, when using only the Northern portion of the stream \cite{Newberg2010} find that the orbit is best fit to a Milky Way potential which has a mass enclosed within 60 kpc of about $2.6 \times 10^{11} \ M_{\odot}$. 

We assemble our Orphan Stream data in two parts. In both cases the positions and proper motions are from Gaia DR2, but accurate individual distances and radial velocities have been measured for disparate sets of stars. 

To compile the first subsample, we begin with a list of stream members that have accurate distance measurements: the Orphan stream RR Lyrae from \cite{Koposov2019}. We fit a polynomial to the radial velocity track information from \cite{Koposov2019} and use it to predict radial velocities to our stream members based on their location along the stream. Although the RR Lyrae stars stretch from $\phi_1 \ \sim -78$ to $\sim 123$ degrees, we discard those that have $\phi_1 < 0$, i.e. those in the Southern Galactic hemisphere. We do this because there are no radial velocity measurements in the negative $\phi_1$ section of the stream (see the bottom right panel of Figure~\ref{fig:tracks}), leading our estimates to depend too heavily on the selection of the degree of the polynomial and its fit to the positive $\phi_1$ section of the stream. As an added advantage, this cutoff eliminates the part of the stream that appears to be most strongly affected by the LMC, which is not in our potential model.

To compile the second subsample, we begin with a list of stream members that instead have individual radial velocity measurements: the stream members from \cite{Li2017} with radial velocities from SDSS or LAMOST. Next, we fit a polynomial to the distance track data from \cite{Koposov2019} to find distances for the stars. 

After combining the two data sets, we make an additional cut in $L_z$ - $L_{\bot}$ and $\mu_{\alpha}$ - $\mu_{\delta}$ space, selecting after visual inspection the 129 stars that form a clump in velocity space. More details can be found in Appendix~\ref{ap:orphan}.

\subsection{The Palomar 5 stream}
The tidal streams around the Palomar 5 globular cluster were first found by \cite{Odenkirchen2001}. They detected two symmetrical tails on either side of the cluster, extending in total about $2.6$ degrees on the sky. Subsequent data has allowed the stream to be traced further out and revealed that the two tidal tails are far from symmetric, the tails having distinctly different lengths and star counts. The current known length of the trailing trail is $23$ degrees \citep{Carlberg2012} while that of the leading tail is only 3.5 degrees \citep{Odenkirchen2003}. The reason for the asymmetry is not clear but possible options include perturbations from spiral arms, rotating bar, molecular clouds and dark matter subhaloes \citep{Pearson2017, Amorisco2016, Erkal2017}. Palomar 5 is currently near the apocentre of its prograde orbit, which ranges from pericentre at $7-8$ kpc to apocentre at around $19$ kpc from the Galactic Centre \citep{Odenkirchen2003, GD2006a, Kupper2015}.
The stream has previously been used to constrain the Milky Way potential by \cite{Kupper2015} who found the mass enclosed within Palomar 5 apocentre distance to be $M(< 19)\ \mathrm{kpc} = (2.1 \pm 0.4) \times 10^{11} \ M_{\odot}$ and the halo flattening in the z-direction to be $q_z = 0.95^{+0.16}_{-0.12}$.

As before, we use a combination of sources to get full 6D phase space information for stars in the Palomar 5 stream. We use a list of 27 RR Lyrae member stars with distance estimates from \cite{PW2019} and another 154 members with radial velocity measurements from \cite{Ibata2017}. 
To find the distances for the members from \cite{Ibata2017} and radial velocities for members from \cite{PW2019}, we use the measurements of the individual members in the other set, using the same track-fitting strategy as for the other streams. In other words, we fit a polynomial to the distances from \cite{PW2019} to find distance estimates for the \cite{Ibata2017} members and, similarly, fit a polynomial to the radial velocity data from \cite{Ibata2017} to find radial velocity estimates for the \cite{PW2019} members. As always, the positions and proper motions are from Gaia DR2.

After the two data sets are then joined, a cut in $L_z$ - $L_{\bot}$ and $\mu_{\alpha}$ - $\mu_{\delta}$ space is performed, resulting in the sample of 136 stars. More details can be found in Appendix~\ref{ap:pal5}.

Note that since we have adopted the distances from \cite{PW2019}, our Pal 5 stars are closer than previously reported. \cite{PW2019} find a mean cluster heliocentric distance of $20.6 \pm 0.2$ kpc while previous distance measurements are $\sim 23$ kpc [e.g.][]\citep{Odenkirchen2001, Carlberg2012, Erkal2017}. The main cause for this distinction is that the previously reported distances were computed from distance moduli \citep{Harris1996, Dotter2011} that were not corrected for dust extinction.

\begin{figure*}
\centering
\includegraphics[width=0.49\linewidth,trim={0 10 0 10}, clip]{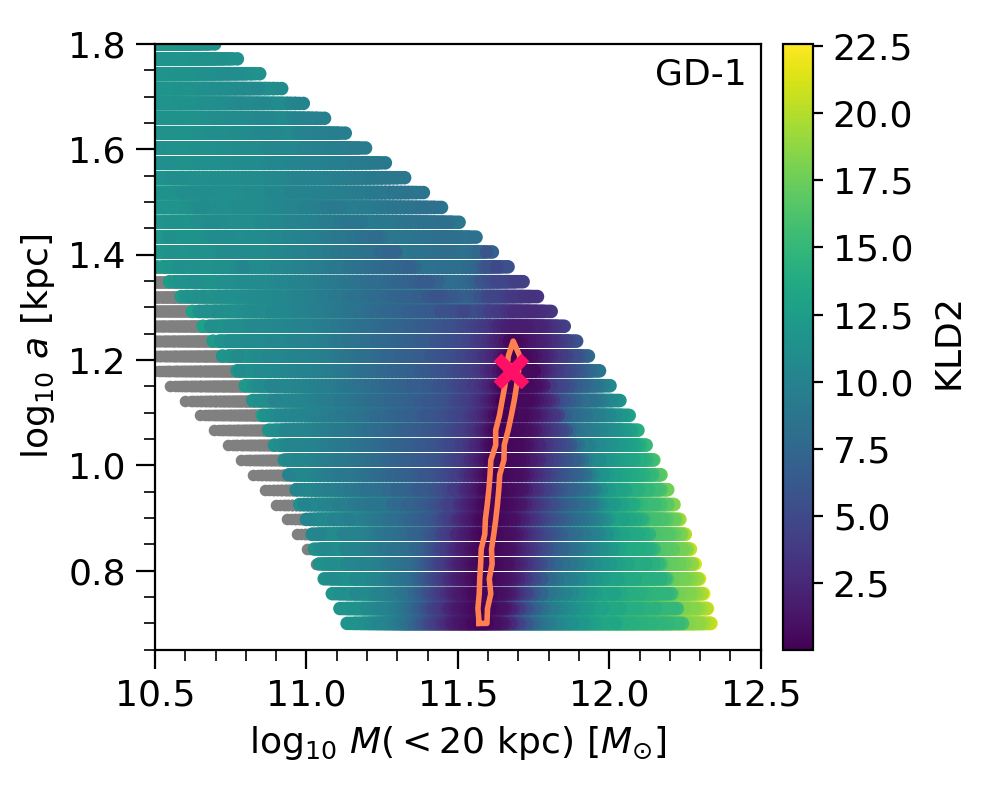}
\includegraphics[width=0.49\linewidth,trim={0 10 0 10}, clip]{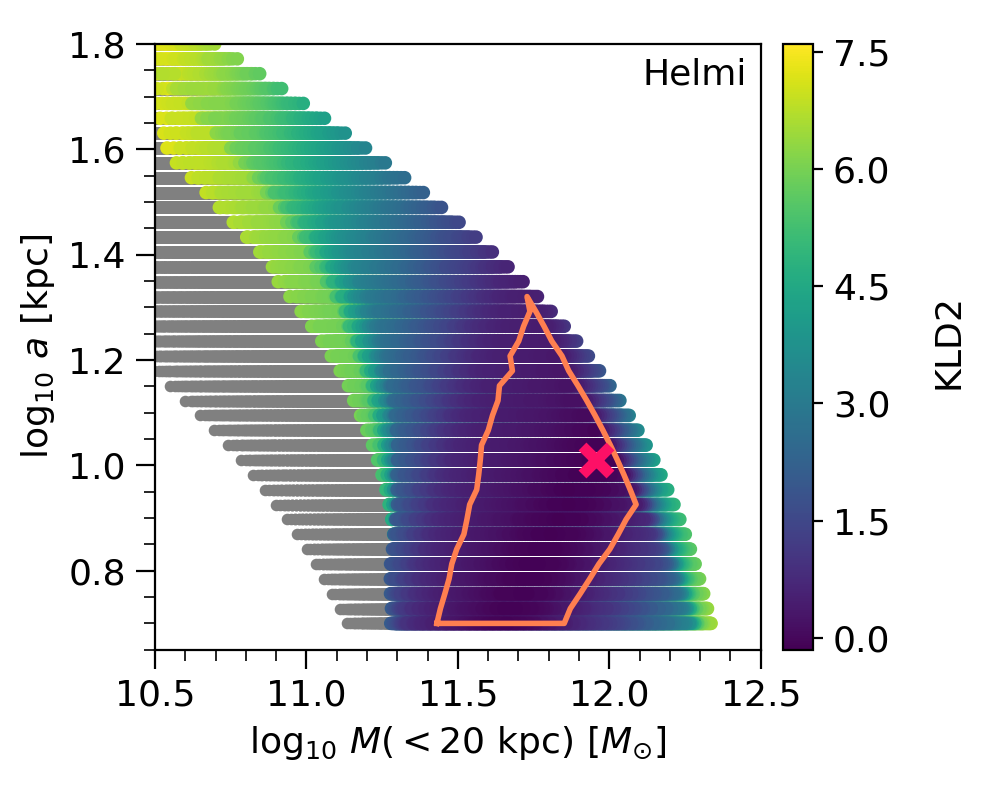}\\
\includegraphics[width=0.49\linewidth,trim={0 10 0 10}, clip]{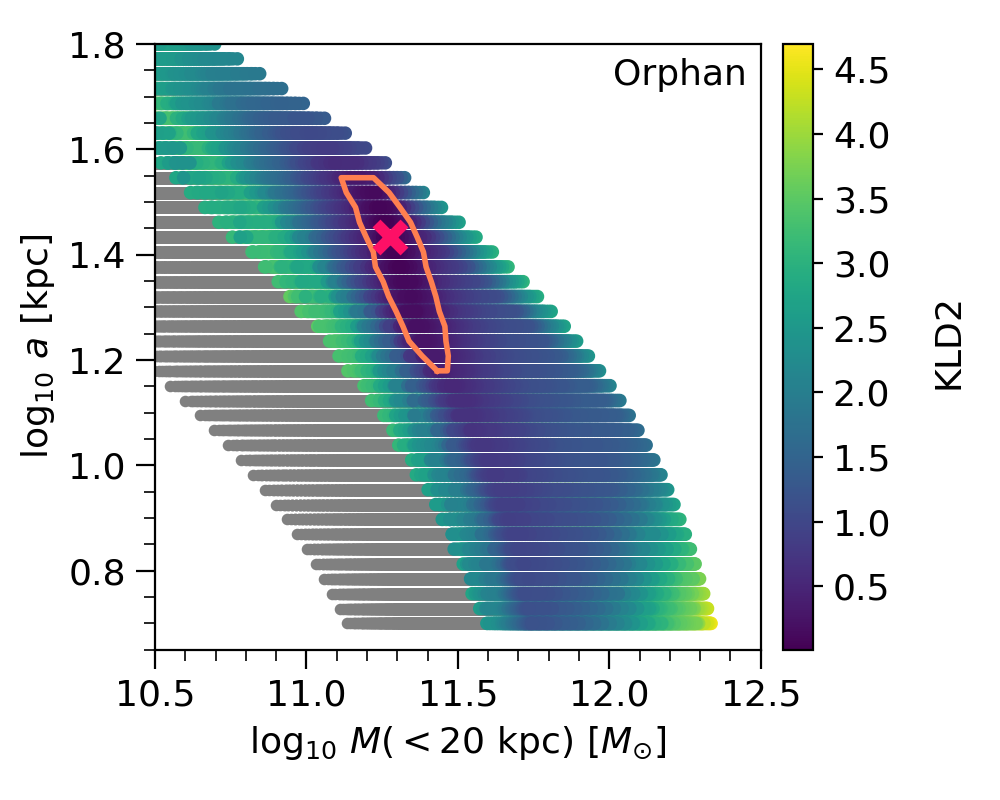} 
\includegraphics[width=0.49\linewidth,trim={0 10 0 10}, clip]{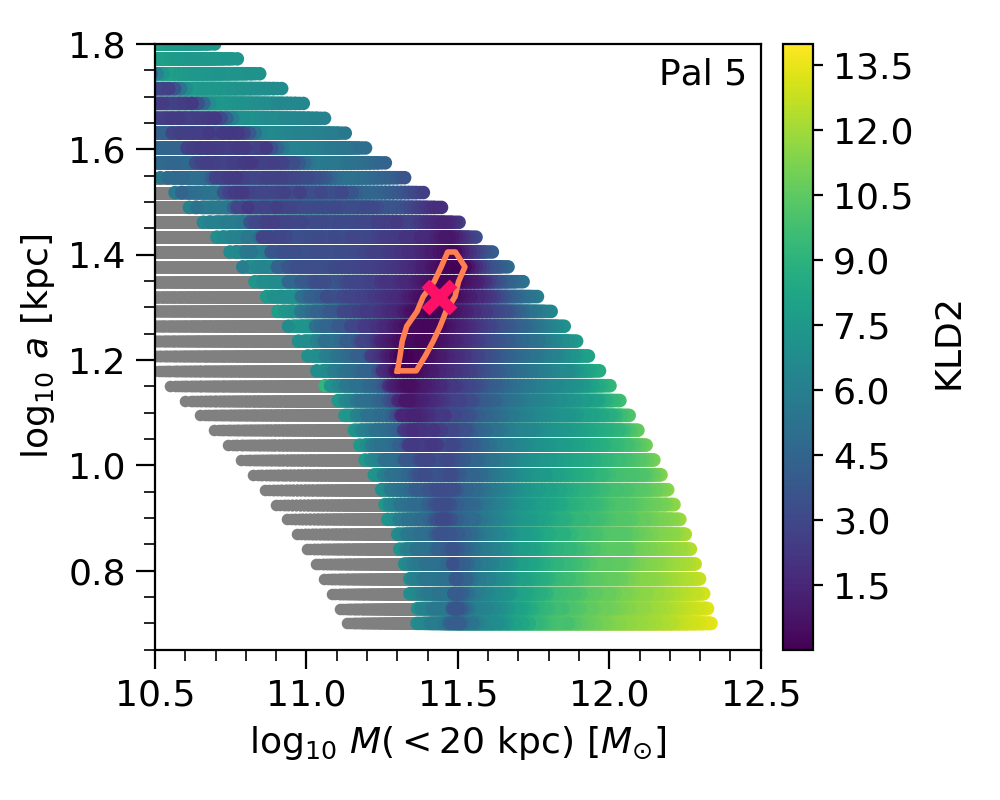}\\
\caption{Individual stream results for the single-component potential in the enclosed mass--scale length space. The best-fit point is marked with a pink cross, the grey points represent potentials that resulted in unbound stars and were therefore discarded, and other points are colour-coded according to their KLD2 value. The orange region shows $1 \sigma$ contours (defined as described in \S \ref{sec:kld2}).}
\label{fig:individual}
\end{figure*}

\subsection{The Helmi stream}
The Helmi stream was first detected by \cite{Helmi1999} as a cluster of 12 stars in the angular momentum space of the local halo, based on Hipparcos measurements. The orbit of the stream was found to be confined within 7 and 16 kpc from the Galactic Centre. While forming a single clump in $L_{\rm z}$ - $L_{\bot}$ diagram, in velocity space the structure separates into two distinct groups: one with positive $v_z$ and the other with negative $v_z$. \cite{Helmi1999} postulated that the two clumps originate from a common dwarf galaxy that has since its disruption reached a highly phase-mixed state. This view explains the observed bimodality of $v_{\rm z}$ as a feature that arises due to the existence of several wraps of the stream near the Solar neighbourhood (as shown in Figure 5 of \citealt{Helmi2008}; see also \citealt{McMillanBinney2008} for further discussion of this phenomenon).

Using Gaia DR2 data complemented by radial velocities from the APOGEE, RAVE and LAMOST surveys, \cite{Koppelman2019} found 523 new members of the Helmi stream within 5 kpc of the Sun selected in $L_z$ - $L_{\bot}$ space.
In this work we include 401 high confidence members that are within 1 standard deviation of the mean radial velocity from the \cite{Koppelman2019} sample with their full 6D phase space information.

\section{Results for a single-component potential}
\label{sec:results1}

In this section we present our results for a single-component St\"ackel potential with both individual and combined stream data sets (Sec.~\ref{sec:ind3par} and Sec.~\ref{sec:com3par}, respectively). The best-fit parameter values (those that maximize KLD1) and uncertainties (derived from KLD2) are summarised in Table~\ref{tab:results}. Here, we focus on the KLD2 distributions, while those for KLD1 can be found in Appendix~\ref{app:B}. The KLD1 values associated with the best-fit parameters are given in Table~\ref{tab:results_kld1}.

\subsection{Results for the individual stream data sets}
\label{sec:ind3par}

Figure~\ref{fig:individual} shows the individual results for GD-1, Helmi, Orphan and Pal 5 on the enclosed mass - scale length plane. The parameter space is colour-coded by their KLD2 values. The smaller the KLD2 value, the more similar the action distribution of that potential is to the action distribution of the best-fit potential. The potentials with values of $\mathrm{KLD2} \leq 0.5$ - marked in Figure~\ref{fig:individual} with orange - is the $1 \sigma$ region, as explained in Section~\ref{sec:kld2}. The grey points stand for discarded potentials, where at least one star is on a dynamically unbound orbit as discussed in Section~\ref{sec:calc_actions}.

Figure~\ref{fig:individual} indicates that the GD-1, Pal 5 and Orphan streams deliver constraints that are much more precise than those from the Helmi stream. 
 
The best-fit values for $M(<20 \ \mathrm{kpc})$ range from $1.89\, \times \, 10^{11}$ to $\sim 9 \times 10^{11} M_{\odot}$ between individual streams, with the Orphan stream returning the lowest and Helmi stream the highest estimates. Although their best-fit values differ by an order of magnitude, their derived confidence intervals are compatible. However, the confidence intervals from the GD-1 stream are in tension with those from Pal 5 and Orphan streams. This is likely a manifestation of the systematic biases affecting single-stream fits.

With the exception of Palomar 5 and, to some extent, the Orphan stream, the individual streams cannot place strong constraints on the scale length of the potential: the best-fit values range from $10.24$ to $27.12$ kpc between individual streams, with the Helmi stream yielding the lowest and the Orphan stream the highest values. All four streams accept $a$ values between $\sim 15$ kpc and $\sim 17$ kpc within a 1$\sigma$ uncertainty level.

The best-fit values for flattening range from $0.88$ to $1.40$, with the lowest estimate belonging to the GD-1 and the highest to the Pal 5 stream. We remind the reader that in the St\"ackel convention flattening is defined as $ \frac{a}{c}$, so $e > 1$ corresponds to a oblate potential while $e < 1$ corresponds to a prolate potential.

The GD-1 stream provides a strong upper limit to the flattening - only accepting prolate shapes ($e \leq 0.95$) - but it is unable to determine the lower limit. The Helmi stream accepts almost all flattening values except the ones corresponding to the most prolate ($e < 0.64$) and most oblate ($e > 1.74$) shapes. The Pal 5 and Orphan streams, on the other hand, provide a lower limit to flattening, both only allowing oblate shapes ($e \geq 1.09$ for Orphan and $e \geq 1.21$ for Pal 5).

\subsection{Results for the combined data set}
\label{sec:com3par}

\begin{figure}
\centering
\includegraphics[width=0.98\linewidth,trim={0 10 0 11}, clip]{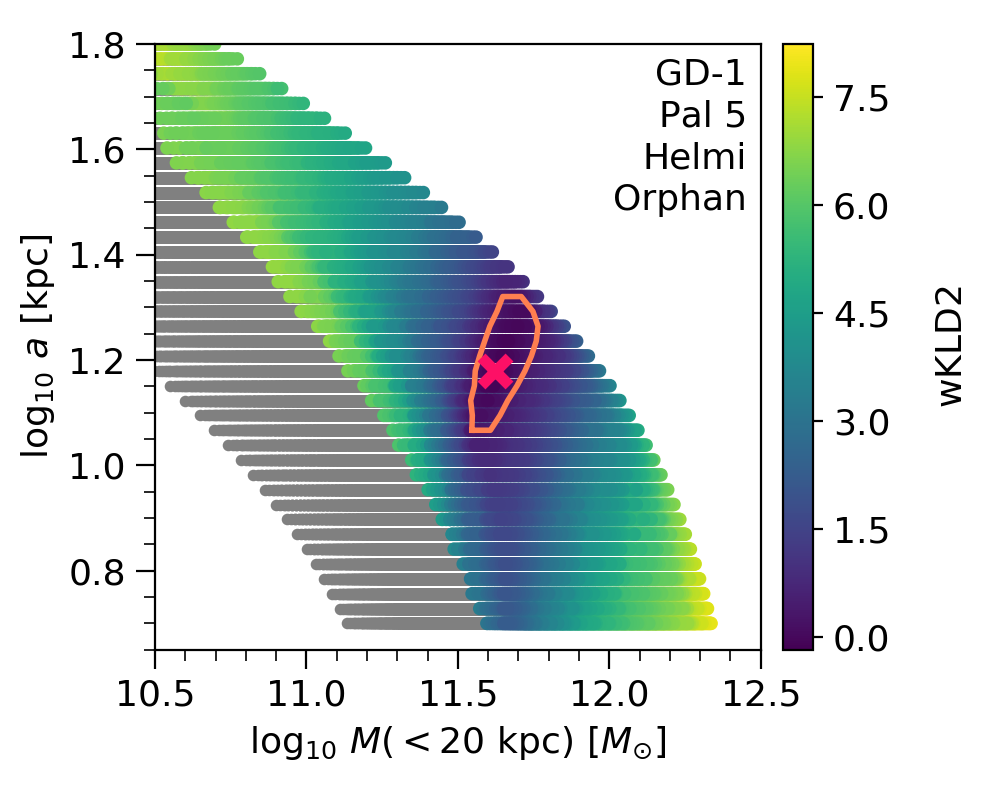}\\
\includegraphics[width=0.98\linewidth,trim={0 10 0 10}, clip]{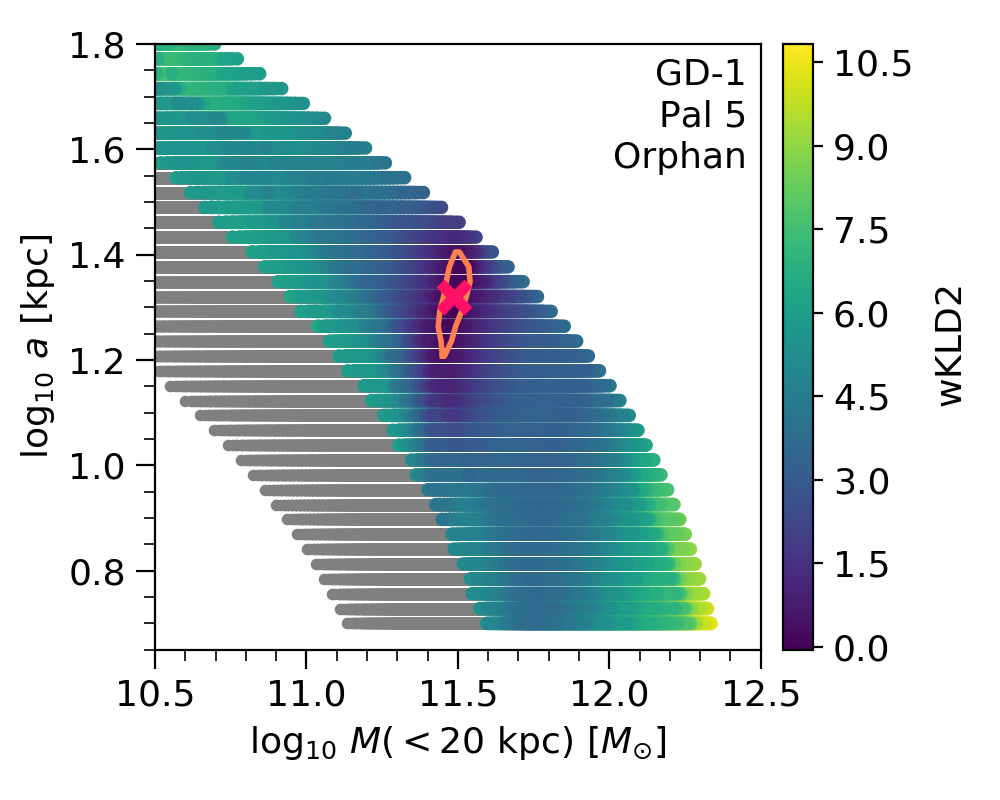}
\caption{As in Figure~\ref{fig:individual}, but showing the weighted combined data results for the single-component potential. {\bf Top:} results when combining all four streams. {\bf Bottom:} results for the combination of GD-1, Orphan and Pal 5.}
\label{fig:combined_all}
\end{figure}

\begin{figure}
\centering
\includegraphics[width=0.99\linewidth,trim={0 30 0 15}, clip]{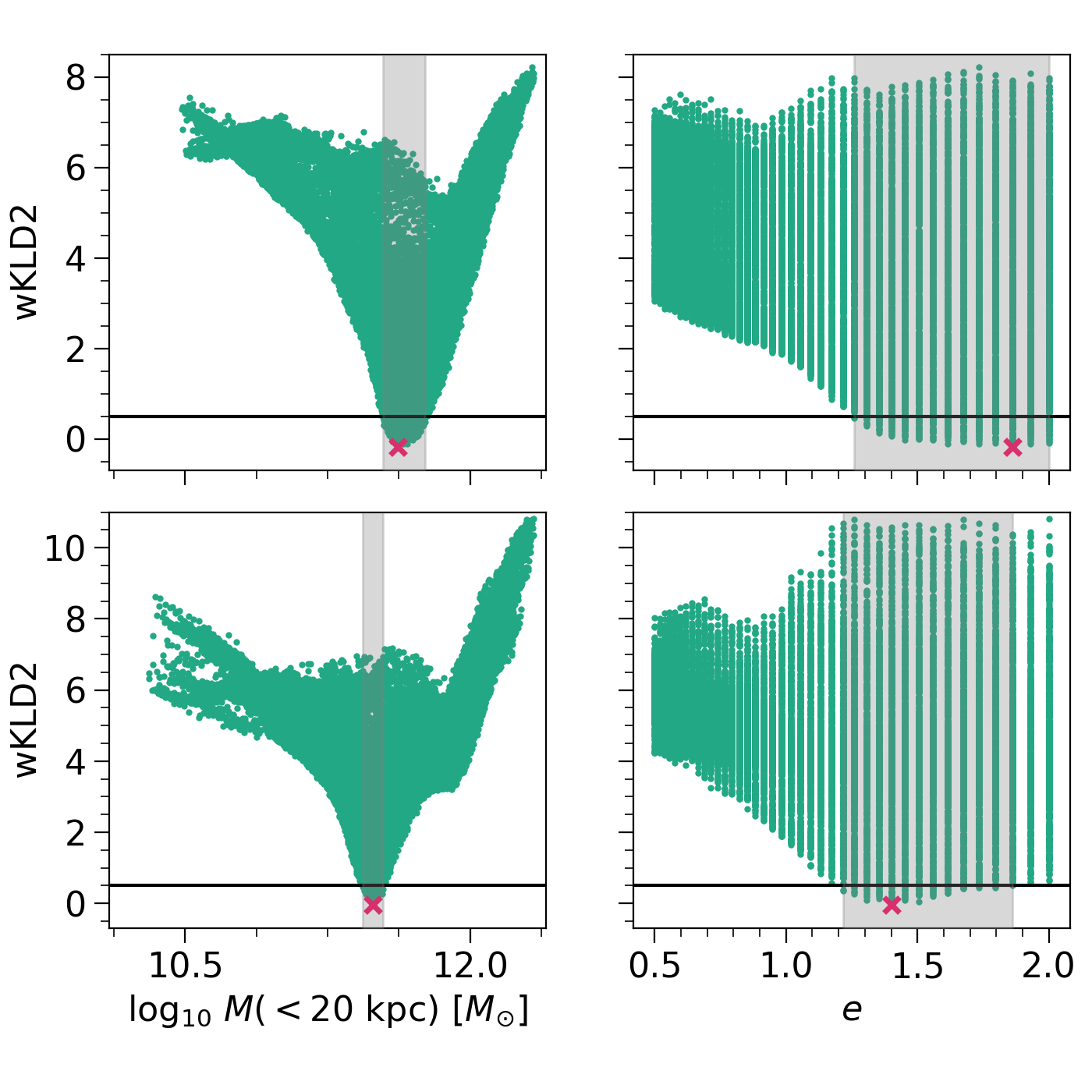}
\caption{The weighted combined data results for a single-component potential: marginalised single parameter distributions. The top panel shows the results from the combination of all four streams and the bottom panel shows the results of the combination of GD-1, Orphan and Pal 5 streams. The green points show the parameter values against the wKLD2 of the potential they belong to. The values of the parameters in the best-fit potentials are marked with a pink cross. The black lines are drawn at wKLD2 = 0.5 which signifies the $1 \sigma$ confidence interval. The light grey bars show the range of values that are accepted with $1 \sigma$ confidence.}
\label{fig:marginal_all}
\end{figure}

\begin{figure*}
\centering
\includegraphics[trim={20 0 45 30}, clip, width=0.99\linewidth]{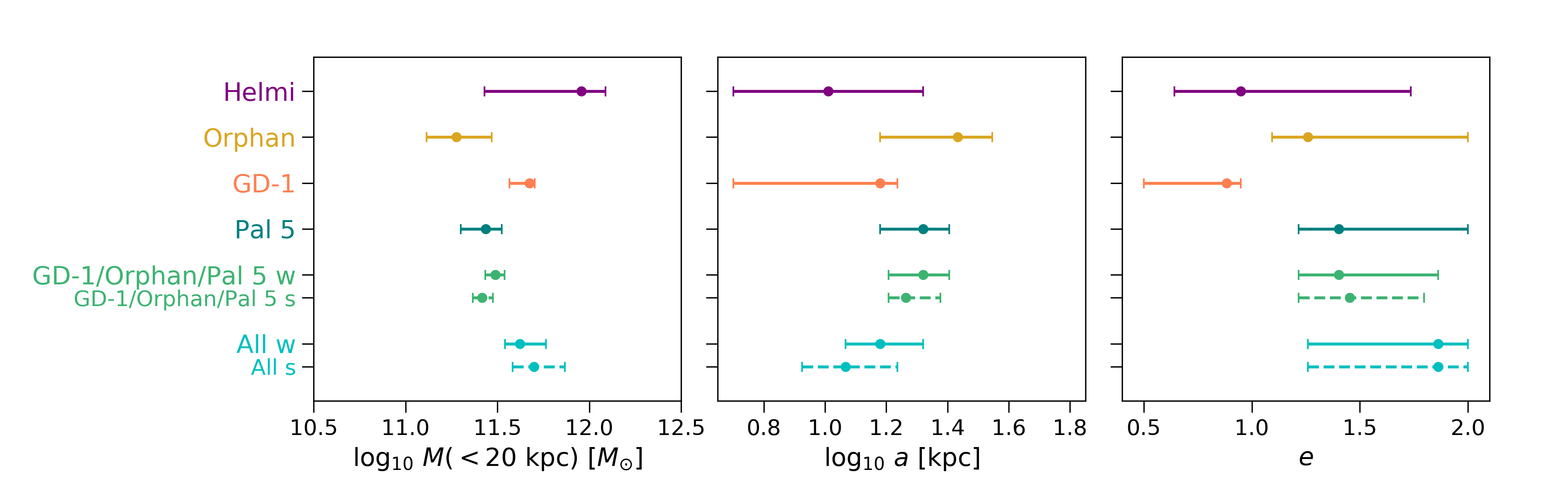}
\caption{Comparison of best-fit parameter values for the single-component potential with their $1 \sigma$ confidence intervals for enclosed mass (left), scale length (middle) and flattening (right). Results for individual streams are labeled with the stream name. Combined results are labeled with ``GD-1/Orphan/Pal 5'' for the three-stream combination and ``All'' for the four-stream combination. In addition, the combined stream labels end with an ``s'' or a ``w'' for the standard and weighted analyses, respectively. We remind the reader that $e > 1$ corresponds to a oblate potential while $e < 1$ corresponds to a prolate potential.}
\label{fig:error}
\end{figure*}

\begin{table*}
\begin{tabular}{ c c c c c c c c}
Streams & $\mathrm{N}_*$ & $M(<20 \ \mathrm{kpc}) \times 10^{11} \ [M_{\odot}]$ & $a \ [\mathrm{kpc}]$  & $e$ & $M_{\mathrm{tot}} \times 10^{12} \ [M_{\odot}]$ \\
\hline
GD-1 & 69               & $4.73^{+0.33}_{-1.05}$ & $15.12^{+2.10}_{-10.11}$ & $0.88^{+0.07}_{-0.38}$ & $2.65^{+0.51}_{-1.88}$ \\
\\
Helmi & 401             & $9.06^{+3.20}_{-6.36}$ & $10.24^{+10.68}_{-5.23}$ & $0.95^{+0.79}_{-0.31}$ & $2.81^{+0.35}_{-2.39}$ \\
\\
Pal 5 & 136             & $2.73^{+0.60}_{-0.74}$ & $20.92^{+4.50}_{-5.80}$ & $1.40^{+0.60}_{-0.19}$ & 
$1.86^{+1.30}_{-1.14}$ \\
\\
Orphan & 117             & $1.89^{+1.05}_{-0.60}$ & $27.12^{+8.05}_{-12.00}$ & $1.26^{+0.74}_{-0.17}$ & $2.35^{+0.81}_{-1.32}$ \\
\\
GD-1/Orphan/Pal 5 standard & 322 & $2.60^{+0.39}_{-0.28}$ & $18.37^{+5.45}_{-2.24}$ & $1.45^{+0.35}_{-0.24}$ & $1.38^{+1.43}_{-0.35}$ \\
\\
GD-1/Orphan/Pal 5 weighted & 322 & $3.08^{+0.39}_{-0.35}$ & $20.92^{+4.50}_{-4.79}$ & $1.40^{+0.46}_{-0.19}$ & $2.09^{+1.07}_{-1.00}$ \\
\\
GD-1/Helmi/Orphan/Pal 5 standard & 723 & $5.01^{+2.37}_{-1.18}$ & $11.66^{+5.56}_{-3.23}$ & $1.86^{+0.14}_{-0.60}$ & 
$1.30^{+1.86}_{-0.44}$ \\
\\
GD-1/Helmi/Orphan/Pal 5 weighted & 723 & $4.18^{+1.63}_{-0.70}$ & $15.12^{+5.80}_{-3.46}$ & $1.86^{+0.14}_{-0.60}$ & 
$1.47^{+1.69}_{-0.55}$ \\
\hline
\end{tabular}
\caption{The individual and combined stream results for a single-component potential. Best-fit parameters are given with their $1\sigma$ confidence intervals. We remind the reader that $e > 1$ corresponds to a oblate potential while $e < 1$ corresponds to a prolate potential.}
\label{tab:results}
\end{table*}

The weighted combined data results are shown in Figure~\ref{fig:combined_all}: the combination of all four streams on the top panel and the combination of GD-1, Orphan and Pal 5 on the bottom panel. As a reminder, the weighted results incorporate the knowledge of stream membership by (a) calculating the stream-weighted versions of KLD1 and KLD2 (the latter is marked on the relevant figures as wKLD2), and (b) by artificially separating streams in action space during the density estimation. This is in contrast to the standard results which assume no knowledge of the stream membership. In this section, we discuss the weighted results only. The standards results will be discussed in the subsequent section. 

Constraints from analysing combined data sets are tighter than those yielded by individual stream data sets. The $1\sigma$ constraint on the enclosed mass for the combination of GD-1, Orphan and Pal 5 data sets is much tighter than that obtained by combining all four streams' data sets.
The possible reasons are discussed in more detail in Section~\ref{sec:en_mass_disc}.

The two different analyses also give results that are somewhat inconsistent with each other: we find $4.18^{+1.63}_{-0.70} \times 10^{11} \ M_{\odot}$ when analysing the combination of four streams' data and $3.08^{+0.39}_{-0.35} \times 10^{11} \ M_{\odot}$ for the GD-1, Orphan and Pal 5 data sets, i.e. the inclusion of Helmi stream data pulls the consensus result to a higher enclosed mass.

In contrast, the scale lengths of the two sets of combined data are in good agreement within the errors: we find $15.12^{+5.80}_{-3.46}$ kpc and $20.92^{+4.50}_{-4.79}$ kpc, for the combination of four- and three-stream data sets, respectively. 

Although the flattening parameters of the two combinations are consistent with one another within their $1\sigma$ intervals, their best-fit values vary from $1.86$ for the combination of four streams to $1.40$ for the combination of GD-1, Orphan and Pal 5. These high best-fit values, which correspond to oblate potentials, are likely driven by the Pal 5 and Orphan streams, which individually disfavour the lower values of $e$. Both Pal 5 and Orphan individually accept values of flattening above $1.2$, and the $1\sigma$ confidence intervals of the combined sets clearly reflect this, as both find a lower limit of $\sim 1.2$. 

The combined stream results therefore show virtually no improvement over the individual results. We therefore conclude that we can only weakly constrain the flattening parameter with our data and this one-component potential.

We do not expect the single-component potential to be a good representation of the Milky Way's actual gravitational field for streams whose orbits intersect the disc, such as the Helmi and Pal 5 streams. Although we give results for these streams and use them in some combined fits with this model, we caution that the best-fit parameters should not be interpreted as representing a particular component of the Milky Way's structure. The best-fit models tend to respond to the need for a more concentrated central component (i.e. the disc) than the model will allow by increasing the total mass, leading to larger circular velocities at large radii.

Figure~\ref{fig:marginal_all} shows the total distributions of wKLD2 for the enclosed mass and flattening parameters that result from the analysis of the combined data sets (upper panel: four stream data set; lower panel: three-stream data set). The black horizontal lines are at wKLD2 = 0.5, the value that corresponds to the $1\sigma$ confidence interval for a single-component potential. The quoted $1\sigma$ confidence intervals correspond to the parameter range on the x-axis in each panel where $ \mathrm{wKLD2} \leq 0.5$ (shown with a grey shaded region). 
It is clear, in both cases, that the confidence interval of the enclosed mass (a combination of the scale radius and total mass parameters) is well defined. In contrast, the flattening is only weakly constrained.

Figure~\ref{fig:error} summarises the enclosed mass, scale length and flattening results with their $1\sigma$ confidence limits as presented in Table~\ref{tab:results}. For comparison, the results from both the standard and weighted KLD analysis for combinations of streams are shown. 
We find that the difference between the best fit values of the standard and weighted methods is not significant: the results are consistent within their $1 \sigma$ confidence limits. Nevertheless, the use of the stellar membership knowledge clearly affects the results. In the case of the four combined streams, this has the expected effect of lowering the best-fit enclosed mass value: the influence of the Helmi stream, that contains the largest number of stars in our sample, has now been off-set. The opposite happens for the combination of GD-1, Orphan and Pal 5 combination, where the best-fit enclosed mass increases, because we counteract the fact that the Orphan and Pal 5 members outnumber those of GD-1 in our sample. In addition, no appreciable shift is found for the best-fit scale length and flattening between the two approaches.

\section{Results for a two-component potential}
\label{sec:5par}

\begin{figure*}
\centering
\includegraphics[width=0.49\linewidth,trim={0 10 0 10}, clip]{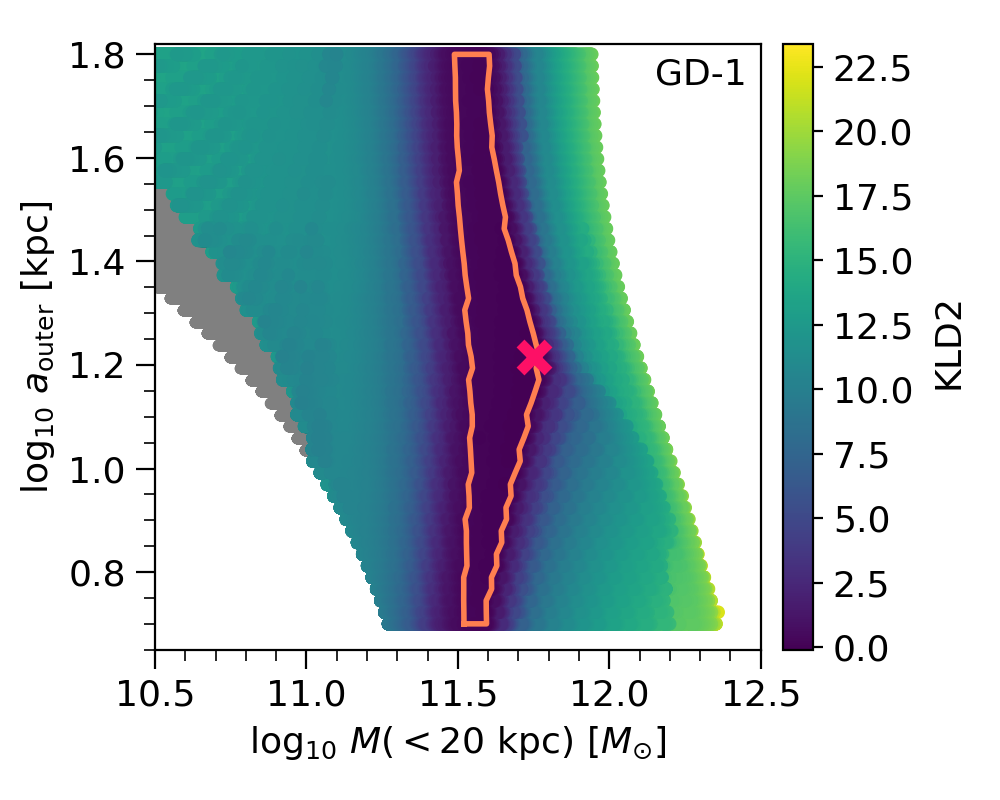}
\includegraphics[width=0.49\linewidth,trim={0 10 0 10}, clip]{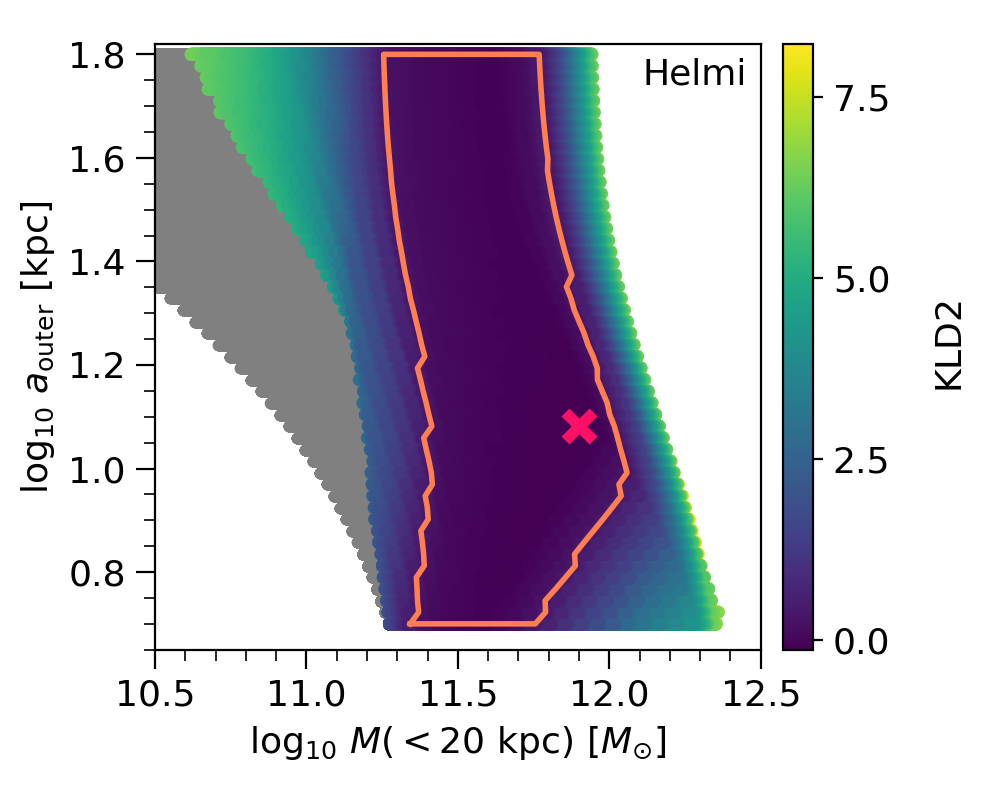}\\
\includegraphics[width=0.49\linewidth,trim={0 10 0 10}, clip]{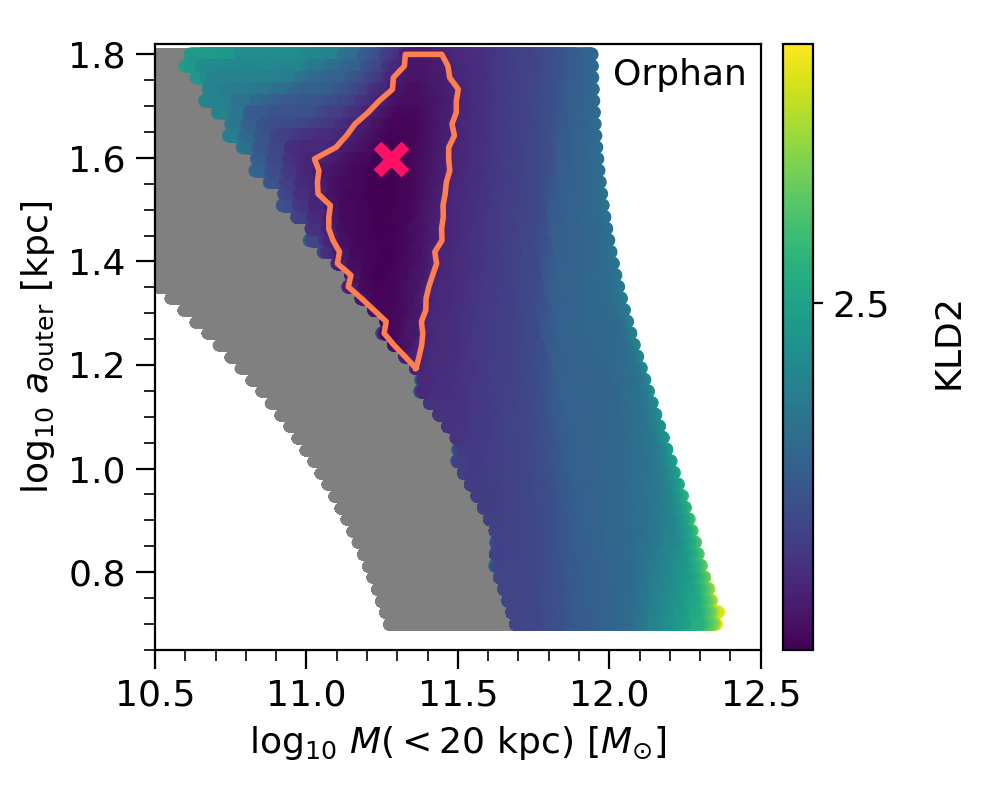}
\includegraphics[width=0.49\linewidth,trim={0 10 0 10}, clip]{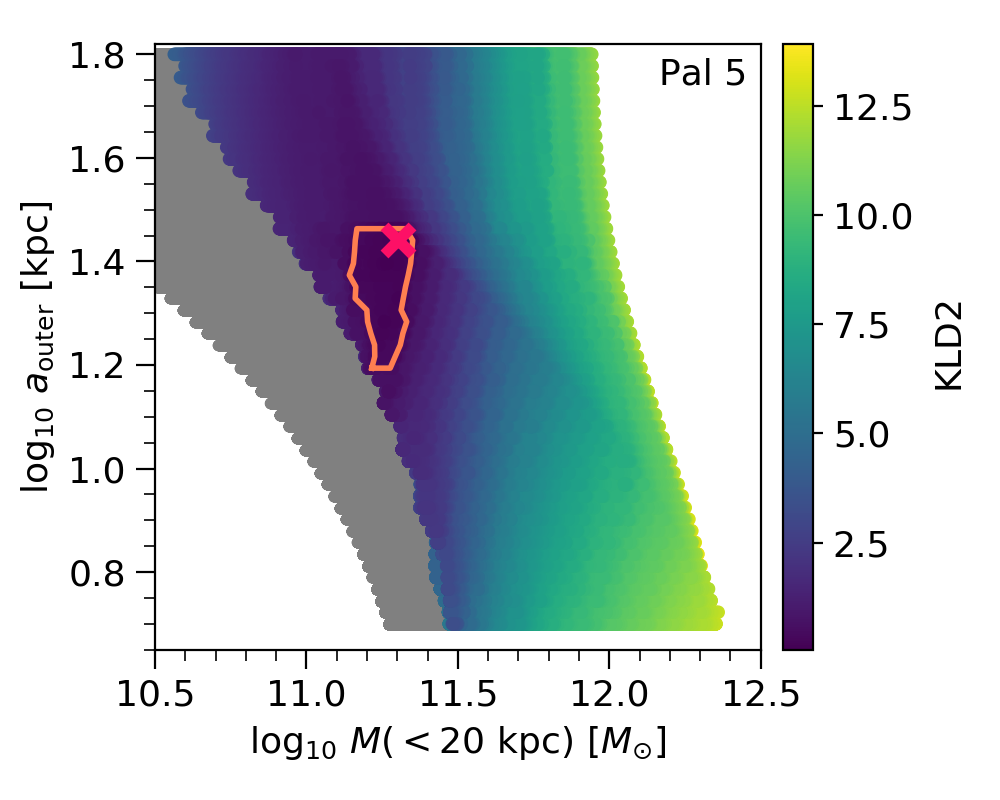}\\
\caption{As in Figure~\ref{fig:individual}, but showing individual stream results for the two-component potential.}
\label{fig:individual_5par}
\end{figure*}

In this section we present the results of fitting the streams with the two-component St\"ackel potential. The best-fit parameter values are summarised in Table~\ref{tab:results_5par}. As for the single-component potential (Section~\ref{sec:ind3par}), we focus on the KLD2 results for the confidence intervals and summarise KLD1 values associated with the best-fit parameters in Appendix ~\ref{app:B} and Table~\ref{tab:results_kld1} and Figures~\ref{fig:actions_comb} and \ref{fig:actions} give examples of a KLD1 distribution, alongside the action distributions produced by two different potentials.
The analysis of the single-component model results for stream combinations showed that the difference between the best fit values of the standard and weighted methods is not significant, so we will only discuss the weighted results from this point onward.

\subsection{Results for individual stream data sets}

Figure~\ref{fig:individual_5par} presents the results of the individual streams on the enclosed mass--$a_{\mathrm{outer}}$ plane. The $1\sigma$ confidence intervals, again marked in orange. GD-1's $1 \sigma$ interval for the enclosed mass forms a relative narrow stripe well within the allowed parameter space of potentials producing bound orbits for all stars. In contrast, while the Orphan and Pal 5 streams also produce clear confidence intervals for the enclosed mass, their uncertainty regions in the direction of low $a_{\mathrm{outer}}$ are limited by the edge of the allowed parameter space of potentials producing bound orbits for all stars (see the discussion in Section~\ref{sec:validation}).
Finally, the Helmi stream includes within its $1 \sigma$ confidence contour a significant subset of all explored enclosed mass values.

Compared to the single-component case, the best-fit $M(<20 \ \mathrm{kpc})$ values are now in better agreement between individual streams, ranging from $1.91\, \times \, 10^{11}$ to $7.93 \times 10^{11} M_{\odot}$. As with the single-component potential, the Orphan stream returns the lowest and the Helmi stream the highest estimates of enclosed mass. As before, the $1\sigma$ ranges of Pal 5 and GD-1 are in tension, but this is no longer true for the GD-1 and Orphan pair.

The mass estimates of individual streams are in good agreement with the measurements obtained with a single-component potential.

The variation in the best-fit values is the smallest in the case of the Orphan stream, whose best-fit values differ only by $1\%$ between the two models. It is also notable that the best-fit values of the Helmi stream differ only by $12\%$ between the models, in spite of their large error bars. The best-fit values of the GD-1 and Pal 5 streams change by $19 \%$ and $26 \%$, respectively, between the two models.

Pal 5 is therefore the most sensitive to the change of model: this might be because Pal 5 has the smallest pericentre distance relative to the Galactic Centre, where the mass in the new inner component is concentrated. In addition, \cite{Pearson2017} have shown that Pal 5 was likely affected by the Galactic bar on its pericentre passage which might add to its sensitivity to the centrally concentrated mass profile.
While the Helmi stream also has a small pericentre distance, it is not as sensitive to the change in potential model. The possible reasons for this are discussed in Section~\ref{sec:en_mass_disc}.

Although all four streams have a best-fit flattening of the outer component of $\sim 1$, Pal 5 and Orphan data are the only ones that can actually constrain the flattening of the outer component, limiting it to be lower than $1.04$ in both cases. The GD-1 and Helmi streams include the entire allowed range of values within their $1\sigma$ confidence contours. We remind the reader that in the two-component potential we have limited our exploration of the halo to near-spherical and oblate shapes, which corresponds to $e > 1$.

The flattening of the single-component model cannot be directly compared to the flattening of the outer (or the inner) component of the two-component model. In the single-component case the flattening parameter reflects the combined axis ratios of the Galactic disc, bulge and halo. Therefore, we expect the flattening not to be spherical. In the two-component case, the two different axis ratios add more flexibility to our model, but neither of them corresponds fully to the flattening of the single-component model. We can, however, make a qualitative comparison in the case of Pal 5: the single-component model best-fit flattening is $\sim 1.40$, which may be interpreted as a synthesis of the two-component results, where the outer flattening is $\sim 1$, while the inner flattening is $2.55$ (as expected if the latter describes a component that incorporates a disc-like structure).

Although in most cases we cannot constrain the flattening parameter, the fact that all streams prefer a nearly spherical halo could be explained by the limitation of the Staeckel potential. \cite{Batsleer1994} found that to produce flat rotation curves they needed an almost spherical halo. Both the halo and disc potentials are described in the same spheroidal coordinate system (i.e. they must have the same foci) but have independent length scales $a_{\mathrm{inner}}$ and $a_{\mathrm{outer}}$. The halo component has the larger scale compared to which the foci are relatively close together, and the halo thus appears almost spherical. On the other hand, the foci are far apart relative to the smaller scale of the disc potential, giving it a more oblate shape.

None of the other parameters can be strongly constrained by any of the streams. As the enclosed mass is a function of all five potential parameters, we conclude that only combinations of these five parameters, but not their individual values, can be constrained.

\subsection{Results for the combined data set}

\begin{figure}
\centering
\includegraphics[width=0.98\linewidth,trim={0 10 0 10}, clip]{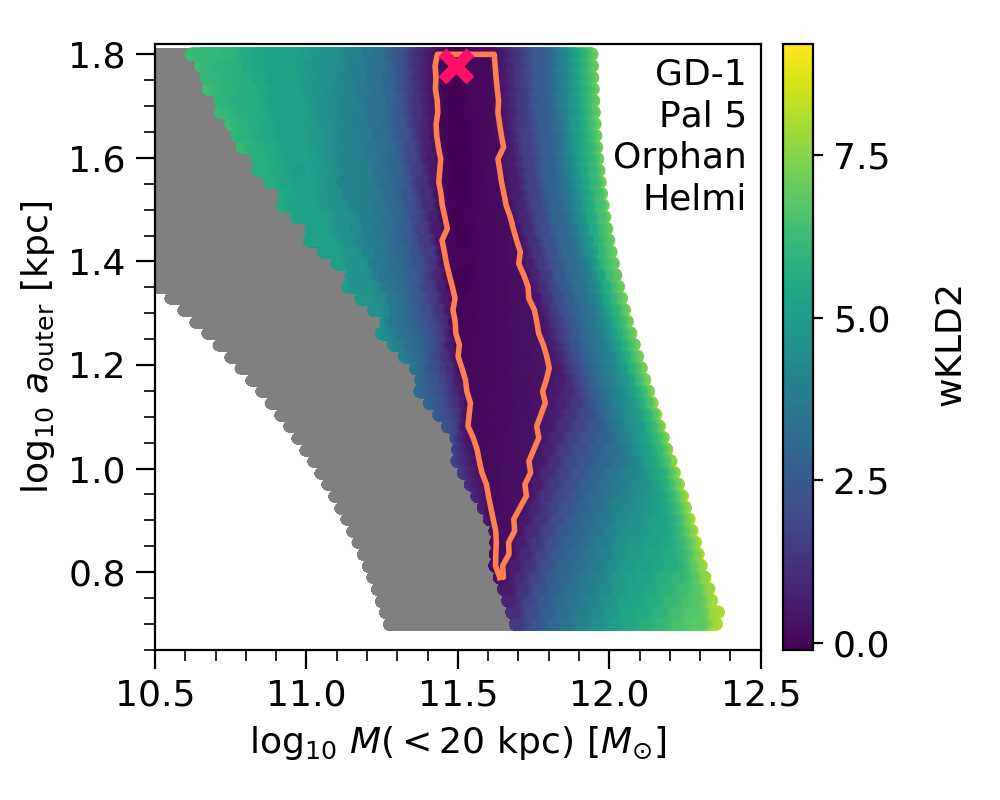}\\
\includegraphics[width=0.98\linewidth,trim={0 10 0 10}, clip]{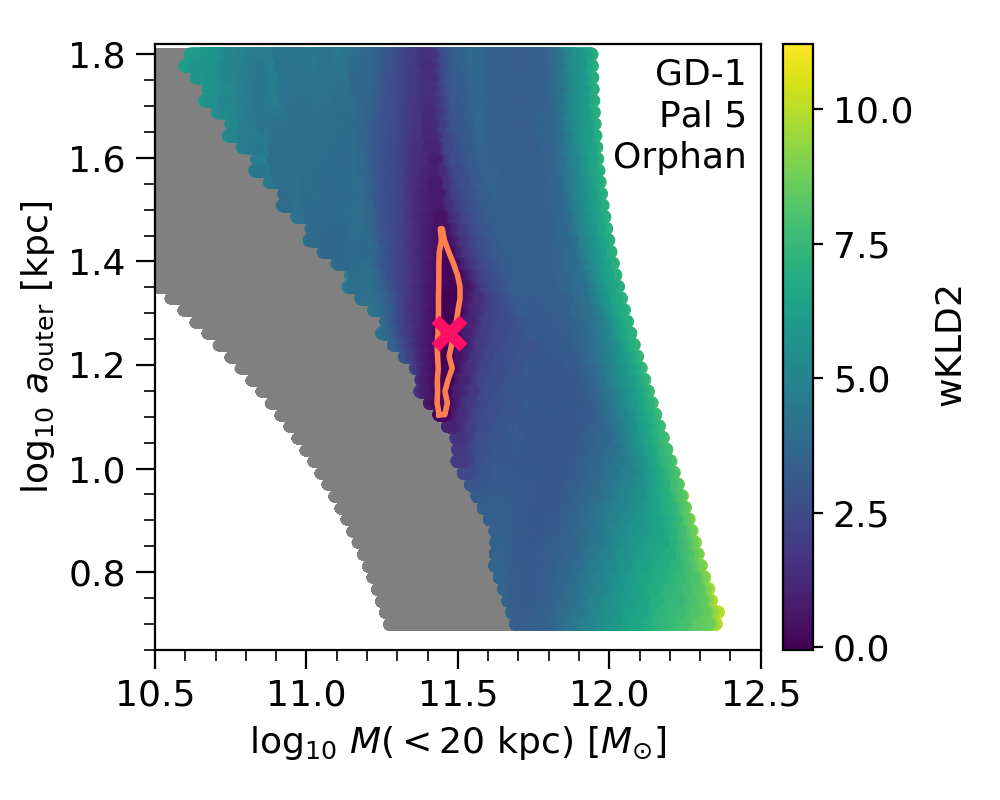}
\caption{As in Figure~\ref{fig:individual}, but showing the weighted combined data results for the two-component potential. {\bf Top:} results for our four-stream data set. {\bf Bottom:} results for the combination of GD-1, Orphan and Pal 5.}
\label{fig:combined_all_5par}
\end{figure}

The results of our analysis of combined stream data sets are shown in Figure~\ref{fig:combined_all_5par} in the enclosed mass - $a_{\mathrm{outer}}$ plane. The top panel shows the results of combining all four streams, while the bottom panel shows the results of combining just GD-1 and Pal 5.

The enclosed mass estimates of the two sets of combined results are consistent with each other within $1\sigma$. We find $M(<20 \ \mathrm{kpc})=3.12^{+3.21}_{-0.46} \times 10^{11} \ M_{\odot}$ for the combination of four streams and $M(<20 \ \mathrm{kpc})=2.96^{+0.25}_{-0.26} \times 10^{11} \ M_{\odot}$ for the combination of GD-1, Orphan and Pal 5. As for the single-component potential, the three-stream combination returns confidence limits that are smaller than the limits for the four-stream combination.
Notably, the mass estimates from combined data sets appear robust against the adopted model for the potential: they are consistent with those obtained with a single-component potential ($\sim 4.18^{+1.63}_{-0.70}$ and $\sim 3.08^{+0.39}_{-0.35}$ $\times 10^{11} \ M_{\odot}$, respectively). The change in the best-fit estimates is $25 \%$ in the case of the four-stream combination and $4 \%$ in the case of the three-stream combination. 

The analysis of the combined data sets again show no significant improvement over the  individual results of Pal 5 and Orphan: the four-stream combination places an upper limit of $e_{\mathrm{outer}} \leq 1.1$ while the combination of GD-1, Orphan and Pal 5 limits it to $\leq 1.07$, both with a best-fit value of $\sim 1$.

The other parameters cannot reliably be constrained with the current data. As seen on Figure~\ref{fig:combined_all_5par}, the limits on the scale length of the outer component extend almost the entire prior range when the data of all four streams is used. The same applies for the scale length of the inner component, the total mass and the mass ratio parameters. In contrast, with the combination of GD-1, Orphan and Pal 5 data, we see a smaller uncertainty region for $a_{\mathrm{outer}}$. However, the lower limit of this region is defined by the the edge of the allowed parameter space of potentials producing bound orbits for all stars. Relaxing this strict constraint would likely increase this region, also allowing lower $a_{\mathrm{outer}}$ (see discussion in Section~\ref{sec:assumptions}). We conclude that it is the combinations of the five model parameters that give a fixed enclosed mass, rather than individual parameter values, that are constrained.

\begin{figure*}
\centering
\includegraphics[trim={35 0 35 30}, clip, width=0.99\linewidth]{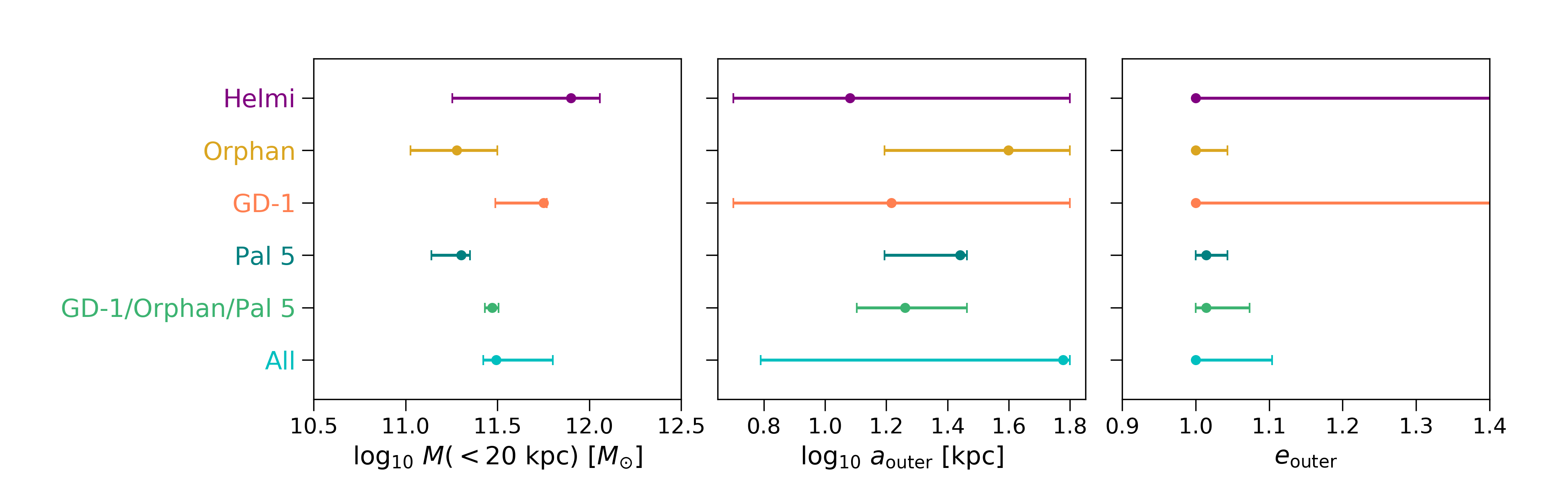}
\caption{Comparison of the best-fit parameter values for the two-component potential with their $1 \sigma$ confidence intervals for enclosed mass (left), scale length (middle) and flattening (right). Results for individual streams are labeled with the stream name. Combined results are labeled with ``GD-1/Orphan/Pal 5'' for the three-stream combination and ``All'' for the four-stream combination. We remind the reader that $e > 1$ corresponds to a oblate potential in our convention.}
\label{fig:error_5par}
\end{figure*}

Figure~\ref{fig:error_5par} summarises the results of the two-component model as given in Table~\ref{tab:results_5par}.

\begin{table*}
\begin{tabular}{ c c c c c c c c}
Streams & $\mathrm{N}_*$ &  $M(<20 \ \mathrm{kpc}) \times 10^{11} \ [M_{\odot}]$ & $a_{\mathrm{outer}} \ [\mathrm{kpc}]$  & $e_{\mathrm{outer}}$ & $a_{\mathrm{inner}} [\mathrm{kpc}]$ & $M_{\mathrm{tot}} \times 10^{12} \ [M_{\odot}]$ & $k$ \\
\hline
GD-1 & 69                    & $5.64^{+0.25}_{-2.56}$ & $16.46^{+46.64}_{-11.44}$ & $1.00^{+0.94}_{-0.00}$ & $4.69^{+0.32}_{-3.69}$ & $3.16^{+0.00}_{-2.65}$  & $0.01^{+0.29}_{-0.00}$\\
\\
Helmi & 401                   & $7.93^{+3.51}_{-6.14}$ & $12.07^{+51.03}_{-7.06}$ & $1.00^{+0.86}_{-0.00}$ & $3.49^{+1.52}_{-2.49}$ & $2.80^{+0.36}_{-2.49}$ & $0.02^{+0.28}_{-0.01}$\\
\\
Pal 5 & 136                  & $2.01^{+0.23}_{-0.63}$ & $27.59^{+1.46}_{-11.97}$ & $1.01^{+0.03}_{-0.01}$ & $5.01^{+0.00}_{-3.08}$ & $2.80^{+0.36}_{-1.97}$ & $0.01^{+0.09}_{-0.00}$\\
\\
Orphan & 117                  & $1.91^{+1.26}_{-0.84}$ & $39.62^{+23.47}_{-24.00}$ & $1.00^{+0.04}_{-0.00}$ & $1.53^{+3.48}_{-0.53}$ & $3.16^{+0.00}_{-1.97}$ & $0.04^{+0.21}_{-0.03}$\\
\\
GD-1/Orphan/Pal 5 weighted & 322     & $2.96^{+0.25}_{-0.26}$ & $18.25^{+10.81}_{-5.54}$ & $1.01^{+0.06}_{-0.01}$ & $3.27^{+1.74}_{-2.27}$ & $1.95^{+1.21}_{-0.89}$ & $0.01^{+0.11}_{-0.00}$\\
\\
GD-1/Helmi/Orphan/Pal 5 weighted & 723 & $3.12^{+3.21}_{-0.46}$ & $59.92^{+3.18}_{-53.75}$ & $1.00^{+0.10}_{-0.00}$ & $2.87^{+2.15}_{-1.87}$ & $3.16^{+0.00}_{-2.33}$ & $0.12^{+0.18}_{-0.11}$\\
\hline
\end{tabular}
\caption{The individual and combined stream results for a two-component potential. Best-fit parameters are given with their $1\sigma$ confidence intervals. Note that $e > 1$ corresponds to a oblate potential in our convention.}
\label{tab:results_5par}
\end{table*}

\section{Validation}
\label{sec:validation}

Several types of different tests of this method with mock data have previously been performed. In \cite{Sanderson2015}, the authors demonstrated that this method works for mock streams integrated in an isochrone potential when also fitting an isochrone potential. In \cite{Sanderson2017}, the authors showed that they could recover a good approximation to a simulated cosmological halo potential when fitting a simple, spherical NFW potential. This was a simpler model than the ones we fit in this work, and had a larger mismatch to the shape of the simulated halo that was being fitted (which was more triaxial than is expected for real halos) and to its radial profile (which was pronouncedly not NFW) than we expect to be the case for the St\"ackel models that we fit here, which have already been shown to accommodate a MW-like rotation curve and allow for a variation in the degree of flattening with radius. 
    
\cite{Sanderson2015} and \cite{Sanderson2017} also studied the effect of observational errors on the performance of this method extensively. They found that including the Gaia errors serves to expand the uncertainty regions slightly, compared to the error-free case, but that otherwise the results are not significantly affected. This is supported by our findings, as we will show in this section.

These works also found that the total number of stars used is less important than the number of different streams represented. Although it is perhaps a little surprising at first glance that we get good results with so few streams (for comparison, \cite{Sanderson2017} used 15 streams and \cite{Sanderson2015} showed that about 20-25 streams are needed for the error bar sizes to converge) \cite{Sanderson2015} also showed that already with 5 they started to get a relatively unbiased answer in those tests (see discussion in their Section 7). Moreover, these tests were performed only with satellite streams which are thicker than the globular cluster streams that we make use of in this work.

In this section we further validate our results by discussing the effect of measurement errors, considering the consequences of cleaning our sample to restrict the analysis to the more informative stars, reviewing our decision to discard potentials that produce unbound stars, exploring the orbits that our results would produce for the individual streams and analysing stream orbital phase information.

\subsection{Tests of fitting assumptions}
\label{sec:assumptions}

\begin{figure*}
\centering
\includegraphics[width=0.98\linewidth, trim={0 0 0 9}, clip]{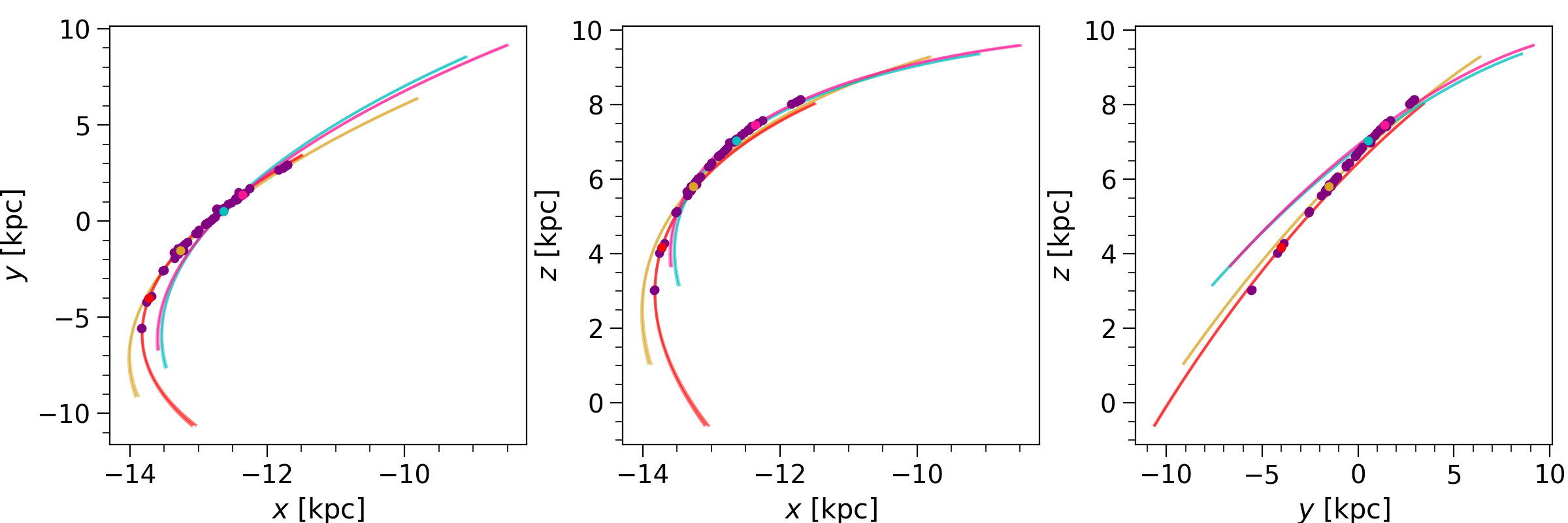}\\
\includegraphics[width=0.98\linewidth, trim={0 0 0 9}, clip]{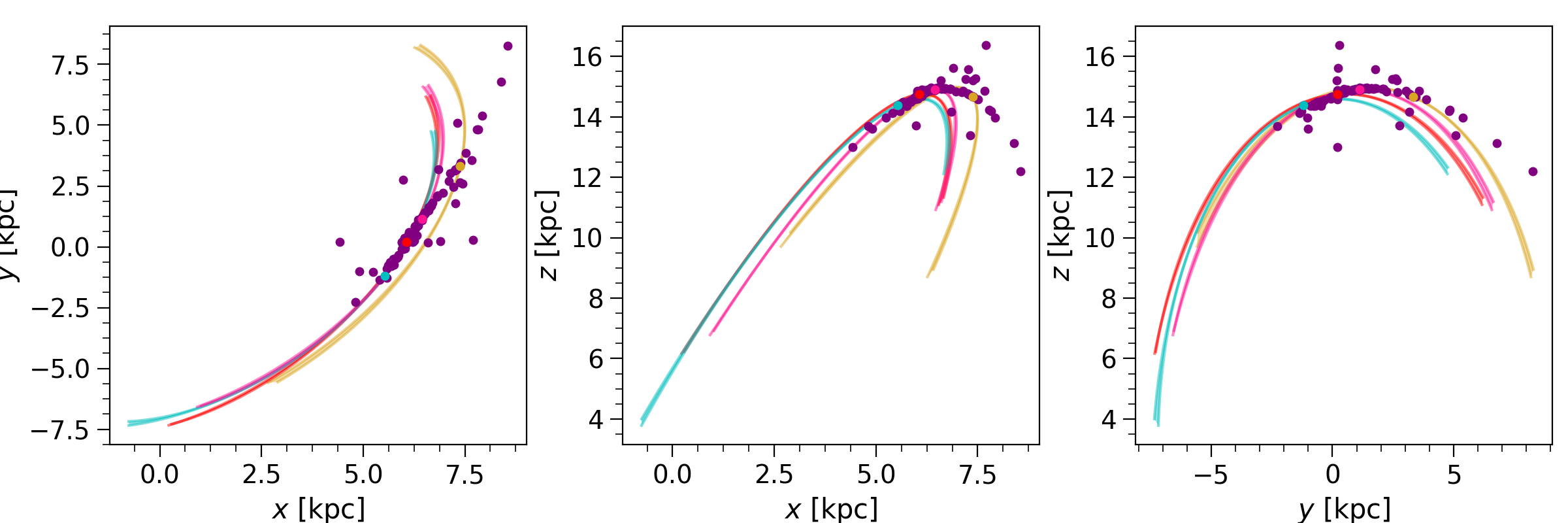}\\
\includegraphics[width=0.98\linewidth, trim={0 0 0 9}, clip]{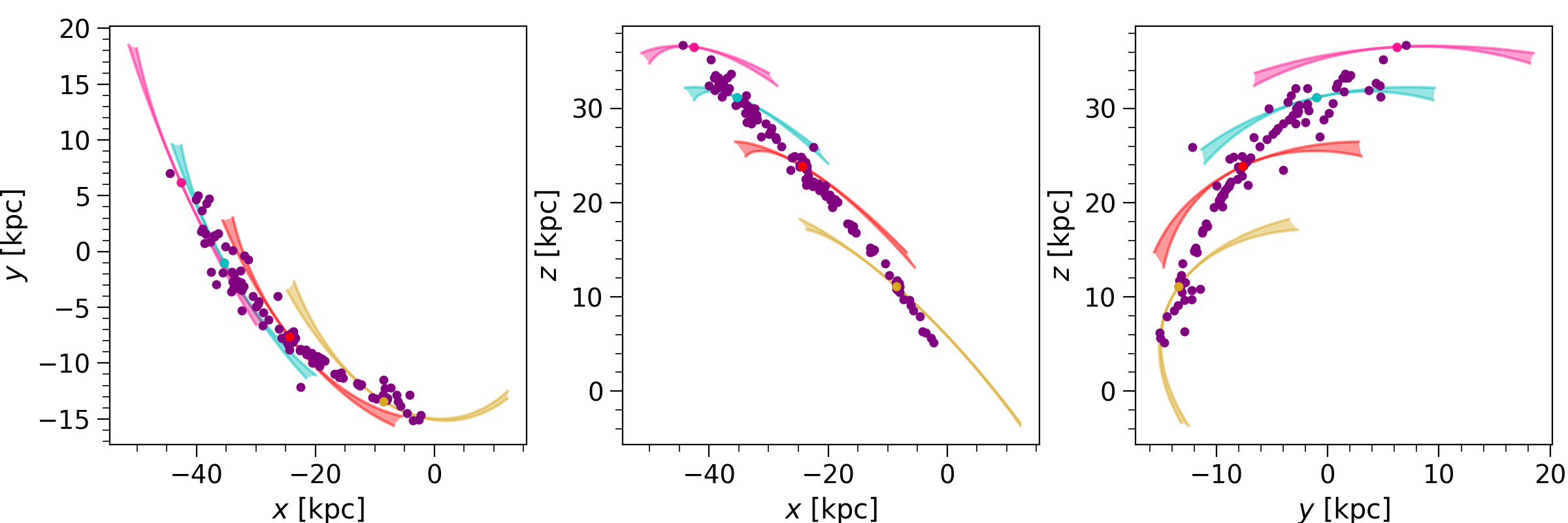}
\caption{ The orbits for the GD-1 (top), Pal 5 (middle) and Orphan (bottom) streams. These shaded regions correspond to the allowed orbits of the stars whose current position has been marked with a dot of the same colour. The edges of the shaded regions are defined by the potentials that produce the highest and the lowest enclosed mass within 20 kpc among those that are within $1 \sigma$ of the best-fit results of the combined analysis of GD-1, Orphan and Pal 5 streams. Other stars in the stream are shown with purple dots. Axes are in the Galactocentric frame.}
\label{fig:orbits}
\end{figure*}

To evaluate the impact that measurement errors would have on our results, we run a test with the GD-1 sample where the input positions and velocities are modified in the following manner.
We draw the new sky positions, proper motions and radial velocities for each GD-1 star from a normal distribution centered at the their measured values with a width determined by the measurement uncertainties. These new values are assigned as the stars' current observables. The estimated distances do not have formal measurement errors, but we evaluate an uncertainty of 0.5 kpc based on the spread of the track measurements as shown in the top left panel of Figure~\ref{fig:tracks}. We again draw the new distances from a normal distribution centered at our estimated distances with the width of 0.5 kpc. We transform the modified observables to Galactocentric $(\boldsymbol{x}, \boldsymbol{v})$ and repeat our analysis for the  two-component potential model. The new result is consistent with the original one. The $1 \sigma$ region for the enclosed mass parameter shifts only slightly compared to the original result: while previously the $1 \sigma$ region encompassed values from $3.08$ to $5.89 \times 10^{11} M_{\odot}$, with the perturbed observables it shifts to include a region from $3.33$ to $6.07 \times 10^{11} M_{\odot}$. The best-fit value itself differs by $\sim 8\%$ from the result quoted in Table~\ref{tab:results_5par}.

To see how our results would change when relaxing the strict condition that no stars must be unbound in accepted potentials, we repeat our analysis for the GD-1 sample allowing for a maximum of 10\% of the stars to be unbound. We find that our results for all parameters are unaffected: enforcing the strict no-unbound-stars rule does not have any impact on the GD-1 results.

When we repeat this analysis for other streams, we find that:
\begin{itemize}
\item The enclosed mass parameter is similarly unaffected in the case of the Orphan and Helmi streams. Pal 5 enclosed mass region however does shift somewhat. While the original $1 \sigma$ region encompasses values from $1.38$ to $2.24 \times 10^{11} M_{\odot}$, the region that also allows 10\% of the stars to be unbound contains values from $1.59$ to $2.83 \times 10^{11} M_{\odot}$. The best-fit value itself differs by $\sim 7\%$ from the result quoted in Table~\ref{tab:results_5par}.
\item The uncertainty regions of Pal 5 and Orphan now reach lower in $a_{\mathrm{outer}}$ as they are no longer limited by the edge of the allowed parameter space (the grey points in Figure~\ref{fig:combined_all_5par}). This confirms once again that we are unable to meaningfully constrain any parameters besides the enclosed mass. 
\end{itemize}

To estimate the effect of cleaning up our stream sample by making selections in angular momentum, we re-analyse the GD-1 stream in the two-component potential without making any cuts to the original sample of 82 stars. The most significant change in the results is that the $1 \sigma$ region for the enclosed mass has now been slightly extended. While the previous $1 \sigma$ region encompasses values from $3.08$ to $5.89 \times 10^{11} M_{\odot}$, the region resulting from the uncleaned sample contains values from $3.05$ to $6.14 \times 10^{11} M_{\odot}$. The best-fit value itself differs by $\sim 6\%$ from the result quoted in Table~\ref{tab:results_5par} which is well within the $1\sigma$ region for both fits. This is consistent with our expectation that minor selections in constants-of-motion space to clean up outliers slightly improves the constraints but does not significantly bias the fit. 
When repeating our analysis with the uncleaned samples of Pal 5 and Orphan streams we find that each stream has at least one star that is unbound across all the trial potentials. We therefore additionally relaxed the no-unbound-stars restriction, allowing for a maximum of 10\% of the stars to be unbound as in the previous paragraph. The combined effect further extends the uncertainty regions while having only a minor effect on the best-fit value. For Pal 5 the uncertainty region now extends from $1.54$ to $2.96 \times 10^{11} M_{\odot}$, only a small increase from when we considered the cleaned sample with 10\%-unbound-stars in the previous paragraph. The best-fit value experiences $\sim 0.5\%$ change compared to the previous case, and a total of $\sim 8\%$ change compared to the original result quoted in Table~\ref{tab:results_5par}. For Orphan, we see the $1\sigma$ region increase from $1.07$ to $3.17 \times 10^{11} M_{\odot}$ in the original case to $0.97$ to $3.42 \times 10^{11} M_{\odot}$. The best-fit value has changed by $\sim 7\%$ from the result quoted in Table~\ref{tab:results_5par}.

In conclusion, none of the above mentioned choices affects our main result, the enclosed mass estimates, by more than 8 percent; all changes are far below $1 \sigma$.

\subsection{Predicted orbits}
Figure~\ref{fig:orbits} shows the results of orbit integration using the results of the combined GD-1, Orphan and Pal 5 analysis. We track 4 stars in each of the GD-1, Pal 5 and Orphan streams. We do not show the orbits for the Helmi stream because the Helmi stream's stars are phase-mixed and do not exhibit coherent stream-like structure in position-space. Each shaded region corresponds to the star whose current position has been marked with a dot of the same colour. The edges of the shaded regions are defined by the potentials that produce the highest and the lowest $M(<20\ \mathrm{kpc})$ among those that are within $1 \sigma$ of the best-fit potential. The Figure confirms that the orbits predicted by our analysis from GD-1 and Pal 5 are locally aligned with the tracks of these streams. 

Even though we see plausible orbits also for Orphan stream's stars, we do not expect perfect alignment here for two reasons. First, the Orphan stream is a remnant of a dwarf galaxy and as such has a larger energy spread than the GD-1 and Pal 5 streams. Second, the Orphan stream has been shown to be perturbed by the LMC and any potential that neglects the influence of the LMC is therefore unlikely to find a perfect fit to the Orphan stream track.

\subsection{Action space and stream orbital phase}
\label{sec:phase}

In this section we discuss the action-space characteristics of our streams and validate the fit with orbital phase information that is not utilized in our fitting method.

\begin{figure*}
\centering
\includegraphics[width=1.\linewidth, trim={0 0 0 0}, clip]{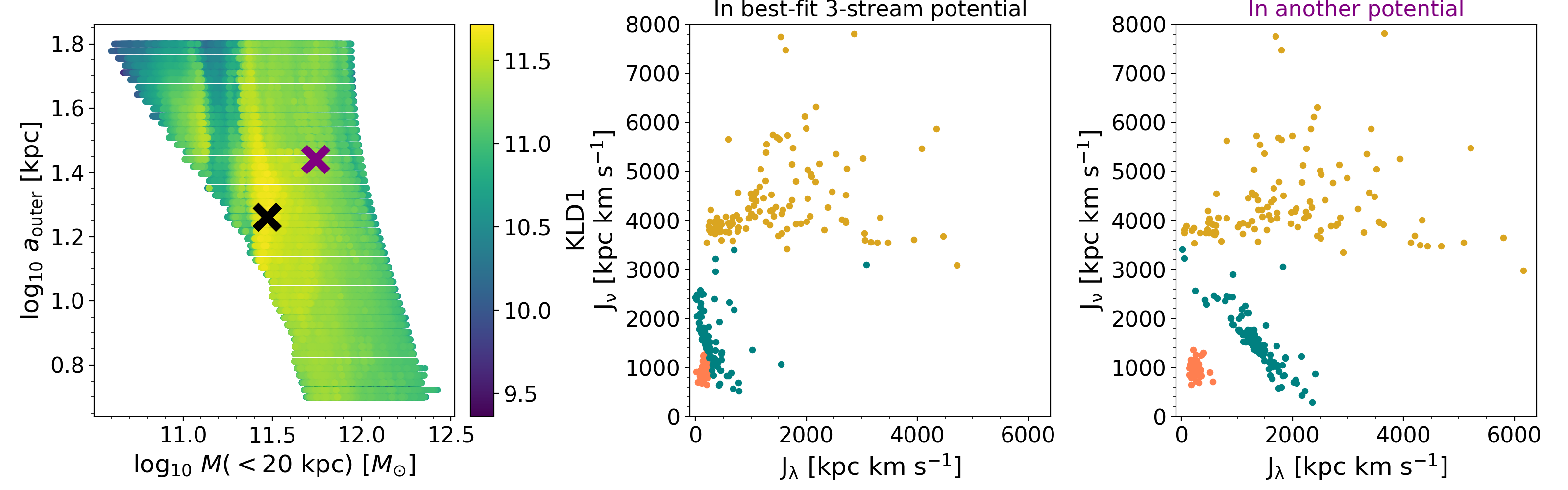}\\
\caption{Comparison of actions of GD-1 (orange points), Pal 5 (teal points) and Orphan (yellow points) produced by two different two-component potentials: the best-fit potential of the three-stream combination and another potential. Left: KLD1 values of the three-stream combination as a function of enclosed mass and scale length of the outer component. The higher the KLD1 value the more clustered the action-space. Center: action distribution of the best-fit potential (black cross in left panel). Right: action distribution of a different potential with KLD1 0.46 lower than the corresponding best fit (purple cross in left panel). All Orphan stars have been shifted up by $2500\ \rm kpc\ km\ s^{-1}$ in $J_{\nu}$ for clarity on both panels.}
\label{fig:actions_comb}
\end{figure*}

In the left panel of Figure~\ref{fig:actions_comb} we show an example of the KLD1 contours as a function of enclosed mass and scale length of the outer component, for the GD-1, Pal 5 and Orphan streams in the two-component model. For each data set, there is a clear single peak in parameter space; the best-fit model is marked with the black cross. To give an idea of the difference in clustering associated with a given difference in KLD1, we also mark another location in parameter space that has KLD1 lower than the best fit (purple cross) for comparison. In the central panel, we show the action distribution corresponding to the best fit models (black cross), which should be the most clustered. For comparison, in the right panels, we show the action distribution corresponding to the potential with lower KLD1 (purple cross), which shows a visibly less clustered action distribution. The centroid of the cluster has also moved, which is expected since the stream will be on different orbits in the two different potentials. 

The middle panel of Figure~\ref{fig:actions_comb} should also be compared with the middle panels of Figure~\ref{fig:actions} which show individual stream best-fit action spaces. 

\begin{figure}
\centering
\includegraphics[width=1.\linewidth, trim={0 44 0 0}, clip]{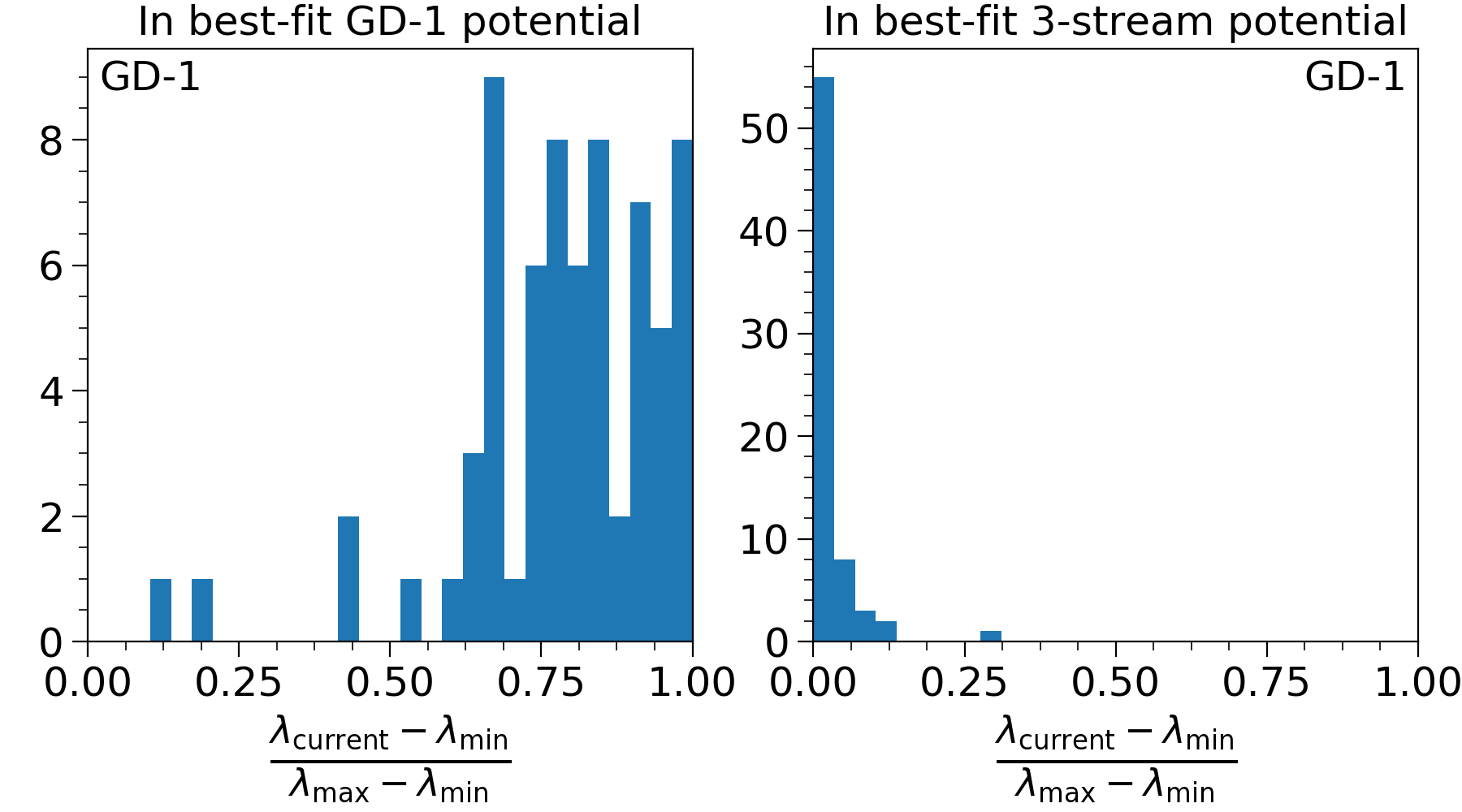}\\
\includegraphics[width=1.\linewidth, trim={0 3 0 0}, clip]{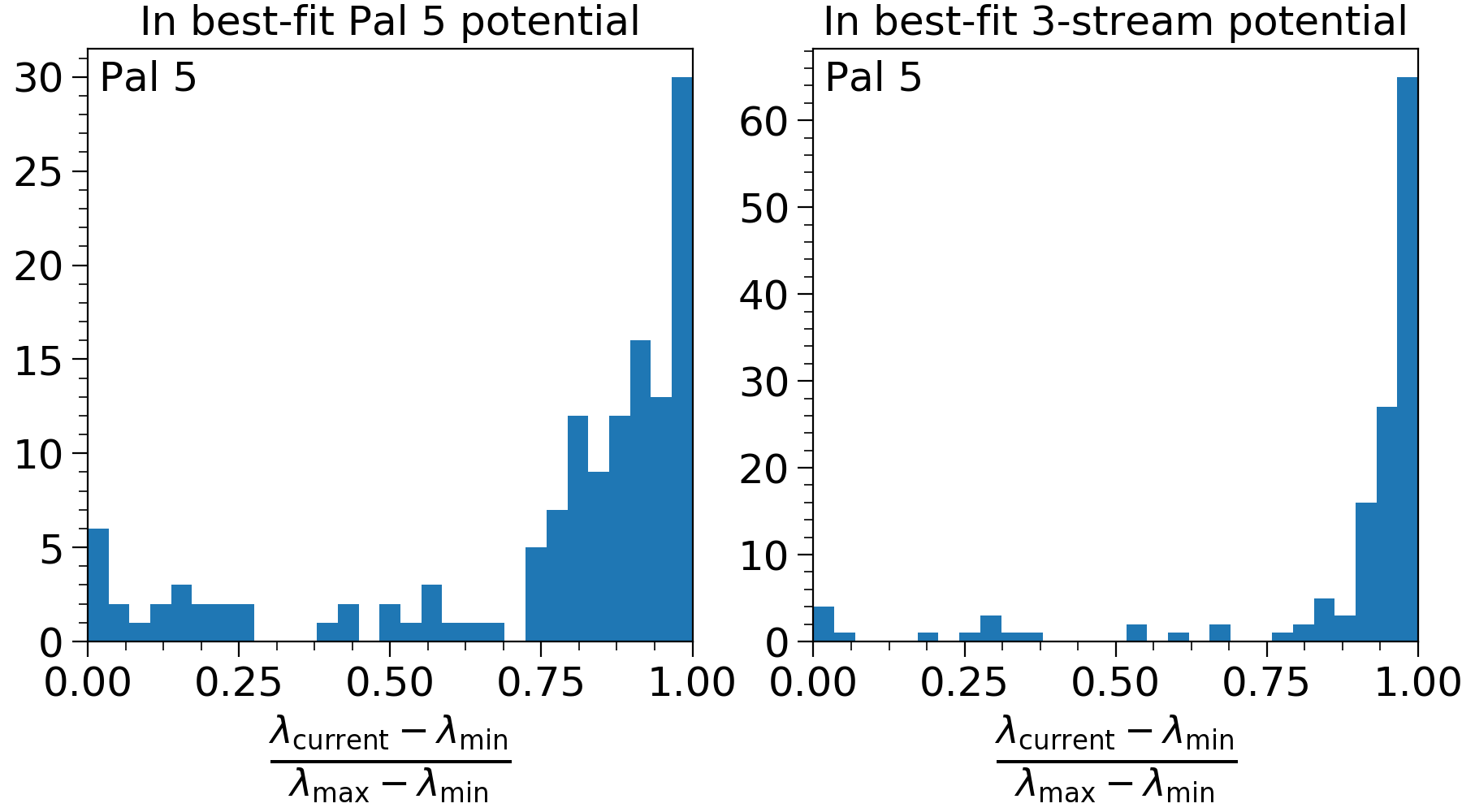}\\
\caption{Histograms of the current $\lambda$ for stars in GD-1 and Pal 5 in different two-component potentials, normalized by the pericentre and apocentre positions of each star's orbit to serve as a proxy for the orbital phase, with pericentre at 0 and apocentre at 1. {\bf Top left panel:} The approximate orbital phase of GD-1 stars assuming the best-fit GD-1 potential. {\bf Top right panel:} The position of GD-1 stars assuming the potential  best-fit to the combined GD-1, Orphan and Pal 5 data. {\bf Bottom left panel:} The position of Pal 5 stars assuming the best-fit Pal 5 potential. {\bf Bottom right panel:} The position of Pal 5 stars assuming the potential best-fit to the combined GD-1, Orphan and Pal 5 data.}
\label{fig:lambda_hist}
\end{figure}

Our fit only makes use of the action coordinates, while the information about the orbital phase of each star, are not used. Figure~\ref{fig:lambda_hist} shows the histograms of the current $\lambda$ for stars in GD-1 and Pal 5 streams in different potentials. The current $\lambda$ positions of the stars have been normalized to lie between the $\lambda_{min}$ and $\lambda_{max}$ positions of each star's orbit in the considered potential. Since $\lambda$ corresponds roughly to the radial direction in our spheroidal coordinate system, the $\lambda_{min}$ and $\lambda_{max}$ can be viewed as the pericentre and apocentre positions of these stars' orbits. Figure~\ref{fig:lambda_hist} then shows the approximate orbital phase of each star in the given potential.
Most of the GD-1 stars are near their apocentre in the best-fit potential obtained using GD-1 alone, in tension with other studies that have found GD-1 to be near its pericentre (see Section~\ref{sec:catalogue}). We also see that there is considerable variation in the phases of the stars in the best-fit GD-1 potential. These issues disappear when we consider the GD-1 stars in the potential that was best fit to the combination of GD-1, Orphan and Pal 5 data. All stars are now near their pericentre, as expected, and there is a very clear agreement between the phases of the stars. 

A similar, although less striking, behavior is evident when we consider Pal 5 stars. If we assume the best-fit Pal 5 potential, most of the stars are near the apocentre, as expected, but there are still quite a few stars near pericentre as well. When switching to the potential that is best-fit for the combination of GD-1 and Pal 5 data, we once again see a better agreement in the orbital phases of the Pal 5 stars. The peak at apocentre is much better defined and there are fewer outliers at all other phases. 

These instances are illustrative of the biases to which fits of individual streams are susceptible. For individual streams, maximal clustering can occur with the wrong potential exactly because we are not including any information about the current phase of the stars. Stars are sorted along the stream based on their energies and this sorting of energies should also be present in action space: stars that have higher energies are on slightly larger orbits, and this should be reflected in action space. However, since we are neglecting the phase information (i.e. we do not know how far apart the stars are), we can inadvertently lose this expected sorting of energy and allow the formation of clusters that are smaller than the ones produced by the true potential. 
However, the potential at which such a biased solution occurs will be different for each different stream. Fitting multiple streams simultaneously prevents these individual biases from being confused with the true potential, since no individually biased solution is preferred by more than one stream.

An example of this can be seen in Figure~\ref{fig:actions}, where we show a comparison of the action space corresponding to individual-stream and three-stream best-fit potentials and the energy gradient present in the formed clusters.  

The most prominent example is given by the GD-1 stream, for which maximal clustering is reached when all its stars are compressed near $J_{\lambda} \sim 0$ (first row middle panel of Figure~\ref{fig:actions}). As discussed above, this is caused by a high enclosed mass that forces the stars to be positioned at the apocentres of their orbits. However, we also see that there is no clear energy gradient in the action-space cluster, leading us to suspect that what we are selecting is not actually the true potential. Turning to the action distribution of GD-1 in the best-fit three-stream potential (first row right panel of Figure~\ref{fig:actions}), we see a clear improvement: first of all, the stream is no longer confined to near $J_{\lambda} \sim 0$ (which is unreasonable as we expect the orbit of GD-1 to have some radial variation) and, second, the energy gradient along the action-space cluster is now evident.
The same effect can be observed in the case of the Orphan stream (third row in Figure~\ref{fig:actions}). It is, however, not as obvious in the case of Pal 5 stream (second row in Figure~\ref{fig:actions}). The latter is in line with expectation, as the action space of the Pal 5 stream changes the least between the two potentials and the KLD1 value also changes the least between these potentials. That is to say, the KLD1 value of the best-fit three-stream potential (purple cross in Figure~\ref{fig:actions}) also has a fairly high value in the Pal 5-only analysis, while the difference is greater for both Orphan and GD-1. 

\section{Comparison and discussion}
\label{sec:discussion}
In this section, we put our results into a broader context. We first compare our enclosed mass estimates to those obtained previously with different techniques, applied to the same stellar streams (Section ~\ref{sec:en_mass_disc}). Next, we compare our results with other, more general mass estimates (see Figure~\ref{fig:enc_mass}). Then we comment on the overall potential shape by qualitatively comparing our inferred Galactic rotation curve with recent data \citep{Eilers2019}, and with the curve obtained from the widely used potential from \cite{Bovy2015} (Section~\ref{sec:rotcurv}).

\subsection{Enclosed mass estimates}
\label{sec:en_mass_disc}

Our most precisely constrained parameter is the enclosed mass at 20 kpc from the GD-1 stream $5.64^{+0.25}_{-2.56}$ ($4.73^{+0.33}_{-1.05}$)\,$\times\, 10^{11} M_{\odot}$, the Pal 5 stream $2.01^{+0.23}_{-0.63}$ ($2.73^{+0.60}_{-0.74}$)\,$\times\, 10^{11} M_{\odot}$, the Orphan stream $1.91^{+1.26}_{-0.84}$ ($1.89^{+1.05}_{-0.60}$)\,$\times\, 10^{11} M_{\odot}$ and their combined data sets $2.96^{+0.25}_{-0.26}$ ($3.08^{+0.39}_{-0.35}$)\, $\times\, 10^{11} M_{\odot}$. Here we quote our best-fit values with $1 \sigma$ uncertainties for the two-component (single-component) model. The single-component accepted ranges are smaller and almost entirely contained within the two-component model accepted ranges. The relative change in best-fit values between the two models is $19\%$, $26\%$, $1\%$ and $4\%$, respectively.

Our two-component model enclosed mass estimates should be compared with recent mass measurements performed on the same streams with independent techniques.
With the orbit-fitting method, \cite{Malhan2019} find a mass of $M(< 20 \ \mathrm{kpc}) = 2.5 \pm 0.2 \times 10^{11} \ M_{\odot}$ with GD-1 data. Our best-fit GD-1 measurement is more than twice as high as this. However, as explained in Section~\ref{sec:phase}, the measurement we yield with GD-1 data using our method is a biased one that places all GD-1 stars incorrectly at apocentre. A more meaningful comparison would be with our results from the three-stream combined data. This is in agreement with the \cite{Malhan2019} measurement on the low mass end.

With their streakline modelling \cite{Kupper2015} obtain $M(< 19\ \mathrm{kpc}) = (2.1 \pm 0.4) \times 10^{11} \ M_{\odot}$ using Pal 5 data. This should be compared with our $M(< 19\ \mathrm{kpc}) = 1.81^{+0.21}_{-0.53} \times 10^{11} \ M_{\odot}$ when only Pal 5 data is used or with $M(< 19\ \mathrm{kpc}) = 2.71^{+0.20}_{-0.20} \times 10^{11} \ M_{\odot}$ from the combined three-stream result.
The latter is in agreement with results obtained by \citet{Kupper2015} on low mass end, while the Pal 5 only result shows much greater agreement.

\cite{Erkal2019} fit the Orphan stream data using realistic stream models generated by the modified Lagrange Cloud Stripping technique and also include the influence of LMC in their Milky Way potential. Their result of $M(< 50 \ \mathrm{kpc}) = 3.80 ^ {+0.14}_{-0.11} \times 10^{11} \ M_{\odot}$ should be compared with our Orphan-only measurement of $M(< 50 \ \mathrm{kpc}) = 6.78 ^ {+2.00}_{-2.23} \times 10^{11} \ M_{\odot}$. Our measurement is considerably higher than that of \cite{Erkal2019}. We speculate that the cause for this could at least partly be the fact that we neglect the effect of the LMC.

In contrast to the other streams, the Helmi stream stars fail to find a preferentially high clustered configuration in action space for our potential models, resulting in very poor parameter determination. 
There are several reasons for the strikingly weak parameter constraints provided by the Helmi stream.
First, all stars in our Helmi stream sample are within the 6D {\em Gaia} volume, which is likely only a small segment of the whole, phase-mixed stream. A spatially limited ``local sample'' like this produces a subtly biased subset in action space \citep{McMillanBinney2008} that could be interfering with the fit for this stream.
Given its confinement to the 6D {\em Gaia} volume, the Helmi stream also has the most limited range of galactocentric distances of any of the streams we include in our data set. This makes it most susceptible to degeneracies between total mass and scale radius, as discussed in \citet{Bonaca2014} and \citet{Sanderson2016}.
Second, the progenitor of the Helmi stream is a dwarf galaxy, which are generally hotter (in terms of their velocity dispersion) than streams from globular clusters. This means that the Helmi stream stars naturally occupy a larger volume in action space at lower density, compared to streams from globular cluster origin such as GD-1 or Pal 5, and thus have a lower maximum value of the KLD. This in turn places weaker constraints on the potential parameters.
Finally, as discussed in \S\ref{sec:catalogue}, of the four streams we consider in this work, three are selected by standard observational cuts, where interlopers will likely have quite different actions \citep[for an illustration, see][]{DonlonNewberg2019}. The Helmi stream is however identified and selected as a cluster in angular momentum space so any interlopers remaining are likely to overlap with the true stream members in action space, also increasing the minimum size of the action-space cluster. 

\begin{figure*}
\centering
\includegraphics[width=0.99\linewidth, trim={10 10 10 10}, clip]{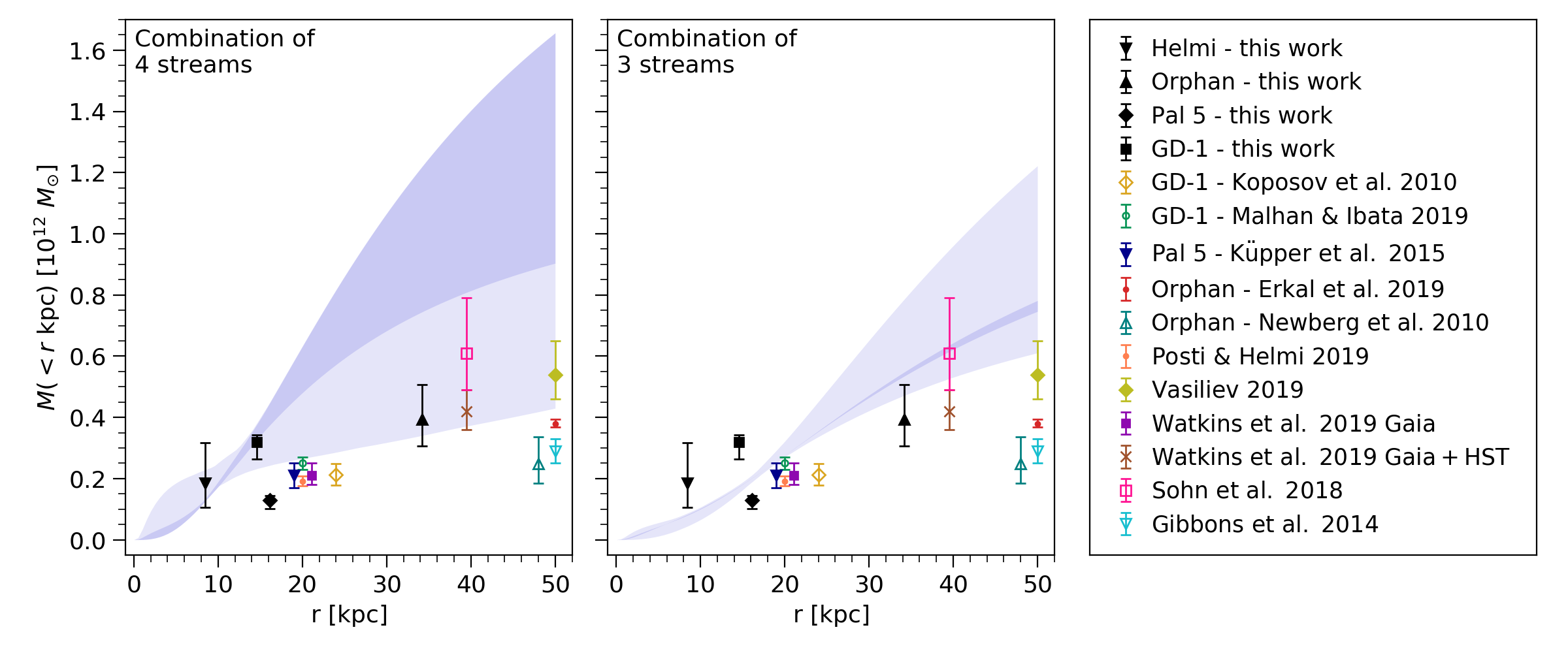}\\
\caption{Comparison of our two-component potential results with previous measurements of the enclosed mass at different radii. The light blue shaded area represents the combined results for all four streams (left panel) and for GD-1, Orphan and Pal 5 (right panel). The darker shaded regions show the subset of the potentials that are compatible with the current measurements of the Local Standard of Rest velocity. The black markers signify individual stream results at their respective average radii. The coloured markers show measurements of the enclosed mass by other authors. The markers showing the results of \protect\cite{Koposov2010} and  \protect\cite{Newberg2010} are slightly offset from 20 kpc and 50 kpc, respectively, for clarity.}
\label{fig:enc_mass}
\end{figure*}

A visual comparison of our results with previous enclosed mass measurements is shown in Figure~\ref{fig:enc_mass}. Our mass estimates agree with some of the recent measurements, while generally allowing for masses that are higher than those from other measurements. We speculate that these systematics are likely due to insufficiencies in our St\"ackel model of the potential and to the limited phase space explored by the data set of 4 streams.

\subsection{The Galactic rotation curve}
\label{sec:rotcurv}
\begin{figure*}
\centering
\includegraphics[width=0.97\linewidth, trim={0 124 0 49}, clip]{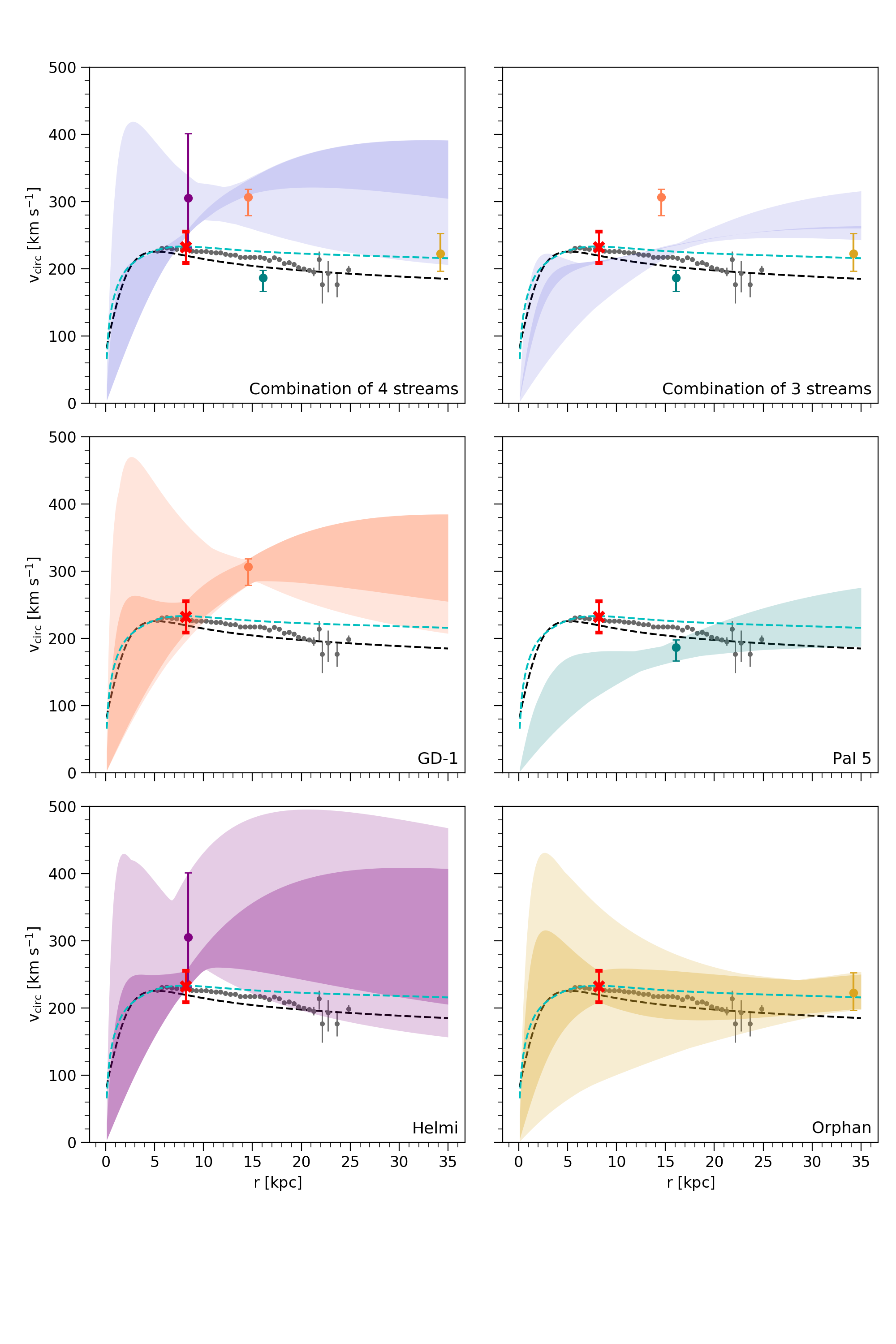}
\caption{Rotation curves corresponding to the results of the two-component potential model. The lighter shaded region shows the rotation curves for potentials within $1\sigma$ of the best-fit for each data set. The purple, orange, teal and yellow data points correspond to the results of the Helmi, GD1, Pal 5 and Orphan streams, respectively: they show the rotation velocity of the best-fit potential at the mean Galactocentric distance of the stars in that stream. The darker shaded region shows the subset of the rotation curves that are compatible with the current measurements of the Local Standard of Rest velocity. For comparison, the dashed black and cyan lines are the rotation curves from the \texttt{galpy} \texttt{MWPotential2014} \protect\citep{Bovy2015} and \protect\cite{McMillan2017}, respectively, and the grey dots represent the data from \protect\cite{Eilers2019}.}
\label{fig:vcirc}
\end{figure*}

To check the global performance of our results, we calculate the resulting rotation curves for our two-component potentials (shaded areas in Figure~\ref{fig:vcirc}) and benchmark them against the rotation curves of \texttt{galpy}'s \texttt{MWPotential2014} \citep{Bovy2015} (black dashed line) and \cite{McMillan2017} (cyan dashed line) and the data from \cite{Eilers2019} (grey points). The average Galactocentric distance of the stars of each stream are marked with a dot at the rotational velocity curve of their respective best-fit potential. At those locations, we plot ``error bars'' given by the rotation curve values produced by potentials with KLD2 values below 0.5. This is also how the shaded uncertainty regions have been computed for all Galactocentric distances. The darker shaded regions represent the subset of these potentials that also go through the Local Standard of Rest. We have taken the rotational velocity at the Sun's location to be $232$ km s${}^{-1}$ \citep[][and references therein]{Koppelman2018} with an uncertainty of 10\%.

The single component potentials (not shown here) do {\em not} provide a good match to the shape of the Galactic rotation curve: their inner rising slope is too shallow and their peaks are too far out.
The two-component potentials have much more flexibility and are able to produce a realistic rotation curve. However, our uncertainties are large and we note that the overall normalisation remains somewhat high with respect to the data points from \cite{Eilers2019}, which beyond $\ge 10$ kpc are barely included in the lower rim of the shaded region.
When taking into account the additional constraint of the Local Standard of Rest velocity (marked by the red cross), we are able to further resolve the velocity curves by discarding the surplus potentials.

Finally, a related method to the one used here was proposed by \cite{Yang2020}, who utilize the 2-point correlation function as a measure of clustering in the action space. They model the Milky Way potential as a combination of the disc, bulge and halo components and calculate the actions using the St\"ackel fudge approximation. As opposed to this work, only the parameters of the halo component are explored while others are held fixed. They apply their method to $\sim 77000$ halo stars between 9 and 15 kpc with the full six-dimensional phase space information from Gaia DR2. So, although using a similar approach, they apply it to the halo stars rather than individual stellar streams as we do here. In contrast to our work, they find the best-fit circular velocity curve to be $5-10\%$ lower than previous measurements. We think this discrepancy could be due to two factors. First, they fix the mass of the disc and bulge and fit only the halo component, but with a small range of distances the scale radius is very difficult to constrain, leading back to the total mass--scale radius degeneracy. Thus if the data prefer a slightly less massive disc/bulge their fit would naturally lead to a lower circular velocity at higher radius, since the halo component will be less massive at the radii of the fit. Second, \citet{Yang2020} use a cutoff for the 2-point function that effectively limits the action-space size, and therefore the mass, of clusters in the distribution to structures comparable to or more massive than Gaia-Enceladus, so their clustering analysis uses essentially a completely different set of stars from ours. Therefore we do not consider it too surprising that their result differs somewhat from ours.

\section{Conclusions}
\label{sec:conclusions}
In this paper, we apply the action-space clustering method to known stellar stream data for the first time, and obtain constraints on the Galactic potential. Specifically, we consider members of the GD-1, Pal 5, Orphan and Helmi streams, both individually and in combination. The motivation for a simultaneous fit for multiple streams lies in obtaining more stringent, and above all robust, constraints on the mass profile over a range of Galactic distances. 

Our conclusions are as follows:

\begin{itemize}

\item The most \emph{precise} constraints on the parameters are obtained with the GD-1, Orphan and Pal 5 streams. In contrast, the Helmi stream allows a much wider range of models. We speculate why this is in section~\ref{sec:en_mass_disc}.

\item Even when combining the streams that yield the most precise constraints, the only parameter we can robustly constrain is the enclosed mass, which we calculate at 20 kpc (corresponding to the mean distance for stars in our sample) for comparison across all our fits. For a two-component potential model we find for the GD-1 stream $M(< 20\ \mathrm{kpc}) = 5.64^{+0.25}_{-2.56}$, for the Orphan stream $M(< 20\ \mathrm{kpc}) = 1.91^{+1.26}_{-0.84}$, for the Pal 5 stream $M(< 20\ \mathrm{kpc}) = 2.01^{+0.23}_{-0.63}$ and for their combination $M(< 20\ \mathrm{kpc}) = 2.96^{+0.25}_{-0.26}$. The combination of all four streams in our sample yields $M(< 20\ \mathrm{kpc}) = 3.12^{+3.21}_{-0.46}$. Our best enclosed mass results are consistent with recent measurements, obtained with the same streams.

\item We have shown that fits from individual streams can lead to biases when using the action-clustering method and discussed the causes for this in Section~\ref{sec:phase}. We have also shown that these biases are canceled by the simultaneous analysis of multiple streams. 

\item Some additional bias in the fits to individual streams and possible tension in their combinations could be introduced by the insufficiencies in the St\"ackel model. However, \citet{Sanderson2015} showed in tests with mock streams that even if the model is identical to the form of the true potential, different streams still show different biases and parameter estimations generally do not overlap until a sufficient number of streams is included in the sample. Therefore, these inconsistencies cannot be solely traced back to the use of the St\"ackel potential. What additional bias the use of the two-component St\"ackel potential introduces when modeling a MW-like galaxy will be investigated in future work. In particular, we intend to test the method on streams found in cosmological-hydrodynamical simulations from the FIRE suite .
  
\end{itemize}

Future developments of this work should incorporate more data from known streams, spread over a large range of Galactocentric distances and with reliable 6D information for stellar members. Moreover, our procedure, as tested here, can more broadly be applied to a large ensemble of halo stars with 6D information without knowing stream membership, as originally intended by \cite{Sanderson2015}. Data to do this over a sufficiently broad distance range are within reach in the near future, thanks to upcoming spectroscopic surveys, such as WEAVE \citep{weave2012}, 4MOST \citep{deJong2019}, DESI \citep{DESI2019}, SDSS-V \citep{Kollmeier2017} and H3 \citep{H32019}, that will complement the increasingly precise next data releases of the ESA Gaia mission.

\section*{Acknowledgements}
We thank the referee for thorough and though-provoking comments that led to substantial improvements in the paper. We also thank Tim de Zeeuw for helpful comments and discussion. 
This work has made use of data from the European Space Agency (ESA) mission {\it Gaia} (\url{https://www.cosmos.esa.int/gaia}), processed by
the {\it Gaia} Data Processing and Analysis Consortium (DPAC,
\url{https://www.cosmos.esa.int/web/gaia/dpac/consortium}). Funding for the DPAC has been provided by national institutions, in particular the institutions participating in the {\it Gaia} Multilateral Agreement.
This project was developed in part at the 2019 Santa Barbara Gaia Sprint, hosted by the Kavli Institute for Theoretical Physics at the University of California, Santa Barbara. This research was supported in part at KITP by the Heising-Simons Foundation and the National Science Foundation under Grant No. NSF PHY-1748958.

\section*{Data availability}

The data underlying this article are available in the article and in its online supplementary material.

\appendix

\section{Data collection}
\label{app:A}
We compile the 6D phase space data for three of the four streams used in this work (GD-1, Orphan, and Palomar 5) from various literature sources. In this Appendix we describe in detail the data assembly process for each individual stream (\S~\ref{ap:gd-1}--\ref{ap:pal5}). To convert from the assembled 6D Heliocentric observables to Galactocentric Cartesian coordinates, we further make the following assumptions:
\begin{itemize}
    \item The Sun is located at a distance of $8.2 \ \mathrm{kpc}$ from the Galactic Centre in the direction of the negative X-axis and $27 \ \mathrm{pc}$ above the Galactic plane in the direction of the positive Z-axis \citep{Chen2001}.
    \item The Sun's peculiar velocity is $(11.1, 12.24, 7.25) \ \mathrm{km\ s^{-1}}$ \citep{Schonrich2010}.
    \item The velocity of the local standard of rest is $232 \ \mathrm{km \ s^{-1}}$ \citep[][and references therein]{Koppelman2018}.
\end{itemize}

\subsection{GD-1 data}
\label{ap:gd-1}
We combine data from four literature sources to create a list of identified GD-1 members with full 6D phase space information. \cite{Koposov2010} performed spectroscopic measurements for 23 GD-1 members at the Calar Alto observatory. They also provide stream track distances, i.e. estimates of constant distance for certain intervals of the stream's track. They divide the GD-1 stream into 6 sections based on the stream-aligned longitude coordinate $\phi_1$ and derive a distance to each by isochrone fitting using SDSS photometry. \cite{Willett2009} list a further 48 high-confidence GD-1 members in their Table 2. The table includes individual radial velocities as measured by the SEGUE survey, but no individual distances. 
Table 1 in \cite{Li2017} contains another 20 GD-1 stars with radial velocity information from LAMOST. We select only the 11 stars that were flagged by the authors as high confidence members, i.e. candidates with confidence level 1. In addition, we use the stream track distances from \cite{Li2018} Table 1, determined by fitting isochrones (i.e. using the same strategy as \citealp{Koposov2010}) to 18 different regions along the GD-1 stream. In total, this makes 82 candidate GD-1 members with measured radial velocities. For these stars we assemble the full phase space information in the following way:

\begin{enumerate}
\item The SDSS identifiers given in \cite{Koposov2010} Table 1 and \cite{Willett2009} Table 2 are matched with the corresponding Gaia DR2 identifiers using Simbad.
\item The Gaia DR2 identifiers for these two data sets are used to acquire Gaia DR2 position and proper motion coordinates for each star.
\item The candidate members in the third data set, stars with confidence level 1 in \cite{Li2017} Table 1, are cross-matched with the Gaia DR2 catalogue in TOPCAT using the LAMOST right ascension and declination with a 1 arcsec search radius to find their Gaia DR2 positions and proper motions.
\item LAMOST radial velocities have been shown to be underestimated by $4.5\ \mathrm{km\ s^{-1}}$ \citep{Anguiano2018}. We correct for this by adding $4.5\ \mathrm{km\ s^{-1}}$ to the quoted LAMOST radial velocities. 
\item Using the transformation matrix in the Appendix of \cite{Koposov2010} the stream aligned longitude coordinate $\phi_1$ is calculated using Gaia DR2 right ascension and declination for all stars. 
\item A polynomial of degree 2 is fitted to the combined stream track distance data from \cite{Li2018} Table 1 and \cite{Koposov2010} Table 3 using the inverted measurement uncertainties as weights.
\item The polynomial fit is then used to find distances to each star based on their $\phi_1$ values (see the top left panel of \ref{fig:tracks}).
\end{enumerate}
Finally, we make a further cut by discarding the 13 stars that are not part of the central clump in $L_z$ - $L_{\bot}$ and $\mu_{\alpha}$ - $\mu_{\delta}$ space (see Figure~\ref{fig:cut}).

\subsection{Orphan Stream data}
\label{ap:orphan}

We draw from two literature sources to build our Orphan stream members list. 

\cite{Koposov2019} identify 109 likely Orphan stream members amongst the RR Lyrae stars in Gaia DR2, which are listed alongside their heliocentric distances in Table 5. The stream track distances defined according to the distribution of the RR Lyrae members are  given in Table C2. In addition, the stream radial velocity track (given in the Galactic standard of rest) derived using likely Orphan stream members found in the SDSS data is presented in Table 3. We collect the 6D phase space information for these 109 members in the following way:
\begin{enumerate}
\item We use TOPCAT to cross-match the stars in Table 5 of \cite{Koposov2019} with the Gaia DR2 catalogue, using a 1 arcsec search radius.
\item We fit a polynomial of degree 2 to the radial velocity track information from Table 3 in \cite{Koposov2019} using the inverted measurement uncertainties as weights. 
\item The transformation matrix in Appendix B of \cite{Koposov2019} is used to find $\phi_1$ for all stars.
\item There are no radial velocity track measurements in the negative $\phi_1$ part of the stream, leading the fit in that region to be unreliable. Therefore, we neglect the 52 stars with $\phi_1 < 0$.
\item The polynomial function is used to find a radial velocity estimate for the remaining 57 stars based on their $\phi_1$ values (see the bottom right panel of \ref{fig:tracks}).
\item Finally, since the fit to the radial velocities is done in the Galactic standard of rest frame, we transform the $v_{gsr}$ assigned to the stream members back to the heliocentric rest frame using the solar reflex motion that \cite{Koposov2019} adopted for their transformation. This is necessary to maintain a uniform transformation of all stars from heliocentric to Galactocentric Cartesian coordinates across all samples, using consistent values for the solar position and velocity.
\end{enumerate}

\cite{Li2017} present 139 Orphan stream members with either LAMOST or SDSS radial velocities. In this work we use only the 82 highest confidence members, i.e. candidates with confidence level 1. We find the full phase space map for these stars using the following steps:
\begin{enumerate}
\item We use TOPCAT to cross-match the SDSS or LAMOST right ascension and declination provided for the stars in \cite{Li2017} with the Gaia DR2 catalogue, again using a 1 arcsec search radius.
\item As for GD-1, LAMOST radial velocities are corrected by adding $4.5\ \mathrm{km\ s^{-1}}$ to the quoted values.
\item A polynomial of degree 4 is fitted to the points defining the heliocentric distance track in Table C2 of \cite{Koposov2019}. We assume the measurement errors of the RR Lyrae distances from Gaia DR2 to be of order 5\% and use the inverted errors as weights in the fit. 
\item The transformation matrix in Appendix B of \cite{Koposov2019} is used to find $\phi_1$ for all stars.
\item The polynomial function is used to find a distance estimate for each star based on their $\phi_1$ values (see the top right panel of \ref{fig:tracks})
\end{enumerate}
A cross-match between the two data sets using Gaia DR2 source identifiers reveals 2 common stars: for these 2 stars we use distances from \cite{Koposov2019} and radial velocities from \cite{Li2017}.

In summary, our data set consists of 2 stars with both radial velocity and distance measurements, 80 stars with radial velocity measurements and fitted distances, 55 stars with distance measurements and fitted radial velocities. The fitted estimates are consistent with the spread of the measurements both in the case of distances and radial velocities, as can be seen in Figure~\ref{fig:tracks}. Lastly, a cut in in $L_z$ - $L_{\bot}$ and $\mu_{\alpha}$ - $\mu_{\delta}$ space is performed to discard outliers. Our final Orphan sample thus consists of 117 stars (see Figure~\ref{fig:cut}).

\subsection{Palomar 5 stream data}
\label{ap:pal5}

We use 2 literature sources to create a list of Palomar 5 stream members with a full 6D phase space map.

\cite{PW2019} find 27 Palomar 5 stream members in the sample of stars that appear both in the PanSTARRS-1 catalog of RR Lyrae stars \citep{Sesar2017} and the RR Lyrae catalogs of Gaia DR2 \citep{Holl2018}, presented with derived heliocentric distances in Table 2 of \cite{PW2019}. We build a full 6D phase space map for these 27 members using the following steps:
\begin{enumerate}
\item We use the Gaia DR2 source identifiers in Table 2 of \cite{PW2019} to determine Gaia DR2 positions and proper motions for all stars.
\item We calculate the stream-aligned longitude coordinate $\phi_1$ using the transformation matrix provided in \cite{PW2017}, applied to the Gaia DR2 right ascension and declination, for all stars.
\item The radial velocity track is created from measurements of individual Pal 5 stream members in Table 2 of \cite{Ibata2017}. We begin by cross-matching the table with the Gaia DR2 catalogue in TOPCAT, using the right ascension and declination with a 1 arcsec search radius. We then transform to stream-aligned coordinates using the rotation matrices provided in \cite{PW2017}, and select only the 115 stars that within $\pm 15 \mathrm{km\ s}^{-1}$ of $-55.30\ \mathrm{km\ s}^{-1}$, guided by the fit performed by \cite{Ibata2017}.
\item A line is fitted to the radial velocities of the retained stars. We add the uncertainties of the measurements and the membership probability in quadrature and use the inverted values as weights in the fit.
\item The polynomial function is used to find a radial velocity estimate for each star based on its $\phi_1$ value (see the bottom middle panel of \ref{fig:tracks}).
\end{enumerate}

\cite{Ibata2017} present a sample of 154 members of the Palomar 5 stream alongside their radial velocity measurements in their Table 2. We find the full 6D phase space for these stars in the following way:
\begin{enumerate}
\item A cross-match between the stars in Table 2 of \cite{Ibata2017} and the Gaia DR2 catalogue is performed in TOPCAT using the right ascension and declination with a 1 arcsec search radius.
We find that not all \cite{Ibata2017} stars cross-match to a unique Gaia star: some Gaia stars are the best match for two (or in one case even three) \cite{Ibata2017} stars. If possible, in each pair we select the star that has a smaller angular distance to their Gaia match, and discard the other star. In the cases where both stars in the pair have the same angular distance to the Gaia star, we select one or the other, randomly.

\item Using the transformation matrix provided in \cite{PW2017}, we calculate the stream-aligned longitude coordinate $\phi_1$ using Gaia DR2 right ascension and declination for all stars.
\item The distance track is created from Table 2 in \cite{PW2019} by cross-matching the table with the Gaia DR2 catalogue using Gaia DR2 source identifiers and transforming to stream-aligned coordinates using the rotation matrices provided in \cite{PW2017}.
\item A polynomial of degree 2 is fitted to these distances. The measurement errors on the distances are of order 3\% \citep{Sesar2017}. We add these uncertainties and the membership probabilities in quadrature and use the inverted values as weights in the fit.
\item The polynomial function is used to find a distance estimate for each star based on their $\phi_1$ values (Figure \ref{fig:tracks}, top middle panel).
\end{enumerate}

\begin{figure}
\centering
\includegraphics[width=1.\linewidth,trim={0 0 0 10}, clip]{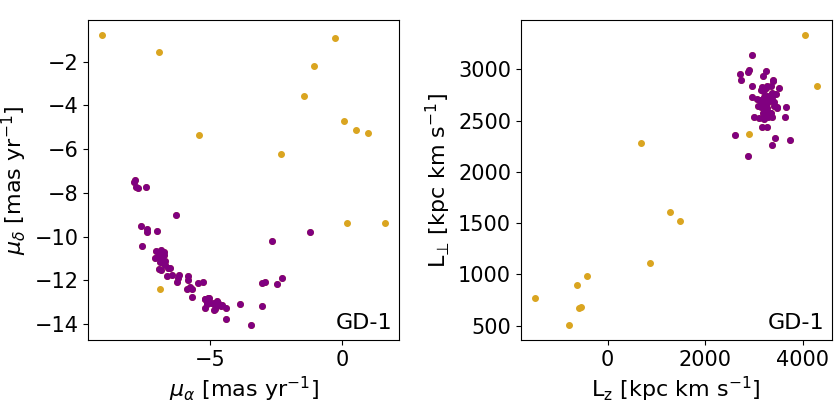}\\
\includegraphics[width=1.\linewidth,trim={0 0 0 5}, clip]{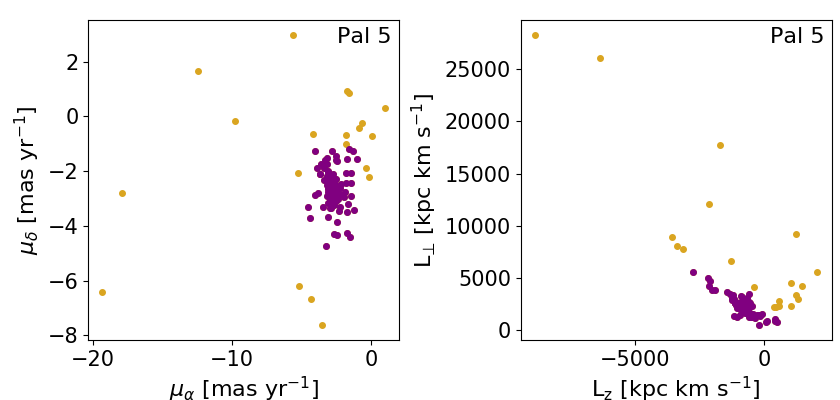}\\
\includegraphics[width=1.\linewidth,trim={0 0 0 5}, clip]{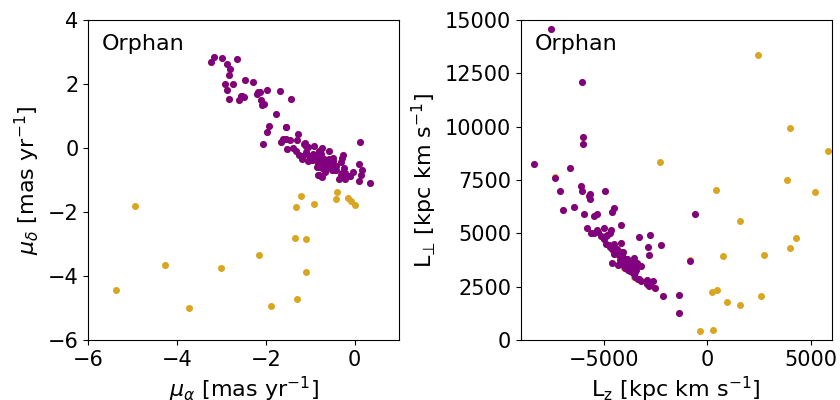}
\caption{Cuts performed to reach the final cleaned sample for GD-1, Pal 5 and Orphan streams: discarded stars are shown in yellow and the final sample is shown in purple. {\bf Left panels:} stars in  $\mu_{\alpha}$ - $\mu_{\delta}$ space. {\bf Right panels:} stars in $L_z$ - $L_{\bot}$ space.}
\label{fig:cut}
\end{figure}

After this procedure, the two data sets are  joined (there are no common stars in the two samples). Finally, we discard outliers in $L_z$ - $L_{\bot}$ and $\mu_{\alpha}$ - $\mu_{\delta}$ space. This cut reduces our final Palomar 5 sample to 136 stars (see Figure~\ref{fig:cut}).

This set of 136 stars contains 10 \cite{Ibata2017} stars that originally had a duplicate with the same angular distance to their matched Gaia star, as explained above. Although we use only one set of these duplicates in our work, we find that if we use the alternate set of 10 stars instead, our results for Pal 5 would remain virtually unchanged: for the single-component model the best-fit value changes 6\% while for the two-component model there is no change to the best-fit value. In both cases the range of values that the $1\sigma$ region encompasses remains unchanged. 

\section{KLD1 example}
\begin{figure*}
\centering
\includegraphics[width=1\linewidth,trim={0 0 0 15}, clip]{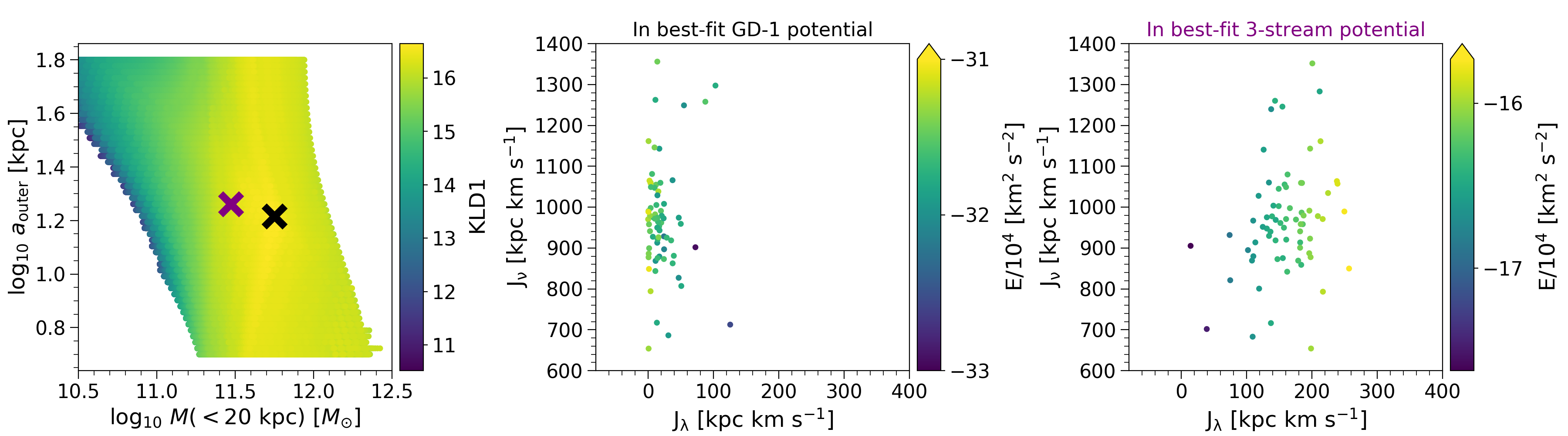}\\
\includegraphics[width=1\linewidth]{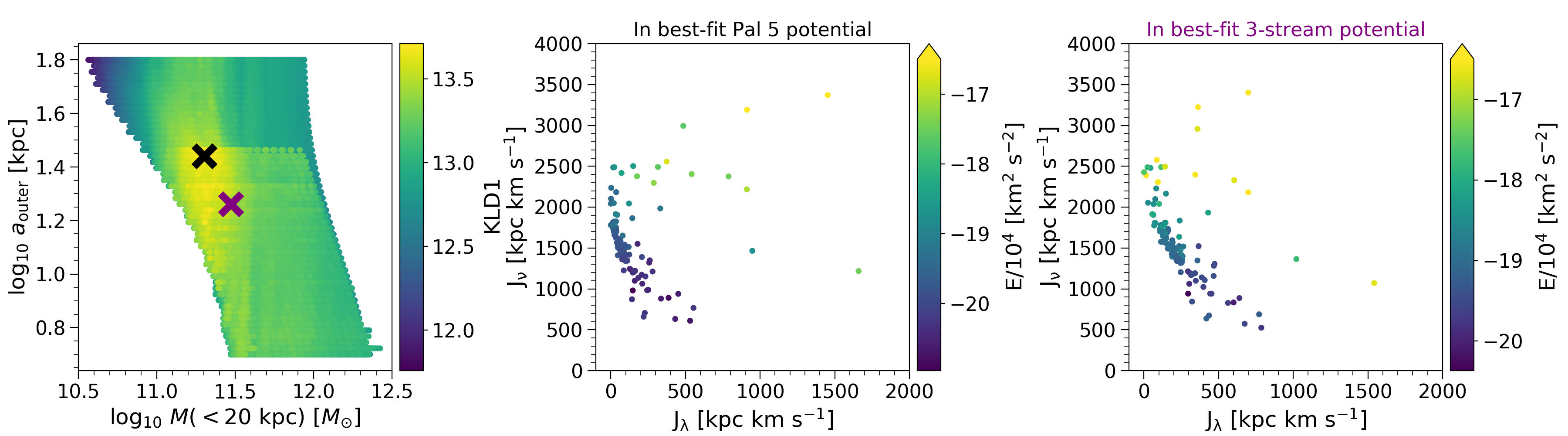}\\
\includegraphics[width=1\linewidth]{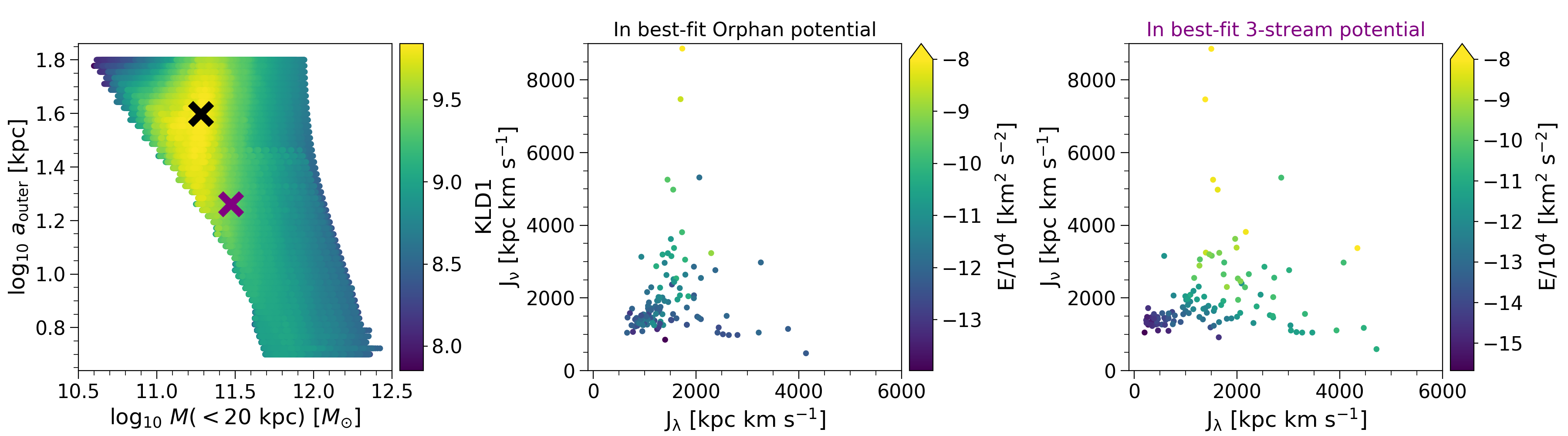}
\caption{Comparison of action space of GD-1 (top panels), Pal 5 (middle panels) and Orphan (bottom panels) produced by two different two-component potentials. The action of stars are coloured based on their energies. {\bf Left:} KLD1 values of corresponding individual stream analysis given as a function of enclosed mass and scale length of the outer component. The higher the KLD1 value the more clustered the action-space.  {\bf Center:} action distribution of the best-fit individual-stream potentials (black cross in left panel). {\bf Right:} action distribution of the same stream in the best-fit three-stream potential (purple cross in left panel). The KLD1 value of the purple cross potential in a single-stream analysis is reduced by 0.57 (for GD-1), 0.35 (for Pal 5) and 0.52 (for Orphan) compared to the KLD1 value of the black cross potential.}
\label{fig:actions}
\end{figure*}

Here, we briefly discuss the determination of the best fit values using the KLD1 (Equations \ref{eq:kld1} and \ref{eq:kld1_w}). Figure~\ref{fig:actions} shows a comparison of the action space of individual-stream best-fit potential and the three-stream best-fit potential. In the left panels we show the KLD1 contours as a function of enclosed mass and scale length of the outer component, for the GD-1, Pal 5 and Orphan streams in the two-component model. The best-fit model is marked with the black cross and the location of the three-stream best-fit potential is marked with a purple cross for comparison. In the central panels, we show the action distribution corresponding to the black cross potential and in the right panels, we show the action distribution corresponding to the purple cross potential. Finally, the stars in action space are coloured based on their energies.  

Table~\ref{tab:results_kld1} summarises the KLD1 values of the best-fit results for all streams and stream combinations. Since these values are per star, they also serve as a measure of how intrinsically clustered each stream's stars are (in the individual fits) as well as how clustered the total distribution is (for the consensus fits). We remind the reader that the higher the KLD1 value, the greater the clustering in action space.

For the individual fits, GD-1 and Pal 5 can achieve the tightest clustering across all models, while the Helmi and Orphan streams are less clustered. GD-1 and Pal 5 are thought to originate from globular clusters (in Pal 5's case the progenitor is known), while Orphan and Helmi are more likely disrupted satellite galaxies. The tighter clustering of the globular cluster streams compared to the satellite streams is consistent with the smaller total phase-space volume of globular clusters compared with satellite galaxies. The Orphan Stream is substantially less clustered than Helmi according to this measure, but this may be partially due to the localized nature of the sample of Helmi stream stars, which likely do not fully sample the phase-space volume occupied by its progenitor.

Differences between the KLD1 of consensus fits and the KLD1 for each individual stream provide some indication of the degree of tension between the best-fit potentials preferred by each stream individually. This is another way of understanding the trade off between the precision of the individual stream fits and the improved accuracy of the combined fits. 

When combining GD-1, Orphan and Pal 5, this tension reduces the information content per star compared to the maximum clustering of both Pal 5 and GD-1, but increases it compared to Orphan. However, for the combined fit from all streams, there is more tension: the individual KLD1 values for three of the streams are much higher than the combination value, but are likely offset in this somewhat by the intrinsically less clustered Orphan Stream.

Finally, the difference between the best-fit KLD1 for different potentials used for the consensus fits indicates how much more (or less) clustering per star is achieved by a change to the model, allowing us to do model comparison. The \citet{akaike1974} Information Criterion is based on this concept, although it is usually expressed in terms of a log-likelihood. Moving from the one- to two-component model produces little to no increase (and sometimes a decrease) for each individual stream and for the two-stream combined model, underlining our finding that most of the additional parameters are not well constrained.

\label{app:B}

\begin{table}
\def\arraystretch{1.4}%
\begin{tabular}{ r c c}
Stream & 1-comp & 2-comp \\
\hline
GD-1 &  17.37 &  16.64\\
Helmi &  13.43 & 13.43\\
Pal 5 & 14.81 & 13.71\\
Orphan & 9.85 & 9.84\\
GD-1/Orphan/Pal 5 standard & 11.83 & -\\
GD-1/Orphan/Pal 5 weighted & 11.95 & 11.72\\
GD-1/Helmi/Orphan/Pal 5 standard & 11.57 & -\\
GD-1/Helmi/Orphan/Pal 5 weighted & 10.99 & 11.06\\
\hline
\end{tabular}
\caption{KLD1 values for the best-fit results from individual and combined streams for single-component (1-comp) and two-component (2-comp) potentials.}
\label{tab:results_kld1}
\end{table}




\nocite{Posti2019}
\nocite{Watkins2019}
\nocite{Sohn2018}
\nocite{Gibbons2014}
\nocite{Vasiliev2019a}

\bibliographystyle{mnras}
\bibliography{streams} 

\bsp	
\label{lastpage}
\end{document}